# Terahertz spin dynamics in rare-earth orthoferrites

**Xinwei Li,**[a,b] **Dasom Kim,**[b] **Yincheng Liu,**[a] **and Junichiro Kono**[b,c,d,*]
[a]California Institute of Technology, Department of Physics, 1200 E California Blvd, Pasadena, USA, 91125
[b]Rice University, Department of Electrical and Computer Engineering, 6100 Main St, Houston, USA, 77005
[c]Rice University, Department of Physics and Astronomy, 6100 Main St, Houston, USA, 77005
[d]Rice University, Department of Materials Science and NanoEngineering, 6100 Main St, Houston, USA, 77005

**Abstract**. Recent interest in developing fast spintronic devices and laser-controllable magnetic solids has sparked tremendous experimental and theoretical efforts to understand and manipulate ultrafast dynamics in materials. Studies of spin dynamics in the terahertz (THz) frequency range are particularly important for elucidating microscopic pathways toward novel device functionalities. Here, we review THz phenomena related to spin dynamics in rare-earth orthoferrites, a class of materials promising for antiferromagnetic spintronics. We expand this topic into a description of four key elements. (1) We start by describing THz spectroscopy of spin excitations for probing magnetic phase transitions in thermal equilibrium. While acoustic magnons are useful indicators of spin reorientation transitions, electromagnons that arise from dynamic magnetoelectric couplings serve as a signature of inversion-symmetry-breaking phases at low temperatures. (2) We then review the strong laser driving scenario, where the system is excited far from equilibrium and thereby subject to modifications to the free energy landscape. Microscopic pathways for ultrafast laser manipulation of magnetic order are discussed. (3) Furthermore, we review a variety of protocols to manipulate coherent THz magnons in time and space, which are useful capabilities for antiferromagnetic spintronic applications. (4) Finally, new insights on the connection between dynamic magnetic coupling in condensed matter and the Dicke superradiant phase transition in quantum optics are provided. By presenting a review on an array of THz spin phenomena occurring in a single class of materials, we hope to trigger interdisciplinary efforts that actively seek connections between subfields of spintronics, which will facilitate the invention of new protocols of active spin control and quantum phase engineering.

**\*Corresponding Author**：Junichiro Kono, E-mail: kono@rice.edu

# 1 Introduction

Physicists have long known that electron spins are the central player of magnetism in solid-state matter. Within a host crystal lattice, spins spontaneously align their orientations to minimize the global free energy, creating a variety of magnetic phases with distinct spin configurations. Understanding the energetics is therefore a key to understanding the properties of quantum magnets in equilibrium and has been the central goal of study since early years. Starting from the beginning of this century, however, the field of modern condensed matter physics has witnessed an explosive growth of interest in spin *dynamics*, that is, the evolution of spins in time in magnetic solids. Such interest comes with the growing awareness of the important roles spin kinetics play in three major contemporary problems (Figure 1) that are significant not only for the fundamental physics of quantum materials but also for practical technological advances.

On the one hand, there is a strong motivation for producing devices that can process information with clock rates that are faster than the current electronic technology by orders of magnitude. Following Moore's law, the compactness of circuit boards has been increased to an unprecedented level, but the information processing speed, that is, the clock rates of computer chips, has remained on the GHz scale for many years.[1] The goal of spintronics research is to harness spin dynamics, whose natural timescale is picoseconds (ps) in antiferromagnets, to build processing units that are not only significantly faster in clock rate but also robust against charge perturbations, low in energy consumption, and smaller in device footprints.[2–5] What further empowers such a vision is the prospect of wave-based computation;[6–10] by using collective spin resonances (i.e., magnons), one can encode, convert, and transport information, and perform logical operations based on coherent wave interference and nonlinear wave interactions.

On the other hand, this is related to a more fundamental venue in condensed matter physics, with the aim to discover and understand exotic phases in quantum materials.[11] Although each contemporary condensed matter problem can be complex in its own way,[12–14] it is not uncommon to find spin dynamics playing a key role in conceptually significant model systems; examples include the Hubbard model and the Kondo model for correlated systems that exhibit high-$T_c$ superconductivity[15] or quantum criticality,[16] and the Ising models[17] (of various flavors) and the Kitaev model[18] for assorted ordered or disordered quantum magnets. Clarifying spin excitations in these complex material classes by spectroscopic means provides a clue for the essential microscopic processes at play.

Opportunities for studying out-of-equilibrium dynamics arise when pulsed laser sources are used to drive quantum magnets impulsively. The interest is fueled by the emergence of the ultrafast laser technology,[19,20] which provides various experimental settings with time resolutions fine enough to resolve the motions of spins *in situ*. Broadly speaking, the technique is able to selectively drive a resonance mode and track in real time the subsequent decay pathways of various degrees of freedom (DOF) back to equilibrium.[21–30] It reveals the fundamental timescales of spin motion and their interaction with the charge and lattice DOF. When the excitation is strong enough to shift the balance of competing energy scales so as to modify the free-energy landscape, a light-induced magnetic phase transition can occur,[31] bringing the grand pursuit of nonequilibrium control of quantum systems,[32] a topic that has long prevailed in atomic, molecular, and optical physics, to a solid-state context. Phenomena that are impossible to be found in thermal equilibrium can thus be investigated, such as dynamical phase transitions,[33] quench dynamics,[34] and nonthermal quantum states.[35]

In recent years, the aforementioned interest in creating efficient spintronic devices, discovering exotic equilibrium orders, and out-of-equilibrium phase engineering has coincided with a few technical advancements proven crucial for their proliferation. High-intensity broadband ultrafast lasers are the ideal instruments for both spectroscopic studies in equilibrium and out-of-equilibrium engineering. One of the most important developments has been made in time-domain terahertz (THz) spectroscopy techniques,[36–38] which closes the THz technology gap, a spectral region that is scientifically important (since it is on resonance with major spin, lattice, and charge excitations) but traditionally hard to access either from the low-frequency (electronics) or high-frequency (photonics) side. In addition, improvements on computational tools have been made to cope with quantum materials with many-body correlation effects;[39–46] some can be even applied to simulate the nonequilibrium state of complex materials, providing significant predictive powers to experiments.

The confluence of interests to address the currently exciting problems (outlined in Figure 1) with the advent of the state-of-the-art experimental and theoretical approaches has led to a rapid expansion in the body of work within the area of THz spintronics and ultrafast quantum materials. The objective of this review article is to discuss selected recent studies that directly address the core subjects of THz spin dynamics, light-spin coupling, and ultrafast magnetism. To put it concretely, we hope to comprehensively review all the *THz-frequency phenomena related to spin dynamics that have been demonstrated so far in rare-earth orthoferrites ($RFeO_3$, R representing a rare-eartch element)*. Four specific subjects are chosen, which are outlined in Figure 1 as four small triangular elements that constitute the larger triangle representing motivations from a broader vision. Before introducing the scope of each section, we first clarify the logic for choosing these topics, especially the reason for confining ourselves to THz-frequency phenomena and $RFeO_3$ systems while discussing spin dynamics.

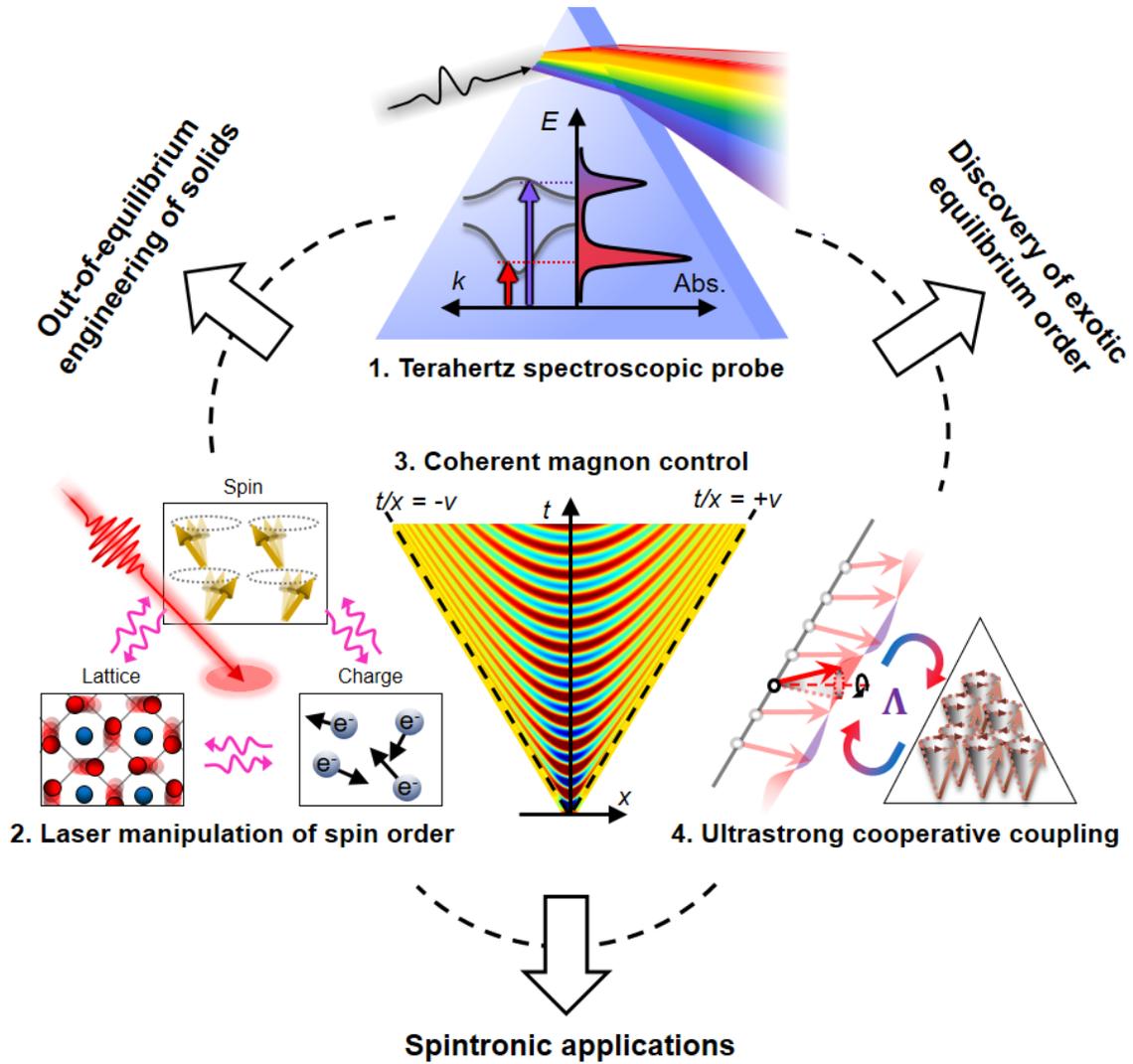

Figure 1. Overview of current scientific and technological interests related to spin dynamics in solid-state materials. Four topics arranged as smaller triangular elements are covered in this review, which are key elements for achieving the three grander goals (three sides of the larger triangle).

Firstly, our focus is on THz-frequency phenomena, or equivalently, picosecond (ps) time-scale dynamics because the THz frequency range coincides with the characteristic energy scale not only for spin–spin exchange interaction in most antiferromagnets, but also for lattice excitations and certain charge excitations within the same host material.[47,48] With the increasing capability of generating and manipulating THz radiation using the current ultrafast technology, all these DOF can be interrogated in a single experimental setting. This brings convenience to investigating dynamical coupling of spins to lattice and charge, which is the reason why the number of spintronics studies in the THz range is increasing rapidly in recent years.[49]

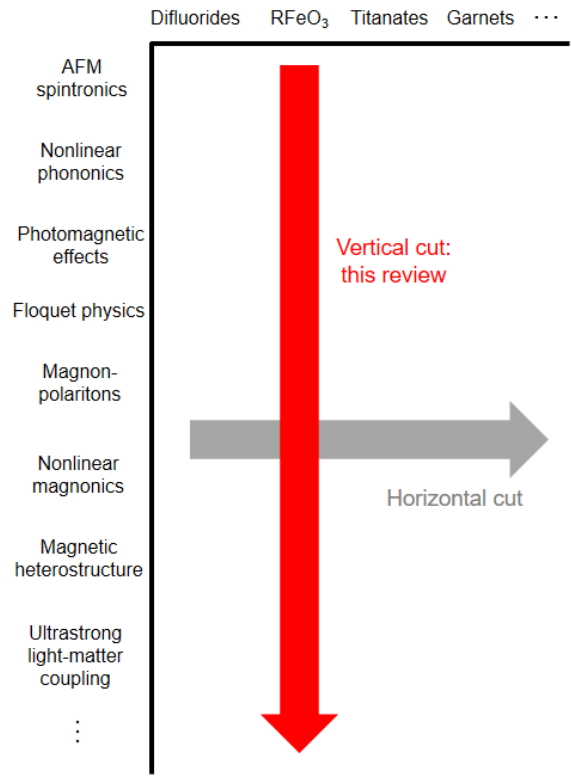

Figure 2. "Phase space" for review articles in spintronics, spanned by the horizontal axis of materials and vertical axis of novel physical phenomena. The current review represents a vertical cut in the phase space.

Secondly, this review focuses on RFeO$_3$ systems exclusively to set a paradigm of a near-exhaustive discussion of *all* THz-frequency dynamical spin phenomena observed so far in a *single* class of materials, even though the principles and phenomena therein can easily go beyond RFeO$_3$ and generalize to a broader range of material classes. As shown in Figure 2, if one defines a "phase space" for the entire research field of spin dynamics, spanned by a horizontal axis representing the array of material systems and another vertical axis listing various physical phenomena, most review articles arrange their discussion in a way similar to making horizontal cuts in the phase space. This means that each of them has a topic orienting to a particular physical phenomenon, and a variety of material systems are mentioned in which the core physics manifests. The method is certainly advantageous in elucidating a newly established physical principle to the community, but its drawback is that the discussion needs to be constantly interrupted by introducing basic properties of various material systems before their connections with the core physics can be elucidated. Our attempt here is to present a review by performing a vertical cut in the phase space, the advantage of which is the emergence of a grand picture encompassing assorted types of interesting physics that can be accessed from one material class. This way emphasizes more on the connections between various physical phenomena, and the background knowledge associated with the material needs to be presented only once. The prerequisite for our approach is that the basic properties of the target material need to be understood well enough for extensive discussions of novel phenomena to build upon, and we believe that $RFeO_3$ is the first class of compounds in the transition metal oxide family that satisfies the criterion based on the past research. Transition metal oxides form the playground of a vast collection of significant concepts in condensed matter physics,[50] including magnetism, strong electron correlations,[51] insulator–metal transitions,[14] charge density waves,[52] orbital order,[53] ferroic order,[54] high-$T_c$ superconductivity,[55] and dimensionality control. The $RFeO_3$ class is related to magnetism and ferroic order among the list but avoids many of the issues that so far remain enigmatic. As fundamental understanding of more complex materials advances, our "vertical-cut" methodology may hopefully find broader usage.

Our discussions are divided into five major sections. Section 2 introduces the background knowledge on the RFeO$_3$ system and experimental methods that have been adopted to study THz-frequency spin dynamics in them. Section 3 focuses on the weak excitation regime of the spin sector by light, namely, using light as a probe to obtain information on the magnetic structure in equilibrium, thermodynamic spin transitions, and elementary excitations that are closely related to the ground states. Section 4 discusses light-induced magnetic phase transitions, where light excitation is no longer a weak perturbation but actively modifies the free-energy landscape of the spin system. Assorted thermal and nonthermal pathways of laser control of magnetism will be examined. In Section 5, our discussions lean slightly towards the application side, which address the issue of how spin excitations in RFeO$_3$ can be useful in the emerging field of antiferromagnetic spintronics and quantum control. Coherent control, polariton devices, magnetic heterostructuring, and nonlinear spin excitations are the major topics in this context. Section 6 provides insight into an emerging interdisciplinary field that merges spintronics with a key concept in quantum optics, that is, ultrastrong light-matter coupling. Viewing dynamic spin–spin interactions from a new angle, the connection between certain magnetic phase transitions in condensed matter systems and the Dicke superradiant phase transition in quantum optics will be clarified. Finally, we end the review by giving an outlook on opportunities that lie in the future by making connections between various intriguing phenomena reviewed in this article.

Due to the broad range of topics that are covered by the current article, we also recommend interested readers to refer to recent excellent reviews, each of which addresses a topic in more depth. These include the topics of ultrafast manipulation of magnetic order,[31,56] photomagnetic effects,[57] ultrafast spectroscopy studies of quantum materials and strongly correlated systems,[58–61] nonthermal laser control of quantum materials,[62] nonlinear phononics,[63,64] Floquet engineering,[65,66] antiferromagnetic spintronics,[4,5,67–69] multidimensional THz spectroscopy,[70] cavity quantum materials,[71,72] and ultrastrong light-matter coupling.[73,74]

## 2 Background knowledge

In this section, we present the background knowledge that is required for understanding the discussions in the following sections. Two aspects will be emphasized. First, in Subsection 2.1, we describe the basic physical properties of the $RFeO_3$ materials, including their crystal and magnetic structures, symmetry classifications, phase diagrams, optical properties, and elementary excitations including phonons and magnons. Second, in Subsection 2.2, we provide an introductory survey of the most widely used ultrafast optical techniques for studying THz spin dynamics. These techniques are used for various time-resolved photomagnetic/magneto-optical experiments as well as THz time-domain spectroscopy experiments; the latter includes THz transmission spectroscopy, emission spectroscopy, and nonlinear spectroscopy using intense THz pulses for selectively driving excitations far from equilibrium.

*2.1 The $RFeO_3$ class*

The $RFeO_3$ class has been extensively studied ever since they were first identified in the 1940s; the first review paper on $RFeO_3$ appeared in 1969.[75] These crystals have an orthorhombically distorted perovskite structure with four molecular units (20 atoms) per unit cell. The lattice structure is described by the centrosymmetric space group $D_{2h}^{16}$-*Pbnm*.[*] The $Fe^{3+}$ and $R^{3+}$ ions occupy the 4(*b*) and 4(*c*) positions, respectively, and are responsible for producing magnetism.

| Bertaut's notation | $\Gamma_1$ | $\Gamma_2$ | $\Gamma_3$ | $\Gamma_4$ | $\Gamma_5$ | $\Gamma_6$ | $\Gamma_7$ | $\Gamma_8$ |
|---|---|---|---|---|---|---|---|---|
| Magnetic group | *mmm* | *mm$\underline{m}$* | *m$\underline{m}$m* | *m$\underline{m}\underline{m}$* | *$\underline{m}$mm* | *$\underline{m}$m$\underline{m}$* | *$\underline{m}\underline{m}$m* | *$\underline{m}\underline{m}\underline{m}$* |
| $Fe^{3+}$ basis | $A_xG_yC_z$ | $F_xC_yG_z$ | $C_xF_yA_z$ | $G_xA_yF_z$ | | | | |

---

[*] We do notice that a small number of papers in the literature have used the *Pnma* setting, but here we follow the main convention of using the *Pbnm* setting (equivalent upon an axis rotation).

| $R^{3+}$ basis | $c_z$ | $f_x c_y$ | $c_x f_y$ | $f_z$ | $g_x a_y$ | $a_z$ | $g_z$ | $a_x g_y$ |
|---|---|---|---|---|---|---|---|---|
| $E$ | + | + | + | + | + | + | + | + |
| $C_{2x}$ | + | + | - | - | + | + | - | - |
| $C_{2y}$ | + | - | + | - | + | - | + | - |
| $C_{2z}$ | + | - | - | + | + | - | - | + |
| $i$ | + | + | + | + | - | - | - | - |
| $i C_{2x}$ | + | + | - | - | - | - | + | + |
| $i C_{2y}$ | + | - | + | - | - | + | - | + |
| $i C_{2z}$ | + | - | - | + | - | + | + | - |
| $R$ | - | - | - | - | - | - | - | - |
| $RC_{2x}$ | - | - | + | + | - | - | + | + |
| $RC_{2y}$ | - | + | - | + | - | + | - | + |
| $RC_{2z}$ | - | + | + | - | - | + | + | - |
| $iR$ | - | - | - | - | + | + | + | + |
| $iRC_{2x}$ | - | - | + | + | + | + | - | - |
| $iRC_{2y}$ | - | + | - | + | + | - | + | - |
| $iRC_{2z}$ | - | + | + | - | + | - | - | + |

Table 1. Magnetic phases in RFeO$_3$ denoted by irreducible representations (irreps).[76] Each phase has an equivalent magnetic group using Shubnikov's conventions. Magnetic modes (basis functions) allowed in each irrep are given in vectors defined in Equations (1) and (2). Transformation properties for each irrep under all point-group symmetry elements of *mmm* are given in "+" (symmetric) and "–" (antisymmetric). *E*: identity operation. $C_{2j}$ (*j* = *x, y, z*): two-fold rotation about axis *j*. *i*: spatial inversion. *R*: time inversion.

To understand the various magnetic phases that appear in this family, it is essential to first introduce Bertaut's analysis of magnetic structures using representation theory.[77] The theory is distinct from the conventional approach of Shubnikov groups[78] to classify magnetic structures, because it does not seek the collection of operations that would leave the symmetry *invariant*. Instead, it outlines the *transformation properties* of a magnetic structure under the symmetry elements of the parent crystallographic group. This enables magnetic structures to be denoted by irreducible representations (irreps) of the crystallographic group. Regarding the space group of $D_{2h}^{16}$-*Pbnm* for RFeO$_3$, the crystallographic point group is *mmm*, which is specified by the character table to possess eight one-dimensional irreps. Eight types of magnetic configurations (or magnetic phases), labeled by $\Gamma_1-\Gamma_8$, can therefore appear. Table 1 displays the transformation properties of each phase under all 16 classical symmetry operations of the parent group,[76] where "+" ("−") indicates that the structure transforms symmetrically (antisymmetrically) under an operation. The configurations $\Gamma_1-\Gamma_8$ have one-to-one correspondence with magnetic groups under Shubnikov's classification, which are displayed in Table 1 as well.

The advantage of representation theory in classifying magnetic structures is the ease with which one can derive basis functions. These functions are written in terms of spin components that transform in the same way as an irrep, and therefore, they give the allowed magnetic modes for a magnetic phase. For the four $Fe^{3+}$ atoms located in 4($b$) sites, whose spins are labled by $S_i$ ($i = 1 - 4$), the basis functions can be well described by the following four vector components:

$$\begin{aligned} F &= S_1 + S_2 + S_3 + S_4, \\ G &= S_1 - S_2 + S_3 - S_4, \\ A &= S_1 - S_2 - S_3 + S_4, \\ C &= S_1 + S_2 - S_3 - S_4, \end{aligned} \quad (1)$$

where $F$ represents the net moment contributed by all four spins in the unit cell, and $G$, $A$, and $C$ correspond to three types of antiferromagnetic (AFM) ordering, distinguished by different ordering wavevectors. Table 1 contains information on which vector components can be basis functions of each irrep. For instance, the fact that $G_xA_yF_z$ ($F_xC_yG_z$) transforms in the same way as the $\Gamma_4$ ($\Gamma_2$) irrep indicates that the $x$ component of $G$ ($F$), the $y$ component of $A$ ($C$), and the $z$ component of $F$ ($G$) can be nonzero in the $\Gamma_4$ ($\Gamma_2$) phase.

One may notice from Table 1 that the $\Gamma_1-\Gamma_4$ phases preserve inversion symmetry, while the $\Gamma_5-\Gamma_8$ phases do not possess inversion symmetry; this can be read from the transformation properties of irreps on the row labeled by the inversion operator "$i$". While three $Fe^{3+}$ spin components can be finite for the $\Gamma_1 - \Gamma_4$ phases, there is no $Fe^{3+}$ component that can be consistent with $\Gamma_5 - \Gamma_8$. This is because the 4($b$) sites (that $Fe^{3+}$ ions occupy) are inversion centers, and the inversion operator would not be able to flip spins on these sites. Instead, $i$-broken phases have to be generated by magnetic ordering of $R^{3+}$ ions. Located on 4($c$) sites, rare earth ions may develop certain spin structures that are allowed in the $\Gamma_5 - \Gamma_8$ configurations. Defining four vectors $f$, $g$, $a$, and $c$ based on the four $R^{3+}$ spins $R_i$ ($i = 1 - 4$) in a unit cell as

$$\begin{aligned} f &= R_1 + R_2 + R_3 + R_4, \\ g &= R_1 - R_2 + R_3 - R_4, \\ a &= R_1 - R_2 - R_3 + R_4, \\ c &= R_1 + R_2 - R_3 - R_4, \end{aligned} \quad (2)$$

enables writing these components in a compact way, as shown in Table 1.

The discussions on the group theory classifications given so far are abstract, but we can now proceed to place them in a concrete context by reading the diagram of the RFeO$_3$ family, the latest version of which was summarized by Li et al.[79] in 2019. Figure 3 shows the temperature-dependent magnetic phases of RFeO$_3$ for R = La to Lu (atomic numbers: 57 – 71) and Y (atomic number: 39); no data is available for R = Pm due to its radioactivity.

For all crystals, the magnetic ordering of $Fe^{3+}$ sets in within the temperature range of 620 – 740 K to form the $\Gamma_4$ phase, featuring AFM ordering along the $x$ and $y$ axes ($G_xA_y$) and a weak net moment along the $z$ axis ($F_z$). The net moment is produced by a small spin canting (canting angle ~ 8.5 mrad) arising from the Dzyaloshinskii-Moriya (DM) interaction. For magnetically inert rare earth ions, $R^{3+}$ = $La^{3+}$, $Eu^{3+}$, $Lu^{3+}$, and $Y^{3+}$, the $\Gamma_4$ phase persists down to the lowest temperature.

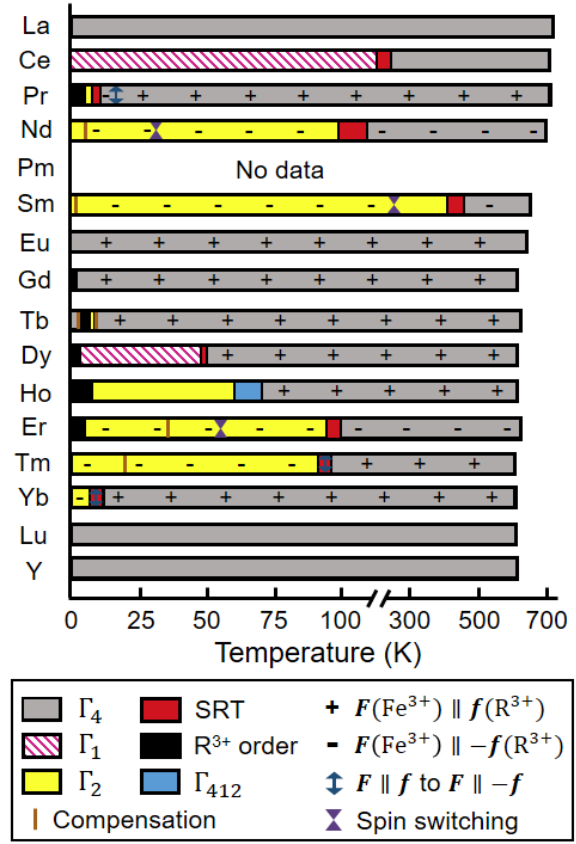

Figure 3. Temperature dependent magnetic phase diagrams for all members of the RFeO$_3$ class,[79] except for R = Pm, for which no data is available due to its radioactivity. SRT: spin reorientation transition. Reproduced with permission from [79].

Magnetic phase transitions arise, however, in crystals where the $R^{3+}$ ions are magnetically active and therefore can interact with the $Fe^{3+}$ spins.[80,81] The transition with the most frequent occurrence is the $\Gamma_4 \to \Gamma_2$ spin-reorientation transition (SRT) with decreasing temperature, during which the spin structure continuously rotates within the x–z plane while their relative angles remain rigid; the transition is of finite width in temperature, and within the intermediate temperature range, components of the $\Gamma_4$ phase ($G_xA_yF_z$) superimpose on those of the $\Gamma_2$ phase ($F_xC_yG_z$), producing the $\Gamma_{24}$ phase. Crystals with R = Pr, Nd, Sm, Er, Tm, and Yb show this transition, among which SmFeO$_3$ has a transition temperature high enough to make itself the only crystal to exhibit the $\Gamma_2$ phase at room temperature. In addition to the continuous-type $\Gamma_4 \rightarrow \Gamma_2$ SRT, there is an abrupt-type $\Gamma_4 \rightarrow \Gamma_1$ SRT observed first in DyFeO$_3$[82,83] (at 50 K) and later in CeFeO$_3$[84,85] (at 240 K). Since the magnetic modes of $A_xG_yC_z$ allowed in the low-temperature $\Gamma_1$ phase does not possess a net moment ($\boldsymbol{F} = 0$), a sharp decrease of magnetization has been observed across the transition. Moreover, HoFeO$_3$ adopts a more complicated sequence of SRTs as $\Gamma_4 \rightarrow \Gamma_{24} \rightarrow \Gamma_{12} \rightarrow \Gamma_2$,[86–89] featuring an intermediate temperature range within which two metastable configurations, $\Gamma_{24}$ and $\Gamma_{12}$, coexist.

Magnetic interactions between $R^{3+}$ and $Fe^{3+}$ not only lead to assorted SRTs of the $Fe^{3+}$ spin configuration, but also cause $R^{3+}$ to be polarized by the exchange field supplied by the $Fe^{3+}$ sublattice. When $R^{3+}$ is paramagnetic, its polarization is always consistent with the irrep that describes the $Fe^{3+}$ spin structure. For instance, in the $\Gamma_4$ and $\Gamma_2$ phases, where $Fe^{3+}$ spins have a net moment, the $R^{3+}$ spins will also develop a polarized net moment which is either parallel or antiparallel to the $Fe^{3+}$ magnetization. These are labeled "+" (parallel) and "–" (antiparallel) in Figure 3, respectively. For those crystals that have antiparallel alignments (R = Nd, Sm, Er, and Tm), the strong temperature dependence of the $R^{3+}$ moment leads to a compensation point,[90] that is, a temperature at which the $R^{3+}$ net moment exactly cancels the $Fe^{3+}$ moment.

At even lower temperatures, typically below 10 K, $R^{3+}$ develops ordering for R = Pr, Gd, Tb, Dy, Ho, and Er. This creates an interesting situation because the $R^{3+}$ order parameter does not have to be compatible with the $Fe^{3+}$ phase, leading to further symmetry breaking. For ErFeO$_3$, $Er^{3+}$ spins develop a $c_z$ mode when they order, and according to Table 1, the $c_z$ mode transforms according to the $\Gamma_1$ irrep, which is different from the $\Gamma_2$ phase of the $Fe^{3+}$ spins (before the transition occurs). The low-temperature magnetic phase of ErFeO$_3$ is therefore a $\Gamma_{12}$ phase,[91,92] whose magnetic group (2m) is an index-two subgroup of either $\Gamma_1$ (mmm) or $\Gamma_2$ (*mmm*). More interesting physics occur when $R^{3+}$ ordering transforms with an inversion-symmetry-breaking representation $\Gamma_j$ with $j \in \{5,6,7,8\}$; the resultant phase that takes both $Fe^{3+}$ and $R^{3+}$ into account would be $\Gamma_{ij}$ with $i \in \{1,2,3,4\}$ and $j \in \{5,6,7,8\}$, which is a noncentrosymmetric phase capable of exhibiting appealing properties that would otherwise be forbidden in centrosymmetric groups. The low-temperature phases of $GdFeO_3$, $DyFeO_3$, and $TbFeO_3$ fall into this category. Their details will be covered in Subsection 3.3 while discussing magneto-electric effects, multiferroicity, and electromagnons.

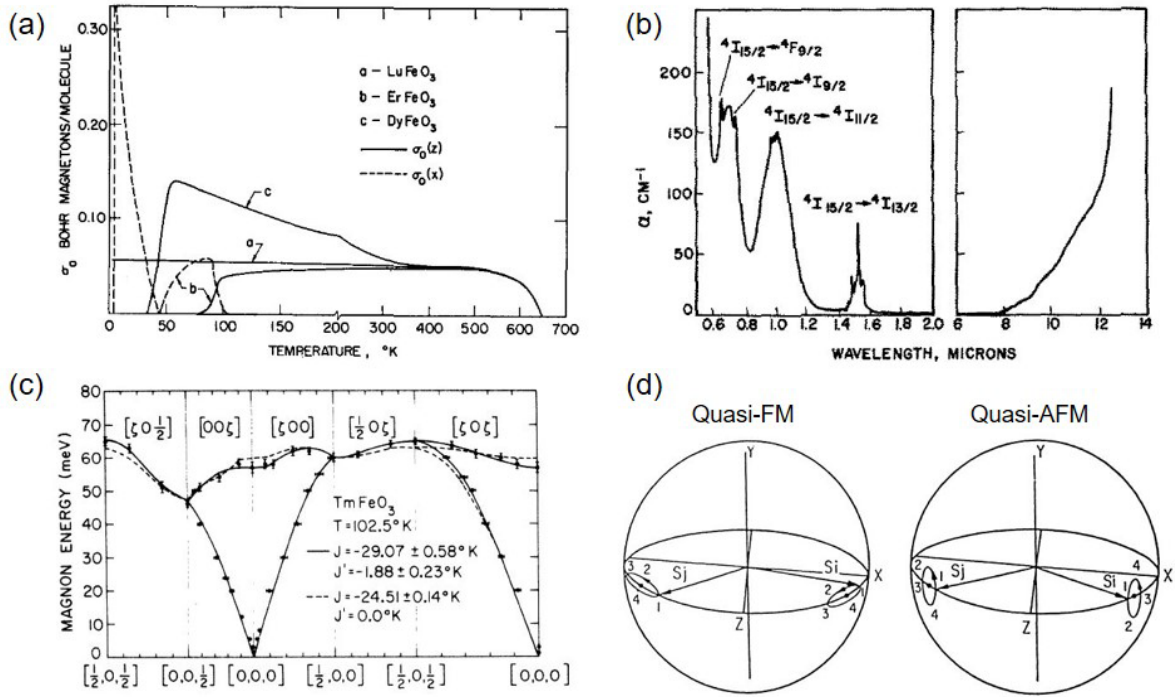

Figure 4. Basic physical properties of $RFeO_3$ crystals. (a) Temperature-dependent magnetization of $LuFeO_3$, $ErFeO_3$, and $DyFeO_3$ along the z axis and x axis.[75] (b) Optical absorption spectrum of $ErFeO_3$.[93] (c) Spin-wave dispersion mapped by inelastic neutron scattering.[94] (d) Spin oscillations in the quasi-FM and quasi-AFM modes.[95] Each spin evolves in the sequence of 1→2→3→4→1… Spins are synchronized by the number label. Reproduced with permission from [75], [93], [94], [95].

A variety of experimental methods have been applied to the RFeO$_3$ family in the last century to characterize their magnetic, electronic, and optical properties. Once a crystal is grown, x-ray scattering and Laue diffraction measurements are typically carried out first to examine the crystal quality.[96] After that, the most straightforward method to study magnetic phase transitions is magnetization measurements. Since the $\Gamma_1$, $\Gamma_2$, and $\Gamma_4$ phases are distinguished by their magnitude and orientation of the net moment, a SRT can be readily identified in magnetization curves. Figure 4(a) shows representative temperature-dependent magnetization curves for LuFeO$_3$, ErFeO$_3$, and DyFeO$_3$.[75] In LuFeO$_3$, the $\Gamma_4$ phase persists through the entire range below the Néel temperature of Fe$^{3+}$, which manifests as a finite but nearly temperature-independent $z$-axis magnetization. Upon cooling, the continuous-type $\Gamma_4 \to \Gamma_2$ SRT in ErFeO$_3$ induces a finite $x$-axis magnetization at the expense of $z$-axis magnetization between 80 K and 100 K. The compensation behavior caused by antiparallel moment alignment between Er$^{3+}$ and Fe$^{3+}$ in the $\Gamma_2$ phase manifests as a temperature point (45 K) with zero $x$-axis magnetization. In DyFeO$_3$, the $z$-axis magnetization diminishes to zero when the abrupt-type $\Gamma_4 \to \Gamma_1$ initiates, in agreement with the fact that the $\Gamma_1$ phase is incompatible with any net moment. The more complicated multistage SRT in HoFeO$_3$ does not show salient features in magnetization that can be distinguished from the SRT in ErFeO$_3$, but Young's modulus and magnetostriction measurements[86] are able to capture the detailed features. The R$^{3+}$-ordered phases typically appear as anomalies on magnetization curves at low temperatures,[82] but scattering experiments[92,97] are certainly more advantageous to precisely determine the magnetic structure.

Regarding the electronic properties, all crystals within the RFeO$_3$ family are categorized as charge-transfer insulators, rather than Mott insulators, under the framework developed by Zaanen, Sawatsky, and Allen.[98,99] This means that the charge gap is governed by the energy difference between the $p$ bands of oxygen anions and the unoccupied upper Hubbard band of $Fe^{3+}$; the assignment is also corroborated by density functional theory computations.[100] In order to describe charge excitations more clearly, Figure 4(b) shows an optical absorption spectrum for an $ErFeO_3$ crystal;[93] other crystals within the same family share much similarity. Intense absorption bands are seen on both sides of a transparent window (1.3 μm to 8 μm). On the short wavelength side, the peaks centered at 1.1 μm, 0.7 μm, and 0.5 μm arise from the $^6A_1 \rightarrow {}^4T_1$, $^6A_1 \rightarrow {}^4T_2$, and $^6A_1 \rightarrow {}^4E + {}^4A$ transitions of $Fe^{3+}$, respectively. Giant absorption is present for wavelengths shorter than 0.5 μm, and is assigned to the charge-transfer transition from oxygen to $Fe^{3+}$.[101] Within the transparent window, an absorption band centered at 1.5 μm is identified to be excitations across spin-orbit coupled atomic states of $Er^{3+}$ ($^4I_{15/2} \rightarrow {}^4I_{13/2}$). For longer wavelengths, the absorption coefficient onsets starting from 8 μm are due to infrared-active phonon excitations. Among the 60 phonon modes, 24 are Raman-active and 36 are infrared-active. Although computations reveal the eigen-modes and frequencies of all infrared-active phonons,[102] many of them remain unobserved except for the Fe-O bond stretching mode at around 18 μm.[103] On the other hand, Raman-active phonons are much better characterized experimentally,[104,105] showing excellent agreement with the theory.

Finally, we discuss magnetic excitations that can be developed on top of the magnetic ground states of $RFeO_3$, whose understanding is crucial for the major topic of this review article, i.e., THz spin dynamics. Within a unit cell that consists of four $Fe^{3+}$ and four $R^{3+}$ ions, we focus on a simple model where $R^{3+}$ is assumed to be paramagnetic. This means that magnetic excitations majorly involve $Fe^{3+}$; cooperative magnetic excitations involving both $R^{3+}$ and $Fe^{3+}$ in the low-temperature $R^{3+}$-ordered phase is, up till today, still an active topic. Shapiro et al.[94] used inelastic neutron scattering measurements to map out the dispersion relations of spin waves, and observed an optical magnon branch appearing in the energy range from 50 – 70 meV, as shown in Figure 4(c). This is the exchange mode of $Fe^{3+}$ in the four-sublattice description. Focusing on the high-energy sector, an isotropic Heisenberg Hamiltonian including both the nearest-neighbor ($J$) and next-nearest neighbor ($J'$) fits this branch well, enabling quantitative extraction of $J$ and $J'$.[106]

On the other hand, an acoustic magnon branch, which softens to a near-zero (not exact zero) frequency when approaching the zone center, also appears. In an attempt to accurately describe this branch, especially its low-energy behavior at the zone center, Herrmann[107,108] developed a two-sublattice model, where sublattices A and B arise from the original four $Fe^{3+}$ ions in a unit cell as

$$S^A = S_1 + S_3,$$
$$S^B = S_2 + S_4. \tag{3}$$

This way, the C-type AFM component in Equation (1) is canceled ($C = 0$), maintaining the G-type AFM and the net moment that can describe the SRTs. The spin Hamiltonian reads

$$H = J_{Fe} \sum_{n.n} \hat{S}_i^A \cdot \hat{S}_{i'}^B - D_y^{Fe} \sum_{n.n} (\hat{S}_{i,z}^A \hat{S}_{i',x}^B - \hat{S}_{i',z}^B \hat{S}_{i,x}^A) \\ - \sum_{s=A,B} \sum_i [A_x (\hat{S}_{i,x}^s)^2 + A_z (\hat{S}_{i,z}^s)^2 + A_{xz} \hat{S}_{i,x}^s \hat{S}_{i,z}^s], \tag{4}$$

where the first, second, and third terms are isotropic exchange interaction, DM interaction, and magnetic anisotropy, respectively. The isotropic exchange $J_{Fe}$ is the dominant energy scale, stipulating the G-type AFM structure $G = S^A - S^B$; the DM interaction provides spin canting that produces the net moment $F = S^A + S^B$; and the anisotropy energies, $A_x$ and $A_z$, specify the orientations of the $F$ and $G$ vectors with respect to the crystal axis. The equilibrium configuration determined by Equation (4) is the $\Gamma_4$ ($\Gamma_2$) phase, with $G \parallel x$, $F \parallel z$ ($G \parallel z$, $F \parallel x$), when $A_x > A_z$ ($A_z > A_x$).

The anisotropy terms are precisely the cause of gaps in the acoustic magnon branch (which is usually considered to be gapless in analogy to acoustic phonons) at the zone center in the inelastic neutron spectra. Indeed, an equation of motion calculation using Equation (4) gives two magnon modes,[107,108] named quasi-ferromagnetic (quasi-FM) and quasi-antiferromagnetic (quasi-AFM) modes, respectively, at

$$\hbar\omega_{FM} = \{24J_{Fe}S^2[2(A_x - A_z)]\}^{1/2}, \quad (5)$$
$$\hbar\omega_{AFM} = \{24J_{Fe}S[6D_{Fe}S\tan\beta + 2A_xS]\}^{1/2},$$

where $S = 5/2$ is the spin quantum number of $Fe^{3+}$, and $\beta$ is the canting angle. The spin trajectories of these modes are shown in Figure 4(d).[95] The quasi-FM mode features the precession of the $F$ vector where the relative angle between the two sublattices remains fixed. On the other hand, the quasi-AFM mode periodically modifies the canting angle, but the $F$ vector remains pointing along its equilibrium orientation, with its amplitude oscillating in time.

Equation (5) provides an important clue for the underlying driving force of a $\Gamma_4 \to \Gamma_2$ SRT, by showing that $\hbar\omega_{FM} \to 0$ when $A_x - A_z \to 0$. Across a $\Gamma_4 \to \Gamma_2$ SRT, anisotropy switches from $A_x > A_z$ to $A_z > A_x$, and the critical softening of the quasi-FM mode that evolves hand in hand with the anisotropy energy drives the phase transition. The quasi-FM and quasi-AFM modes at the zone center are excitable both by neutron scattering and Raman scattering, so their frequencies have been well studied since early days.[94,95,109–111] Softening of the quasi-FM frequency across a $\Gamma_4 \to \Gamma_2$ SRT has indeed been observed. On the contrary, the $\Gamma_4 \to \Gamma_1$ SRT in $DyFeO_3$ is accompanied by softening of the quasi-AFM mode.[95]

The quasi-FM and quasi-AFM modes are the key players in the THz spintronic studies of $RFeO_3$ materials. In addition to having finite neutron and light scattering cross-sections, group theory analysis also stipulates that they also respond linearly to THz-frequency light radiation through the magnetic-dipolar interaction.[109] Section 3 below delineates various types of THz probe of magnon modes. In the next Subsection, we introduce the major ultrafast optical techniques that enable these studies.

*2.2 Ultrafast optical techniques for probing THz spin dynamics*

The study of interaction between light and spins has a long history. In recent years, however, the advancements of ultrafast laser spectroscopy technologies have redefined the way to investigate THz-frequency spin phenomena. Reconstruction of magnetic dynamics occurring in a material system in real time can be readily implemented by a stroboscopic pump-probe experimental setup. The sample is first excited by a pump pulse, and analysis of the probe pulse that arrives at the sample at a variable time delay enables one to infer the evolution of the excited state of the sample. Since the THz frequency range has timescales on the order of ps, the fs-scale pulse width provided by current ultrafast lasers provide enough temporal resolution.

In this Subsection, we review two major types of ultrafast optical techniques for THz studies of spin dynamics. The first is time-resolved photomagnetic/magneto-optical spectroscopy, which is essentially Faraday/Kerr rotation measurements[112] using pulsed lasers (as opposed to continuous wave lasers used in earlier studies), and configure a pump-probe time delay to provide time resolution. Owing to the simplicity of envisioning its experimental setup, emphasis of discussion will be on the theoretical framework of tensor analysis that provides the predictive power for the experiments. The second is THz time-domain techniques, including THz time-domain spectroscopy, THz emission spectroscopy, and nonlinear THz spectroscopy that allows driving spins far away from equilibrium. The working principles of these methods and their advantages will be reviewed.

Probe methods based on spin-resolved photoemission[113,114] and x-ray scattering[115–117] possess advantages in ultrafast spintronics for certain compounds. However, for the RFeO$_3$ family that we focus on here, optical techniques present themselves as not only the most suitable but also the easiest approaches.

*2.2.1 Photomagnetic/magneto-optical effects*

In a material, either the development of magnetic order or the application of an external magnetic field can affect its optical properties in various ways. These processes are generally known as magneto-optical phenomena and have long been applied to the study of magnetism; notable examples are Faraday and Kerr rotations. On the other hand, intense and properly tailored light impinging on a material can directly influence the states and dynamics of spins and actively modify magnetic properties. These processes representing inverse effects of magneto-optical phenomena are generally known as photomagnetic effects. To gain microscopic insights into how photomagnetic and magneto-optic effects are categorized, and how they are related to each other, one needs to use the tensor description of light-spin interaction, as described below.

When a RFeO$_3$ crystal illuminated by light with frequency $\omega$, the total Hamiltonian is given by

$$H = H^0 + H^{MO}.$$

Here $H^0$ is the general spin Hamiltonian of RFeO$_3$, while $H^{MO}$ arises from light-spin interactions and can be expanded to lowest order in the electric field of light, $\boldsymbol{E}(\omega)$, as[57]

$$H^{MO} \approx -[\varepsilon_{ij}^0 E_i(\omega)E_j^*(\omega) + \alpha_{ijk} E_i(\omega)E_j^*(\omega)F_k(0) + \alpha'_{ijk} E_i(\omega)E_j^*(\omega)G_k(0)$$
$$+ \beta_{ijkl} E_i(\omega)E_j^*(\omega)F_k(0)F_l(0) + \beta'_{ijkl} E_i(\omega)E_j^*(\omega)G_k(0)G_l(0) + \beta''_{ijkl} E_i(\omega)E_j^*(\omega)F_k(0)G_l(0)],$$

(6)

where $\boldsymbol{F} = \boldsymbol{S}^A + \boldsymbol{S}^B$ and $\boldsymbol{G} = \boldsymbol{S}^A - \boldsymbol{S}^B$ are the FM and AFM vectors [Equations (1) and (3)], respectively, $\varepsilon_{ij}^0$, $\alpha_{ijk}$, $\alpha'_{ijk}$, $\beta_{ijkl}$, $\beta'_{ijkl}$, $\beta''_{ijkl}$ are tensor coefficients, and the Einstein summation rule is implied (and throughout this paper).

| Property tensor | Contribution to $H^{MO}$ | Magneto-optical effects | | Photomagnetic effects | |
|---|---|---|---|---|---|
| $\alpha_{ijk}$ (axial, i) | $\alpha_{ijk} E_i E_j^* F_k$ | $\varepsilon_{ij}^a = \alpha_{ijk} F_k$ | Magnetic circular birefringence (Faraday/Kerr rotation), Magnetic circular dichroism | $H_k^{PM} = \alpha_{ijk} E_i E_j^*$ | Inverse Faraday effect |
| $\alpha'_{ijk}$ (axial, i) | $\alpha'_{ijk} E_i E_j^* G_k$ | $\varepsilon_{ij}^a = \alpha'_{ijk} G_k$ | | $h_k^{PM} = \alpha'_{ijk} E_i E_j^*$ | |
| $\beta_{ijkl}$ (polar, i) | $\beta_{ijkl} E_i E_j^* F_k F_l$ | $\varepsilon_{ij}^s = \beta_{ijkl} F_k F_l$ | Magnetic linear birefringence (Cotton-Mouton effect), Magnetic linear dichroism | $H_l^{PM} = \beta_{ijkl} E_i E_j^* F_k$ | Inverse Cotton-Mouton effect |
| $\beta'_{ijkl}$ (polar, i) | $\beta'_{ijkl} E_i E_j^* G_k G_l$ | $\varepsilon_{ij}^s = \beta'_{ijkl} G_k G_l$ | | $h_l^{PM} = \beta'_{ijkl} E_i E_j^* G_k$ | |
| $\beta''_{ijkl}$ (polar, i) | $\beta''_{ijkl} E_i E_j^* F_k G_l$ | $\varepsilon_{ij}^s = \beta''_{ijkl} F_k G_l$ | | $H_k^{PM} = \beta''_{ijkl} E_i E_j^* G_l$ $h_l^{PM} = \beta''_{ijkl} E_i E_j^* F_k$ | |

Table 2. Property tensors, their contributions to the Hamiltonian, and their resulting magneto-optical and photomagnetic effects, adapted from [57]. Frequencies in the parenthesis are abbreviated. The electric field $\boldsymbol{E}$ is at frequency $\omega$, while the vectors $\boldsymbol{F}$, $\boldsymbol{G}$, and effective fields $\boldsymbol{H}^{PM}$ and $\boldsymbol{h}^{PM}$ are all at zero frequency. Reproduced with permission from [57].

The permittivity tensor, which governs the optical property, then reads

$$\varepsilon_{ij} = -\frac{\partial^2 H}{\partial E_i(\omega) \partial E_j^*(\omega)} = -\frac{\partial^2 H^{MO}}{\partial E_i(\omega) \partial E_j^*(\omega)}, \tag{7}$$

and one can see that all terms in Equation (6) except for the first term led by $\varepsilon_{ij}^0$ are related to magnetism. According to the Onsager principle, the total permittivity tensor can be separated into two parts $\varepsilon_{ij} = \varepsilon_{ij}^s + \varepsilon_{ij}^a$, where the symmetric part $\varepsilon_{ij}^s = \varepsilon_{ji}^s$ and the antisymmetric part $\varepsilon_{ij}^a = -\varepsilon_{ji}^a$ are even and odd under time reversal, respectively. Since $\boldsymbol{F}$ ($\boldsymbol{F}^2$) and $\boldsymbol{G}$ ($\boldsymbol{G}^2$) are odd (even) under time reversal, the terms that contribute to the total Hamiltonian in Equation (6) can be categorized based on whether it gives rise to either $\varepsilon_{ij}^s$ or $\varepsilon_{ij}^a$. These are given in Table 2.

The property tensors $\alpha_{ijk}$ and $\alpha'_{ijk}$, both giving rise to the asymmetric part, $\varepsilon_{ij}^a$, only contribute to the off-diagonal elements of the $\varepsilon_{ij}$ tensor ($\varepsilon_{ii}^a = 0$). This leads to the eigen-vector of light to take a circularly polarized basis in the medium, giving rise to magnetic circular birefringence, commonly observed as Faraday or Kerr rotations, and magnetic circular dichroism. The proportionality between $\varepsilon_{ij}^a$ and $\boldsymbol{F}$ (or $\boldsymbol{G}$) dictates a sign change of these effects upon spin reversal. On the other hand, $\beta_{ijkl}$, $\beta'_{ijkl}$, and $\beta''_{ijkl}$ contribute to the symmetric part $\varepsilon_{ij}^s$, which is able to take the diagonal components within $\varepsilon_{ij}$. The associated magneto-optical effects manifest in the linearly polarized basis – i.e., magnetic linear birefringence (Cotton-Mouton effect) and magnetic linear dichroism – and do not change sign upon spin reversal.

Photomagnetic effects arise from the same Hamiltonian, Equation (6), and the influence of the optical field on magnetism can be viewed as effective magnetic fields at zero frequency:

$$\boldsymbol{H}^{\text{PM}}(0) = -\frac{\partial H^{\text{MO}}}{\partial \boldsymbol{F}(0)}$$
$$\boldsymbol{h}^{\text{PM}}(0) = -\frac{\partial H^{\text{MO}}}{\partial \boldsymbol{G}(0)},$$

(8)

where $\boldsymbol{H}^{\text{PM}}(0)$ and $\boldsymbol{h}^{\text{PM}}(0)$ are the effective fields exerted on $\boldsymbol{F}$ and $\boldsymbol{G}$, respectively; Table 2 summarizes various forms of them that are associated with the five property tensors. Most notably, the effective field $H_k^{\text{PM}}(0) = \alpha_{ijk} E_i(\omega) E_j^*(\omega)$, which is associated with $\alpha_{ijk}$, the same tensor that induces the Faraday effect, can be further simplified for isotropic media. Replacing $\alpha_{ijk}$ by the minimal expression of $\alpha \cdot \varepsilon_{ijk}$, where $\alpha$ is a scalar constant and $\varepsilon_{ijk}$ is the Levi-Civita tensor (reflecting the axial nature of $\alpha_{ijk}$), we obtain the effective field as

$$\boldsymbol{H}^{\mathrm{PM}}(0) = \alpha(\boldsymbol{E}(\omega) \times \boldsymbol{E}^*(\omega)), \tag{9}$$

which resembles the form of the inverse Faraday effect (IFE), stating that circularly polarized light imposes an effective magnetic field along its propagation direction on the medium. Following the same logic, effective fields associated with the fourth-rank polar tensors $\beta_{ijkl}$, $\beta'_{ijkl}$, and $\beta''_{ijkl}$ can be generated by linearly polarized light, representing the inverse Cotton-Mouton effect (ICME).

If the fields $\boldsymbol{H}^{\mathrm{PM}}$ and $\boldsymbol{h}^{\mathrm{PM}}$ are generated by a pulsed light field through photomagnetic effects, they will be transient in nature (no longer at zero frequency) and will thus be able to excite coherent magnon oscillations in a similar way to how a THz magnetic field pulse excites magnons though the Zeeman torque. The IFE and ICME have been firmly established in RFeO$_3$ through observation of coherent magnons launched by the laser-induced effective fields. The way to distinguish photomagnetic effects from trivial thermal effects is to see if the phase of coherent magnons can be manipulated by the polarization of pump light. In the case of the IFE, altering the helicity of pump light results in sign reversal of the $\boldsymbol{H}^{\mathrm{PM}}$ pulse, and therefore, a 180-degree phase flip in coherent magnons [Figure 5(a)]. For the ICME, the magnon phase is controlled by the orientation of the linear polarization [Figure 5(b)]. Furthermore, as shown in Figure 5(c), the general framework that encompasses coherent magnon excitation by the IFE and ICME is the impulsive stimulated Raman scattering (ISRS). The scattering relies on the coexistence of two colors within the light pulse, one at $\omega$ and the other at $\omega - \Omega$, $\Omega$ being the magnon frequency. The $\omega - \Omega$ component stimulates the inelastic scattering of the $\omega$ component to occur, causing it to decompose into another photon at $\omega - \Omega$ and a magnon quantum at $\Omega$.

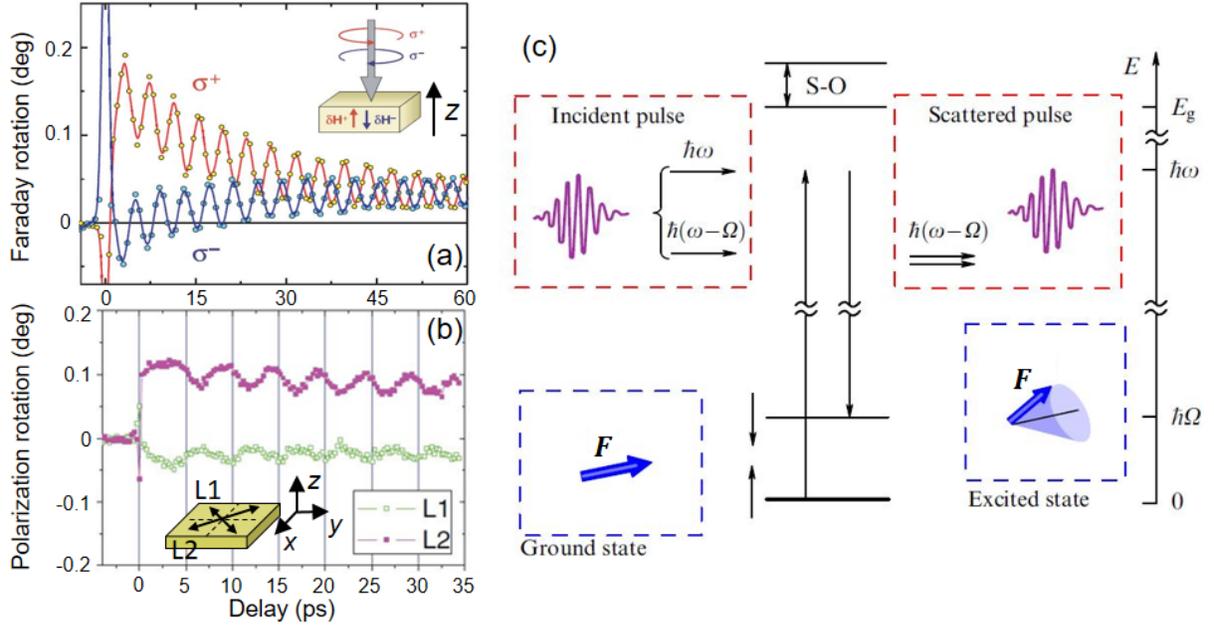

Figure 5. Observation of (a) the inverse Faraday effect[118] and (b) the inverse Cotton-Mouton effect[119] in DyFeO$_3$. The oscillation phases of magnons excited by these photomagnetic effects depend on the polarization of the pump light. (c) Magnon excitation by the two photomagnetic effects can be described by the framework of impulsive stimulated Raman scattering[57]. Reproduced with permission from [118], [119], [57].

In realistic situations, making predictions about the outcomes of pump-probe experiments is rather complex. Specifically, the type of photomagnetic effects that can occur for a particular pumping condition, and the probe configuration that can most efficiently detect the spin dynamics by magneto-optical effects, both depend critically on the material of interest. For RFeO$_3$, based on the knowledge in Table 2, Iida et al.[119] discussed a useful method to address this problem, which we schematically summarize in Figure 6.

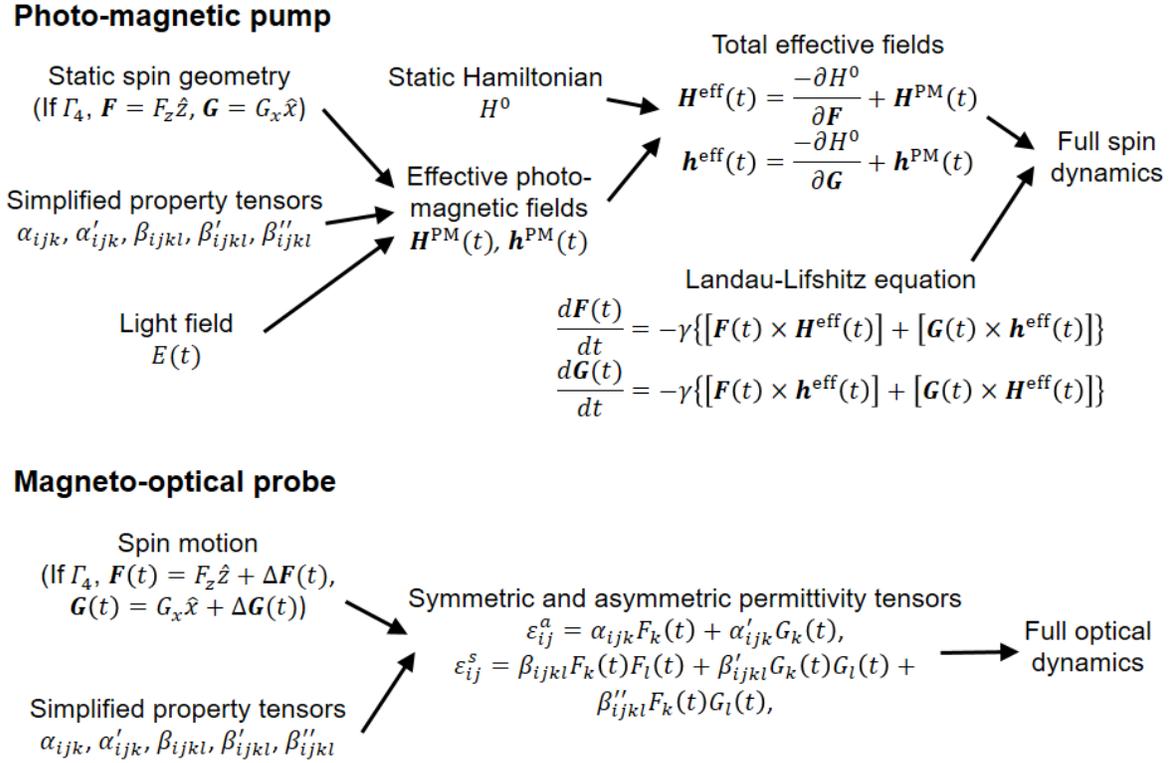

Figure 6. Path of solving for the full dynamics of photomagnetic pump, magneto-optical probe experiments on RFeO$_3$, summarized from the procedure adopted in [119].

When one attempts to predict the spin dynamics excited by a photomagnetic pump, one first calculates effective photomagnetic fields from the known static magnetic geometry, property tensors (Table 2) that are simplified by the magnetic point group symmetries, and the eigenvector of the light field. The effective photomagnetic fields, when combined with intrinsic fields derived from the unperturbed Hamiltonian $H^0$, give the total effective fields (note the time delay dependence of the effective fields due to the transient nature of the pump pulse). The total effective fields can then be used in the Landau-Lifshitz equations to solve for the full time-dependent spin dynamics. After knowing the spin dynamics, one would be interested in detecting them through a magneto-optical probe. This step is easier, as one simply needs to input the time-dependent spin dynamics and the simplified property tensors back into the equation for calculating the optical permittivity tensor (third column in Table 2). The time-dependent optical properties would then be fully determined. Overall, this method is able to provide most qualitative predictions on the pump-probe experiment based on a few symmetry arguments (without resorting to a microscopic model), and is highly efficient and useful.

*2.2.2 THz time-domain techniques*

Provided that the time-resolved photomagnetic/magneto-optical experiments majorly target at studying Raman-active magnetic excitations, one should seek for a complementary ultrafast optical technique that couples to infrared-active spin modes. THz time-domain techniques target at such functionality. They encompass a series of experimental methods, all of which utilize pulsed THz radiation to directly couple to the magnetic excitations through the magnetic-dipolar interaction. However, there is an important aspect the makes them more advantageous compared with the magneto-optical probes.

Since the very beginning of ultrafast magnetism studies, there has existed a controvercy over whether time-resolved mageto-optical signals faithfully represent magnetic dynamics.[120–122] For instance, one may design a Faraday/Kerr rotation experiment to probe $\boldsymbol{F}$ and $\boldsymbol{G}$ vectors, and the attempt would be successful in an equilibrium situation where the axial tensors $\alpha_{ijk}$ and $\alpha'_{ijk}$ have suitable nonzero components (see Table 2). However, in a nonequilibrium situation where the system is excited by a pump pulse before the arrival of the probe pulse, it is unclear whether the probed Faraday/Kerr rotation originates from the actual magnetic dynamics ($\boldsymbol{F}$ and $\boldsymbol{G}$) or from the dynamics of the tensors $\alpha_{ijk}$ and $\alpha'_{ijk}$[121,122] (which is only a trivial optical effect unrelated to magnetism). The difficulty with disentangling the two types of possible contributions has led to confusions in interpreting time-resolved magneto-optical data. However, probing spin dynamics using THz time-domain techniques circumvents such complications and therefore is more direct.[123]

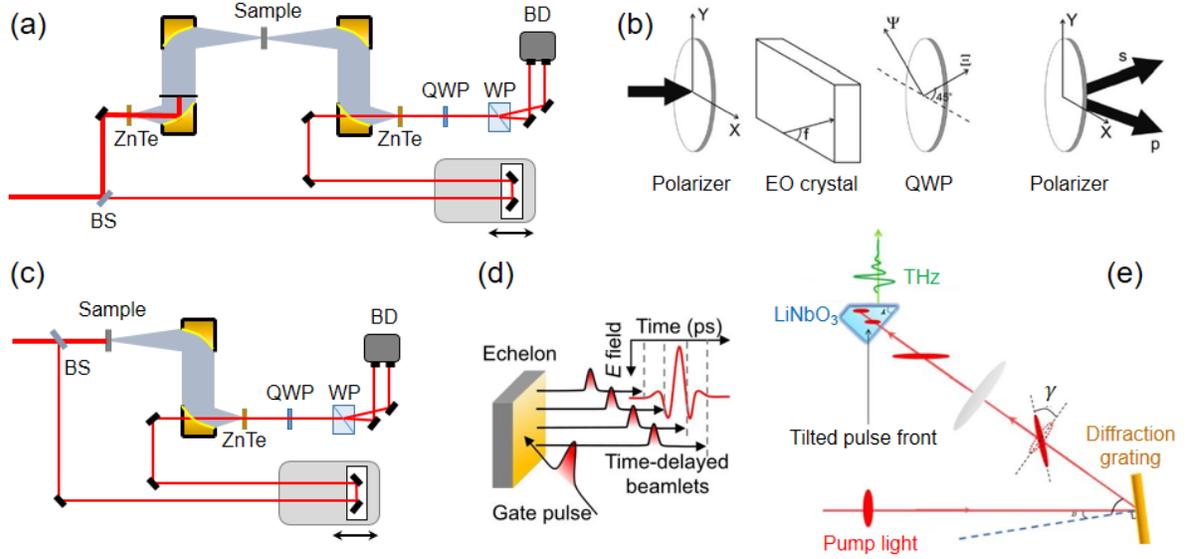

Figure 7. THz time-domain techniques. (a) Layout of a THz time-domain spectroscopy setup configured in a transmission geometry. (b) Zoom-in view of the polarization-sensitive differential detection setup.[36] (c) Layout of a THz emission spectroscopy setup. (d) Function of a reflective echelon used in a single-shot THz spectrometer.[124] (e) Pulse-front-tilt technique for generating intense THz radiation in $LiNbO_3$.[38] BS: beam splitter. QWP: quarter waveplate. WP: Wollaston prism. BD: balanced detector. (b), (d), (e) Reproduced with permission from [36], [124], [38].

THz time-domain spectroscopy is a useful technique to reveal fundamental excitations in solids.[125–138] Its most common layout is in a transmission geometry, as depicted in Figure 7(a), although a reflection geometry is also often used.[139] A pulsed laser beam is first split by a beam splitter into a strong THz generation beam and a weak probe beam. The THz generation beam, upon incident on a nonlinear crystal such as ZnTe, generates pulsed THz radiation through a second-order nonlinear process called optical rectification.[37] The nonlinear polarization

$$P_i^{(2)}(\Omega) \propto \chi_{ijk}^{(2)}(\Omega = \omega_2 - \omega_1; \omega_1, -\omega_2) E_j(\omega_1) E_k^*(\omega_2), \tag{10}$$

gives rise to a radiation field $\boldsymbol{E}_{\text{THz}}(t) \propto \frac{\partial^2 \boldsymbol{P}^{(2)}(t)}{\partial t^2}$. The phase-matching condition requires the phase velocity of the generated THz pulse match the group velocity of the generation pulse:

$$v_{\text{THz}}^{\text{ph}} = v_{\text{opt}}^{\text{gr}}. \tag{11}$$

The THz beam is then collimated and focused onto a sample using a pair of off-axis parabolic mirrors. The THz pulse interacts with the sample by resonantly coupling to its excitations. The transmitted (reflected) wave is then collected and focused onto another piece of nonlinear detection crystal. Electro-optic sampling is carried out by the time-delayed probe pulse that is directed to the same detection crystal, achieving spatial and temporal overlap with the THz pulse. When the THz pulse and the probe pulse co-propagate in the detection crystal, a polarization modulation is induced on the probe pulse, depending on the amplitude and sign of the THz electric field.[36] The modulation is analyzed by a polarization-sensitive differential detection scheme [depicted as a zoom-in view in Figure 7(b)] as a function of probe time delay, mapping out the complete THz electric field waveform in time.

Far-infrared spectroscopy techniques using continuous-wave light sources preceded the advent of THz time-domain spectroscopy and have been applied to $RFeO_3$ crystals to provide crucial insights into their magnetic properties.[87,140–142] However, the phase resolution that is unique to THz time-domain techniques enables precise extraction of Kramers-Kronig compatible real and imaginary parts of index of refraction. Further, time-domain techniques enable one to observe the free induction delay of a resonance directly in time. A THz time-domain spectrometer can be easily configured into a THz emission spectroscopy setup as in Figure 7(c). Here, the nonlinear crystal for THz generation is removed, and the THz radiation emitted from the sample itself under optical excitation is measured. Emission spectroscopy is useful for investigating quasi-FM and quasi-AFM magnon modes in RFeO$_3$ because these modes not only can be excited by optical pulses through Raman scattering, but also can couple with THz radiation through magnetic-dipole interactions.

As THz time-domain techniques gain increasing popularity in the studies of quantum materials, progress is being made currently to incorporate them with extreme experimental conditions such as low temperatures and high magnetic fields. In particular, high magnetic fields can facilitate explorations of exotic quantum properties of materials with high controllability, as has been shown recently using fields up to 30 T.[143] Since such high magnetic fields are typically supplied by pulsed magnets, whose peak field only lasts for a short duration of time (~ 1 ms), conventional electro-optic sampling of THz waveforms by a stage-scan system cannot be used.[144] Nevertheless, single-shot detection schemes have been developed,[145] which are capable of measuring the full THz waveform using just one laser pulse within the very short time window during which the sample is experiencing the peak of the pulsed magnetic field. As shown in Figure 7(d), the key to single-shot detection is to replace the probe delay stage with a reflective echelon mirror,[146] which tilts the pulse front of the optical gate beam by forming time-delayed beamlets; the time delay information is therefore encoded in the pulse-front tilt across the intensity profile of the beam. A fast CMOS camera captures the full image of the gate beam, enabling the extraction of THz waveforms at various time delays.

THz time-domain techniques also nurture the nonlinear THz spectroscopy technique,[147,148] where peak THz electric fields on the order of MV/cm can be generated, leading to an entirely new class of extreme experimental conditions for out-of-equilibrium engineering of quantum materials.[149–154] Being easily resonant with lattice phonons and magnons, strong THz radiation can selectively drive the lattice and/or spin degrees of freedom of the system far out of equilibrium without strong charge excitations (which is usually a significant source of laser-induced heating). We will see the application of such an apparatus in the study of nonlinear magnonics in Subsection 5.4.

In order to achieve higher THz electric field strengths in optical rectification, much effort has been expended on increasing the nonlinearity of the generation crystals. However, at the same time, difficulty arises with regard to maintaining the phase matching condition [Equation (11)], because of the normal dispersions of nonlinear crystals. A scheme of tilted-pulse-front optical rectification has developed to address this challenge.[155–157] As shown in Figure 7(e), the generation beam is tilted in intensity front after being diffracted by a high-efficiency grating. If the tilting angle is $\gamma$, the phase-matching condition in Equation (11) can be achieved as

$$v_{\text{THz}}^{\text{ph}} = v_{\text{opt}}^{\text{gr}} \cos \gamma. \tag{12}$$

When being implemented in nonlinear crystals such as LiNbO$_3$, this relation suggests that the propagation direction of the generated THz wave is different from that of the optical beam. Strong THz electric fields on the order of MV/cm have been generated based on this scheme.

An increasing number of novel schemes for intense THz generation are being developed currently. There are two important highlights. One is the optical rectification from organic crystals – molecular crystals with gigantic nonlinear coefficients (which can be larger by an order of magnitude than inorganic nonlinear crystals).[158–161] By tuning the generation beam to appropriate wavelengths, the phase matching condition can be achieved in the collinear geometry. Another novel scheme to note is carrier-envelope-phase stable difference-frequency generation and optical parametric amplification technology,[162] which generates phase-locked multi-THz (10~30 THz) radiation with strong intensity.[163–166] The same technique is now being expanded to the mid-infrared frequency range,[167,168] dramatically enhancing the capability of quantum control of the lattice sector of quantum materials.

## 3  Probing equilibrium properties

This Section describes various previous studies where THz spectroscopic techniques were used for probing properties of the $RFeO_3$ family in thermal equilibrium, including phase transitions and elementary excitations. There are two objectives that we wish to achieve by discussing these topics. The first objective is to give concrete examples for which the background knowledge presented in the last Section can be put into direct use. We will show how the representation theory in Subsection 2.1 can be used to understand and predict certain THz properties of $RFeO_3$ in equilibrium. The second objective is to lay the foundation for describing the out-of-equilibrium engineering efforts in later Sections. Understanding the THz-frequency signatures of thermodynamic phase transitions facilitates the identification of a similar class of transitions achieved in nonequilibrium scenarios. Furthermore, illustrating fundamental excitations on an individual basis will set the stage for investigating their mutual interactions and dynamical couplings, providing necessary insight for the discussion related to ultrastrong coupling and Dicke cooperativity in Section 6.

We will first consider the temperature- and magnetic-field-induced SRTs in the $Fe^{3+}$ subsystem, particularly the most common $\Gamma_4 \to \Gamma_2$ type transition. The unique advantage of THz time-domain spectroscopy in detecting these transitions is the ability to simultaneously probe magnons' polarization selection rule, frequency shifts, and spin trajectory, all of which are expected to develop anomalies across an SRT. We will then focus on the $R^{3+}$ ion subsystem, highlighting spectroscopic studies of transitions between their crystal-field levels. These levels, with spacings in the THz range, are the major origin of the strong temperature dependence of magnetic properties in the $RFeO_3$ family. We will consider both the non-Kramers type $Tm^{3+}$ and Kramers type $Er^{3+}$. Finally, we will turn to a type of cooperative magnetic excitation that involves both the $R^{3+}$ and $Fe^{3+}$ subsystems, called electromagnons, which only exist in a few low-temperature $R^{3+}$ ordered phases where spatial inversion is broken by magnetic order. Observation of these modes not only signals successful detection of these low-temperature $R^{3+}$ ordered phases but also provides insight into spin-mediated multiferroicity and useful protocols for switching magnetoelectric coupling on/off through magnetic control of spatial inversion symmetry.

*3.1 THz probe of temperature- and magnetic-field-induced spin reorientation transitions*

As shown in Figure 4(c), the spin-wave dispersion in $RFeO_3$ features high-frequency optical branches corresponding to the exchange mode of the four-sublattice model as well as gapped low-frequency acoustic branches whose properties closely link with magnetic anisotropy. Since SRTs in $RFeO_3$ are mainly driven by magnetic anisotropy, probing these acoustic branches provides much information on the SRTs themselves.

THz radiation couples with the zone-center modes of the acoustic branches, which are the quasi-FM and quasi-AFM magnon modes [Figure 4(d)]. These modes respond linearly to THz radiation through the magnetic-dipole interaction.[109] Microscopically, the excitation can be described by a Zeeman torque type interaction $\boldsymbol{T}_i = \gamma \boldsymbol{S}_i \times \boldsymbol{H}_{THz}$, where $\gamma$ is the gyromagnetic ratio, $\boldsymbol{H}_{THz}$ is the magnetic field component of the THz pulse, and $i$ = A or B. The sublattice spins $\boldsymbol{S}_A$ and $\boldsymbol{S}_B$ are tipped to deviate from their equilibrium orientations through the Zeeman torque, initiating precessions about their equilibrium axis. The quasi-FM and quasi-AFM modes hold distinct polarization selection rules: The quasi-FM mode is excited when $\boldsymbol{H}_{THz} \perp \boldsymbol{F} = \boldsymbol{S}_A + \boldsymbol{S}_B$, and the quasi-AFM mode is excited when $\boldsymbol{H}_{THz} \parallel \boldsymbol{F} = \boldsymbol{S}_A + \boldsymbol{S}_B$.

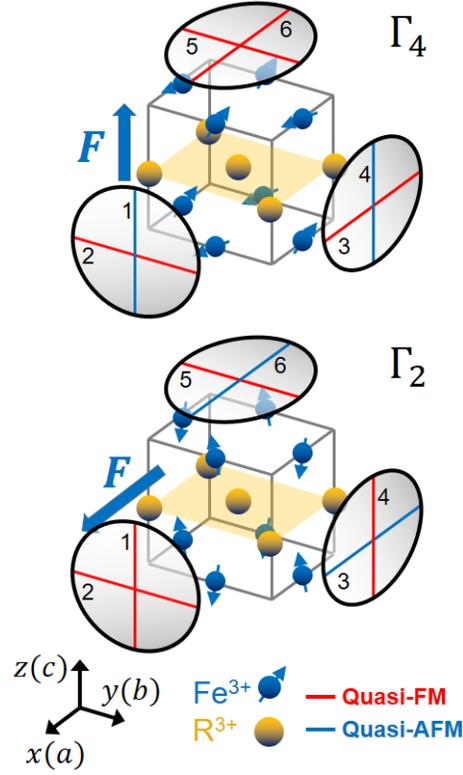

Figure 8. Six measurement configurations, along with the magnon polarization selection rule in the $\Gamma_4$ and $\Gamma_2$ phases. The magnetic field component of the THz pulse $H_{THz}$ is polarized along a line whose color indicates the magnon mode that can be excited. The THz propagation direction is perpendicular to the disk containing the line. Reproduced with permission from [169].

When an SRT occurs, three major features are expected within the THz time-domain signal of the quasi-FM and quasi-AFM modes. First, there is a change in the polarization selection rule that accompanies the direction switching of the $F$ vector. Taking the most common $\Gamma_4 \to \Gamma_2$ SRT as an example, a total of six measurement configurations are possible for linearly polarized THz field interacting with a crystal cut along a major axis; see Figure 8. For each number-labeled configuration, $H_{THz}$ is polarized along a line (marking a major crystal axis), whose color indicates the magnon mode that can be excited, and the THz pulse propagates along the out-of-plane direction of the disk that contains the colored line. For a given measurement configuration, the mode that can be excited changes across a $\Gamma_4 \to \Gamma_2$ transition; for instance, configuration 4 would excite the quasi-AFM mode in the $\Gamma_4$ phase but the quasi-FM mode in the $\Gamma_2$ phase. Second, there is a change of the polarization trajectory of the emitted field of magnons, which reflects the trajectory of spins within the eigenmode. Third, there is a shift of magnon frequency, which is a natural consequence of the fact that the quasi-FM and quasi-AFM frequencies respond sensitively to magnetic anisotropy; see Equation (5).

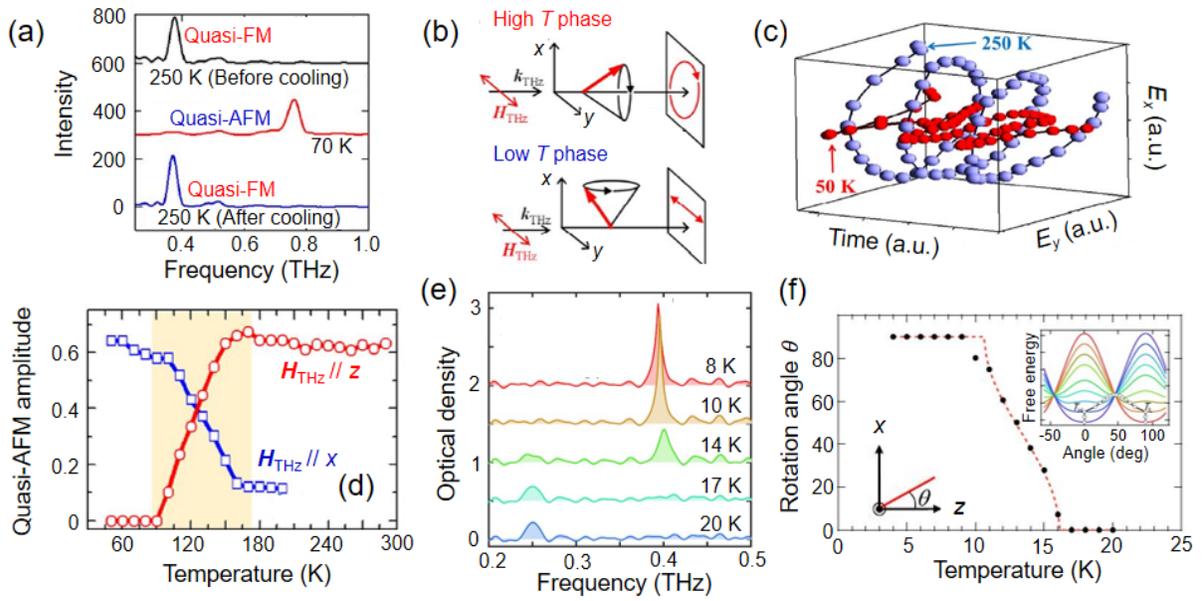

Figure 9. Temperature-dependent SRT probed by THz spectroscopy.[170–172] (a) Switching of the polarization selection rule. (b) Switching of the polarization trajectory of magnon emission. (c) Experimental verification of (b). (d) Continuous spectral weight transfer between two measurement configurations of the quasi-AFM mode amplitude. (e) and (f) Quantifying the rotation angle during the SRT from quasi-AFM to quasi-FM mode spectral weight transfer in z-cut $Dy_{0.7}Er_{0.3}FeO_3$. Inset to (f): calculated free energy landscape at various temperatures (red to blue: low to high temperatures). Reproduced with permission from [170–172].

Yamaguchi et al.[170] have studied the temperature driven $\Gamma_4 \to \Gamma_2$ SRT in $ErFeO_3$. As shown in Figure 9(a), when a THz pulse with $H_{THz} \parallel x$ transmits through a z-cut crystal, corresponding to configuration 6 in Figure 8, it excites the quasi-FM mode at 250 K ($\Gamma_4$) and switches to excite the quasi-AFM mode at 70 K ($\Gamma_2$). When the THz pulse is tailored with $\boldsymbol{H}_{\text{THz}} \parallel y$ (configuration 5 in Figure 8), although both the $\Gamma_4$ and $\Gamma_2$ phases support excitation of the quasi-FM mode, their spin precession trajectories are different [Figure 9(b)]; the oscillations project to a near-circle (elliptical to be exact) pattern to the *x-y* plane in the $\Gamma_4$ phase, but to linear polarization in the $\Gamma_2$ phase. This causes different polarization states of the free induction decay signal that arises from the magnon precession, which is indeed experimentally detected through polarization-resolved measurement of the emission [Figure 9(c)]. This experiment thus demonstrated a unique usage of the time resolution provided by THz time-domain techniques in measuring the magnon free induction decay, which reconstructs the spin trajectories.

While Yamaguchi et al.[170] probed the SRT by only studying two temperature points, one above and the other below the transition temperature, Jiang et al.[171] have investigated the SRT temperature range in detail. Within a *y*-cut NdFeO$_3$ crystal, they detected the amplitudes of the quasi-AFM mode for both the $\boldsymbol{H}_{\text{THz}} \parallel z$ configuration (configuration 4 in Figure 8) and the $\boldsymbol{H}_{\text{THz}} \parallel x$ configuration (configuration 3 in Figure 8), and plotted them as a function of temperature. Figure 9(d) shows that, within the $\Gamma_4 \to \Gamma_2$ SRT temperature range, weight transfer between the two configurations occurs in a continuous manner, pointing to the continuous nature of the SRT.

Suemoto et al.[172] have brought the analysis to a more quantitative level by calculating the rotation angle of the spin structure in the $\Gamma_{24}$ phase from the spectral weight transfer. They used a *z*-cut Dy$_{0.7}$Er$_{0.3}$FeO$_3$ crystal, and aligned $\boldsymbol{H}_{\text{THz}} \parallel x$ (configuration 6 in Figure 8). As shown in Figure 9(e), a continuous-type spectral weight transfer between the quasi-AFM and the quasi-FM mode was observed when the temperature was fine-tuned across the SRT. They were able to quantify the rotation angle as

$$\theta = \arctan(\sqrt{\frac{A_F}{A_{AF}}}), \tag{13}$$

where $A_F$ and $A_{AF}$ are the spectral amplitudes of the quasi-FM and the quasi-AFM modes, respectively. Figure 9(f) shows the temperature dependence of the rotation angle extracted by this method, which shows excellent agreement with a standard second-order phase transition theory incorporating both second-order and fourth-order anisotropy terms in the free energy.[173]

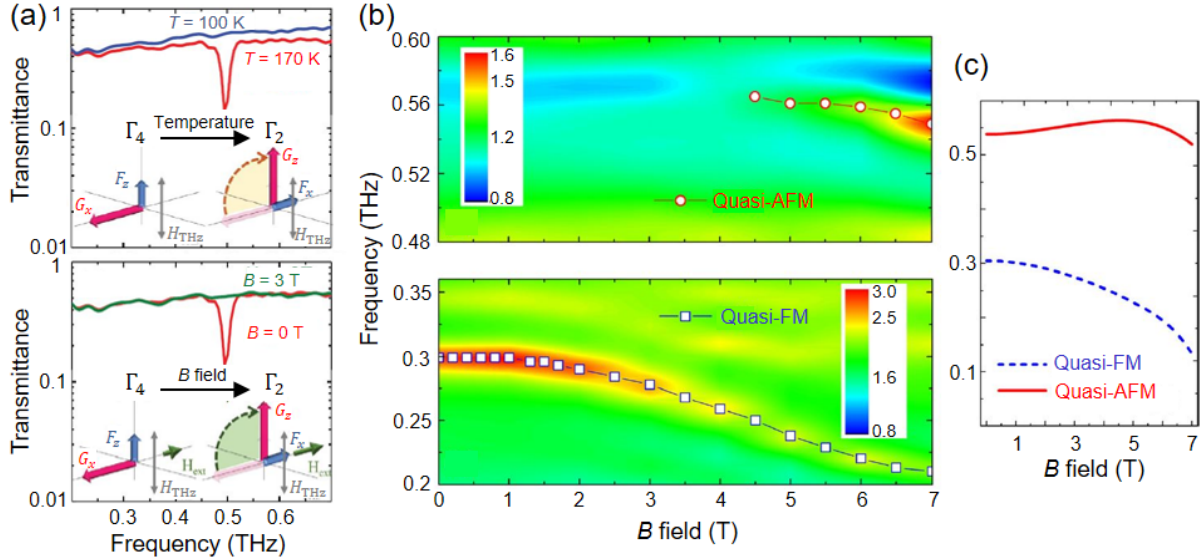

Figure 10. Magnetic-field induced $\Gamma_4 \rightarrow \Gamma_2$ SRT probed by THz spectroscopy.[174,175] (a) Observation of switching of the polarization selection rule in NdFeO$_3$. (b) Field-dependent absorption spectra in YFeO$_3$. (c) Calculated quasi-FM and quasi-AFM frequencies versus magnetic field, to compare with (b). Reproduced with permission from [174,175].

In addition to temperature-driven SRTs, THz time-domain techniques have also been used for probing SRTs driven by an external magnetic field. To perform this study, Jiang et al.[174] aligned $\boldsymbol{H}_{THz} \parallel z$ in a $y$-cut NdFeO$_3$ crystal (configuration 4 in Figure 8). They first demonstrated the polarization selection rule in the temperature-driven SRT. The quasi-AFM mode is detected in the high-temperature $\Gamma_4$ phase but not in the low temperature $\Gamma_2$ phase [Figure 10(a) top]. Then they set the system in the high-temperature $\Gamma_4$ phase, and applied a static magnetic field along the *x*-axis of the crystal. Since the ***F*** vector is prone to align parallel to the static field, a field-driven $\Gamma_4 \rightarrow \Gamma_2$ transition occurs. This again manifests in the polarization selection rule, as shown by the diminishing quasi-AFM mode in the high-field case [Figure 10(a) bottom].

Based on a similar idea, Lin *et al.*[175] have studied a field-driven $\Gamma_4 \rightarrow \Gamma_2$ transition in YFeO$_3$. Since Y$^{3+}$ is nonmagnetic, a $\Gamma_4 \rightarrow \Gamma_2$ SRT would not occur in the temperature-dependent phase diagram of YFeO$_3$ (Figure 3). In a *z*-cut YFeO$_3$ crystal, the polarization of the THz radiation is ***H***$_{THz}$ || *x* (configuration 6 in Figure 8), for which only the quasi-FM mode is expected in the zero-field $\Gamma_4$ phase. However, when a static magnetic field is applied along the *x*-axis, the spin structure will be polarized to approach a $\Gamma_4 \rightarrow \Gamma_2$ transition. Key experimental signatures, including softening of the quasi-FM frequency and increase of the spectral amplitude of the quasi-AFM mode, are observed; see Figure 10(b). Up to the highest field that is applied (7 T), the quasi-FM mode does not soften completely, suggesting that the transition to the $\Gamma_2$ phase is incomplete; this is consistent with prior studies[142,176] that observed the transition at 7.4 T. Nevertheless, a theoretical model taking into account an additional spin-field Zeeman term in the original spin Hamiltonian [Equation (4)] gives field-dependent magnon frequencies that match the experimental results very well [Figure 10(c)].

*3.2 THz probe of crystal-field transitions of rare-earth ions*

For RFeO$_3$ crystals where R$^{3+}$ ions are magnetically active, the energy levels of R$^{3+}$ are crucial for understanding the overall magnetic properties, even within the temperature range where R$^{3+}$ ions remain paramagnetic. The SRTs, compensation behavior, and spin switching are a few notable examples. R$^{3+}$ ions are typically heavy elements where the spin-orbit interaction is significant enough to form atomic states in free space with well-defined quantum number *J*. The confined nature of the 4*f* orbitals stipulates that the crystal electric field only acts as a weak perturbation (weak compared to the spin-orbit interaction) to split these spin-orbit coupled states into crystal-field states (distinguished by the magnetic quantum number *m*). However, the energy scale of crystal-field splitting is on the same order of magnitude as the thermal energy at a few tens of Kelvin. This is the fundamental reason why the magnetic properties of R$^{3+}$ are sensitive to temperature, which, in turn, leads to the rich temperature-dependent magnetic phase diagrams of the RFeO$_3$ family.

In this Subsection, we discuss the use of THz time-domain spectroscopy to study transitions between crystal-field states, known as crystal-field transitions (CFTs), of R$^{3+}$ ions in RFeO$_3$. We will describe how peaks in the absorption spectrum can be assigned to CFTs, whose evolutions with an external magnetic field can be used to quantitatively extract microscopic parameters of R$^{3+}$ sublattices, and how the polarization dependence of the emitted THz field from CFTs can be used to distinguish the dipolar nature of the transitions. While the highlighted work here all used time-domain techniques, some important early studies used continuous-wave far-infrared spectrsocpy to study similar transitions.[140,177,178]

Zhang *et al.*[179] used THz time-domain spectroscopy to study the ground multiplets ($^6$H$_3$) of Tm$^{3+}$ ions in TmFeO$_3$. Here the spin-orbit coupled atomic states are denoted using the spectroscopic notation $^{2S+1}L_J$, where $S$ and $L$ are the total spin and orbital quantum numbers, respectively, and $J = L + S$ is the total angular momentum quantum number. Using an *x*-cut TmFeO$_3$ crystal, the authors aligned $\boldsymbol{H}_{\text{THz}} \parallel z$ (configuration 1 in Figure 8) and obtained the temperature-dependent absorption spectra shown in Figure 11(a). In addition to the sharp lines highlighted by solid markers, which are assigned to the quasi-FM and quasi-AFM modes, a strong absorption band centered at 0.54 THz appears below 70 K, gaining amplitude as the temperature decreases. A constant-temperature cut at 60 K of the colormap gives the spectrum in the top panel of Figure 11(b). To study the polarization selection rule of the transition, the authors also performed measurements using $\boldsymbol{H}_{\text{THz}} \parallel y$ in an *x*-cut crystal and a *z*-cut crystal, corresponding to configurations 2 and 5 in Figure 8, respectively; see the middle and bottom panels of Figure 11(b). Another absorption peak at 1.2 THz is identified.

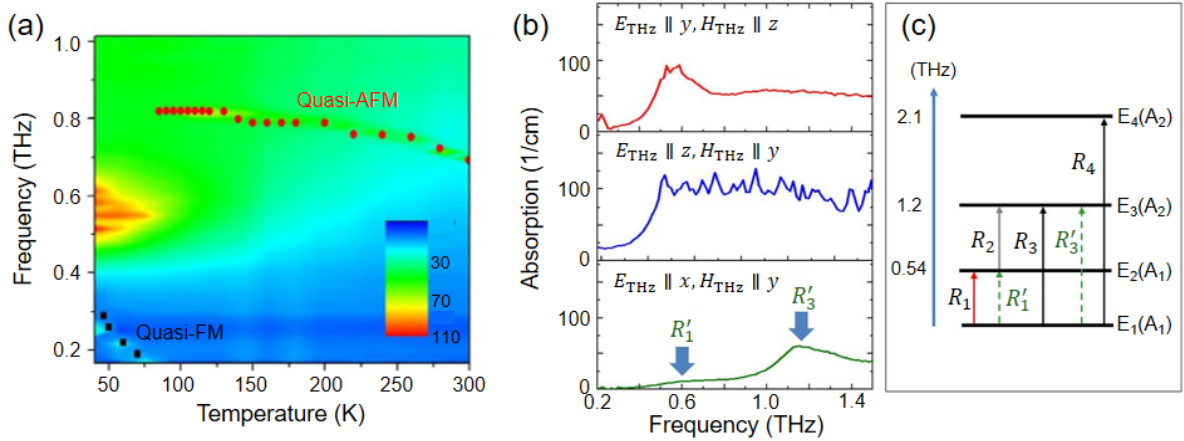

Figure 11. CFTs of Tm$^{3+}$ ions in TmFeO$_3$.[179] (a) Absorption coefficient mapped as a function of frequency and temperature. Red circles: quasi-AFM mode. Blue squares: quasi-FM mode. (b) Constant-temperature cut of absorption coefficient at 60 K for three different measurement configurations. (c) Energy level diagram explaining the THz transitions. Reproduced with permission from [179].

Figure 11(c) displays the energy level structure of Tm$^{3+}$ assigned by Zhang *et al.*[179] to explain the experimental spectra. Tm$^{3+}$, which has even filling in the 4*f* shell, is a non-Kramers-type ion. Since $Tm^{3+}$ ions occupy noncentrosymmetric sites with $C_s$ symmetry, a crystal field generally splits its $^6H_3$ states into singlets, denoted by the two one-dimensional irreps of the $C_s$ group, $A_1$ and $A_2$. The observed absorption peaks centered at 0.54 THz ($R_1$ and $R_1$' peaks) and 1.2 THz ($R_3$ and $R_3$' peaks) can be explained as the $E_1(A_1) \to E_2(A_1)$ and $E_1(A_1) \to E_3(A_2)$ transitions, respectively. The $R_2$ line, assigned to the $E_2(A_1) \to E_3(A_2)$ transition, should in principle appear in the $\boldsymbol{H}_{THz} \parallel y$, $\boldsymbol{E}_{THz} \parallel z$ configuration, but the saturated absorption observed in this case makes this line hard to detect. Further, the fact that the spectral amplitudes of the $R_1$ and $R_3$ lines increase with decreasing temperature is in agreement with the expected trend of the thermal population of the $E_1(A_1)$ ground state, corroborating the assignment of the transitions.

Li et al.[169] used THz time-domain magneto-spectroscopy experiments to study CFTs of $Er^{3+}$ in $ErFeO_3$. Peak splitting and shifting versus magnetic field (applied along different crystal axes) enabled quantitative determination of microscopic parameters of the $Er^{3+}$ ions. Figure 12(a) shows an energy-level diagram of $Er^{3+}$. Since $Er^{3+}$ has odd filling in the 4f shell, the crystal fields remove all but the Kramers degeneracy, forming Kramers doublets $|m = \pm 15/2\rangle$, $|m = \pm 13/2\rangle$, …, where $m$ is the magnetic quantum number. At zero field, time-reversal-symmetry-breaking magnetic exchange interactions between $Fe^{3+}$ and $Er^{3+}$ slightly split the Kramers doublets. By further applying a magnetic field ($B$) along different crystal axes, different patterns of Zeeman splitting occur.

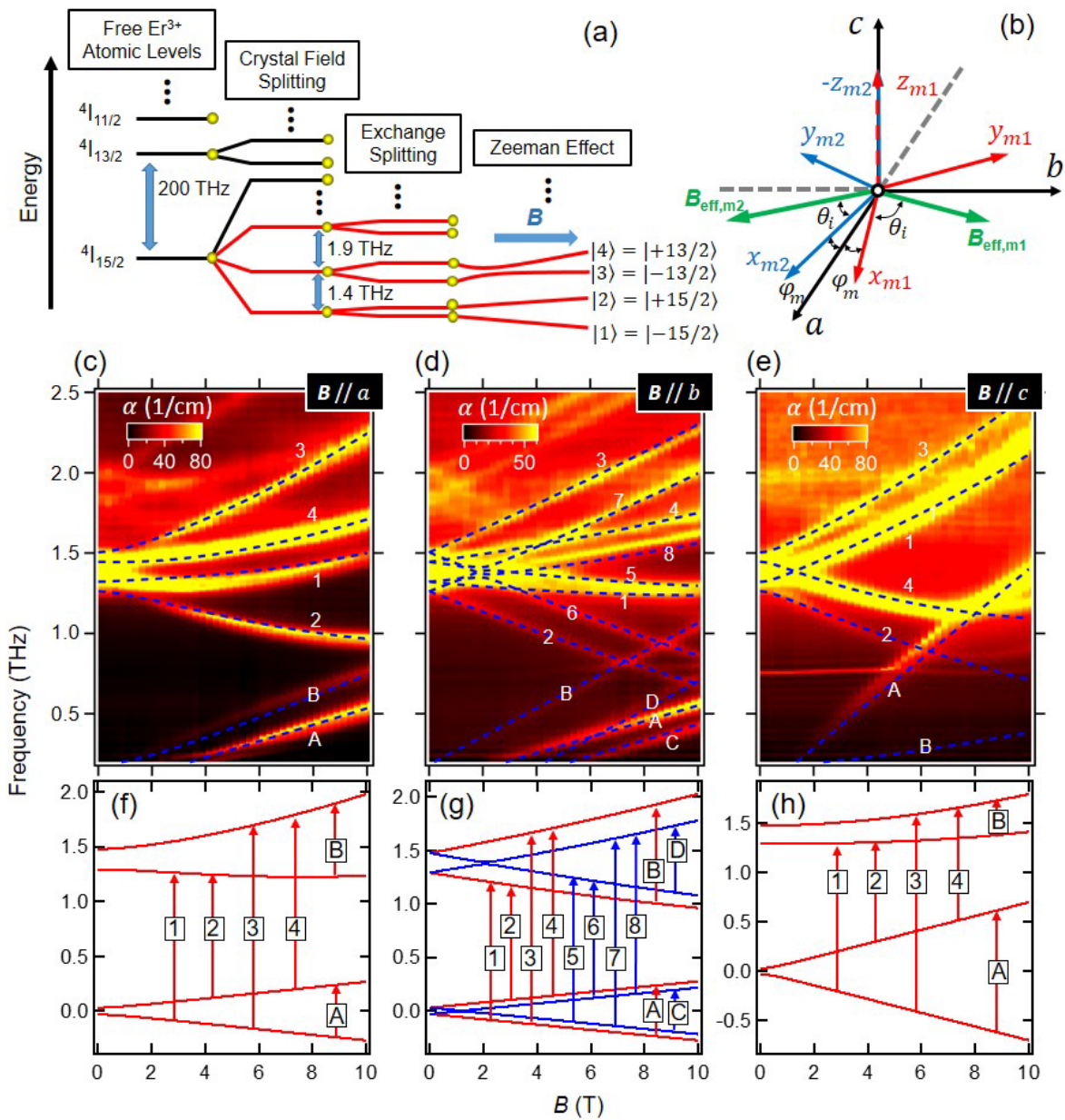

Figure 12. CFTs of $Er^{3+}$ in $ErFeO_3$.[169] (a) Energy level scheme of $Er^{3+}$. Levels shown in red lines are involved in the transitions observed in experiments. (b) The two-sublattice model. Magneto-THz absorption spectra from measurements in (c) configuration 1, (d) configuration 4, and (e) configuration 6. (f), (g), and (h) Energy level calculations from the best fit of experimental spectra using the two-sublattice model, corresponding to the configurations in (c), (d), and (e), respectively. Reproduced with permission from [169].

Figure 12(c), (d), and (e) show experimental absorption spectra for ErFeO$_3$ measured in configurations 1, 4, and 6 in Figure 8, respectively. The sample temperature was 80 K. Generally, the observed absorption lines can be categorized into two groups. One group of transitions, labeled with letters (A, B, …), occur below the THz bandwidth at $B = 0$ and blue-shift with $B$. This group is ascribed to intra-Kramers-doublet transitions, namely, the $|-15/2\rangle \rightarrow |+15/2\rangle$ transition (lines A and C) and the $|-13/2\rangle \rightarrow |+13/2\rangle$ transition (lines B and D). The other group of transitions, labeled with numbers (1, 2, …), emerges from a strong absorption band at ~1.4 THz at $B = 0$ and splits like a "firework" with $B$. This group is ascribed to inter-Kramers-doublet transitions, namely, the $|-15/2\rangle \rightarrow |-13/2\rangle$ transition (line 1, line 5), the $|+15/2\rangle \rightarrow |-13/2\rangle$ transition (line 2, line 6), the $|-15/2\rangle \rightarrow |+13/2\rangle$ transition (line 3, line 7), and the $|+15/2\rangle \rightarrow |+13/2\rangle$ transition (line 4, line 8). In the ***B***//*y* case, the number of lines doubles due to the broken sublattice degeneracy. More inter-Kramers-doublet transition lines centered at 1.9 THz at $B = 0$ can be identified. They are the $|\pm13/2\rangle \rightarrow |\pm11/2\rangle$ transition group.

Because there are two inequivalent Er$^{3+}$ sites in the crystal, the two-sublattice model depicted in Figure 12(b) explains the data well. Defining $j = 1, 2$ to be the Er$^{3+}$ sublattice index, as shown in Figure 12(b), we can see that the site orientations of the two inequivalent sublattices depend on both *m* and *j*, characterized by an angle $\varphi_m$. Furthermore, the Fe$^{3+}$–Er$^{3+}$ exchange coupling is described by an *m*- and *j*-dependent effective field $B_{\text{eff},mj}$ that is experienced by the Er$^{3+}$ ion, characterized by an angle $\theta_m$. Symmetry analysis determines that the local $z_{mj}$ axes have to coincide with the crystal *z* axis, and $B_{\text{eff},mj}$ have to be in the *x*-*y* plane. The Hamiltonian for the crystal-field level *m* and sublattice *j* in a magnetic field is written as the summation of the linear Zeeman term and the quadratic Zeeman term:

$$H_{mj} = H_{\text{linear},mj} + H_{\text{quadratic},mj} = \mu_B \sum_k g_{m,k} \hat{\sigma}_k B_{\text{tot},mj,k} + (\mu_B)^2 \delta_{m,\pm 13/2} \sum_k \Delta D_k (B_{\text{tot},mj,k})^2 \hat{I}, \quad (14)$$

where $\mu_B$ is the Bohr magneton, $k = x, y, z$ is the $Er^{3+}$ local axis index, $g_{m,k}$ are the anisotropic Lande g factors, $\hat{\sigma}_k$ are Pauli matrices, $\boldsymbol{B}_{\text{tot},mj} = \boldsymbol{B} + \boldsymbol{B}_{\text{eff},mj}$ is the net magnetic field experienced by $Er^{3+}$, $\Delta D_k$ are the anisotropic quadratic Zeeman coefficients, and $\hat{I}$ is a 2 by 2 identity matrix. The eigen-energies of all the states can be calculated as a function of $B$ by diagonalizing the Hamiltonian. A single group of parameters can be used to fit all the three experimental spectra in Figure 12(c), (d), and (e); see Table 3 for the parameters obtained from the fits. The energy level schemes of the best fit for the $\boldsymbol{B}//x$, $\boldsymbol{B}//y$, and $\boldsymbol{B}//z$ cases are plotted in Figure 12(f), (g), and (h), respectively. Transition lines are identified, labeled, and plotted correspondingly in Figure 12(c), (d), and (e) as dashed blue lines. Excellent agreement between theory and experiment is achieved.

|  | $\left|m = \pm 15/2\right\rangle$ | $\left|m = \pm 13/2\right\rangle$ |
|---|---|---|
| $\left|\boldsymbol{B}_{\text{eff},mj}\right|$ (T) | 1.19 | 2.1 |
| $\theta_m$ (deg) | 48 | 80 |
| $\varphi_m$ (deg) | 33 | 1 |
| $g_{m,x}$ | 1.9 | 2.5 |
| $g_{m,y}$ | 1.7 | 3.15 |
| $g_{m,z}$ | 5 | 1.2 |
| $\Delta D_x$ (J$^{-1}$) | 1.6×10$^{22}$ | |
| $\Delta D_y$ (J$^{-1}$) | 0.6×10$^{22}$ | |
| $\Delta D_z$ (J$^{-1}$) | 1.7×10$^{22}$ | |

Table 3. Crystal-field parameters for levels $\left|m = \pm 15/2\right\rangle$ and $\left|m = \pm 13/2\right\rangle$ obtained from the fits to the 80 K experimental spectra shown in Figure 12. Reproduced with permission from [169].

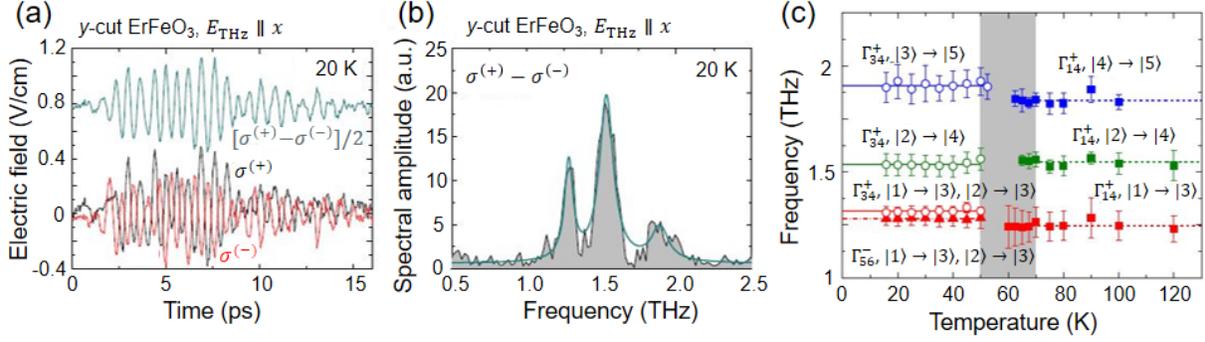

Figure 13. Studying $Er^{3+}$ CFTs in $ErFeO_3$ by THz emission spectroscopy.[180] (a) Pump-helicity-dependent THz emission from y-cut $ErFeO_3$. (b) Fourier transform of the helicity-dependent contribution in (a). (c) Mode symmetry assignments determined from the polarization of emission and pump-helicity dependence. Measurement configuration is the same as (a) and (b). Grey region indicates the SRT temperature range. Crystal-field states are labeled with the same notations as in Figure 12(a). Reproduced with permission from [180].

Mikhaylovskiy et al.[180] have performed polarization-resolved THz emission spectroscopy experiments on $Er^{3+}$ in $ErFeO_3$ to determine the dipolar activity of the CFTs. The energy level diagram is exactly the same as the zero-field diagram in Figure 12(a). A symmetry classification of macroscopic normal modes has been previously performed,[178] dividing CFTs associated with $Er^{3+}$ into either magnetic-dipole active or electric-dipole active in nature. Any dipolar activity can be closely related with the possible polarization state of the emitted THz radiation. Further, although the CFT excitations in this experiment are all created by ISRS processes, the magnetic- and electric-dipole transitions respond to different polarizations of the excitation light. For magnetic-dipole transitions, the effective magnetic field description, given in Sub-subsection 2.2.1, is the dominant mechanism, the most representative of which is the inverse Faraday effect. Figure 13(a) shows THz emission signal along the x axis from a y-cut crystal created by circularly polarized optical pumps with opposite helicities. A helicity-dependent signal is clearly present, whose Fourier transform is shown in Figure 13(b); the peaks appearing in the Fourier transform are assigned to magnetic-dipole-active CFTs.

The ISRS type excitation of electric-dipole-active CFTs, on the other hand, is less straightforward since for centrosymmetric crystals generation of an effective electric field is forbidden via electric-dipole transitions. However, Mikhaylovskiy et al.[180] have proposed a novel mechanism based on virtual magnetic-dipole transitions, $\boldsymbol{E}_{\text{eff}} \propto \chi \boldsymbol{E}_{\text{opt}} \boldsymbol{H}^*_{\text{opt}}$. The following summarizing statement therefore arises: the magnetic-dipole-active THz modes respond to electric-dipole optical transitions, while the electric-dipolar-active THz modes respond to magnetic-dipole optical transitions.[180] Mode assignments based on the pump-helicity dependence and polarization state of the emission were then carried out across the $\Gamma_4 \to \Gamma_2$ SRT, giving Figure 13(c).

*3.3 THz probe of electromagnons*

Electromagnons are magnetic resonances that can be excited by the electric field of light (i.e., electric-dipole-active magnons), which is a type of excitation associated with dynamic magnetoelectric coupling.[181] Their emergence is closely related to magnetoelectric effects and novel ferroic orders,[182–185] and therefore, electromagnons play an important role in the study of electric control of magnetism for functional electronic and spintronic devices.[186,187] From the group theory perspective, spatial inversion symmetry is the most important symmetry element to discuss, because the leading-order magnetoelectric tensors are axial tensors of even rank, which can only have nonzero elements in noncentrosymmetric systems. For multiferroicity to emerge, the system is required to not only be noncentrosymmetric but also belong to a polar group.

In Subsection 2.1, we discussed possible magnetic phases that can arise in RFeO$_3$ using Bertaut's representation theory. We mentioned that, while inversion symmetry is retained in the $\Gamma_1$-$\Gamma_4$ phases, it is broken in the $\Gamma_5$-$\Gamma_8$ phases. Located at 4(*b*) sites as inversion centers, Fe$^{3+}$ alone cannot develop an order that is consistent with $\Gamma_5$-$\Gamma_8$. Therefore, to achieve potential magnetoelectric coupling and multiferroicity in the RFeO$_3$ family, R$^{3+}$ ordering that transforms according to $\Gamma_5$-$\Gamma_8$ is required.

Within the entire family, inversion-symmetry-breaking events by low-temperature magnetic ordering have been demonstrated in GdFeO$_3$,[188] DyFeO$_3$,[97,189] and TbFeO$_3$.[190,191] Although possible observation of room-temperature multiferroicity in SmFeO$_3$ has been reported,[192] symmetry analysis calls the interpretation into question,[193] leading to controvercy regarding whether multiferroicity actually exists in SmFeO$_3$.[194] As shown by the phase diagram in Figure 3, GdFeO$_3$ adopts a $\Gamma_4$ configuration from the Néel temperature of Fe$^{3+}$ (661 K) all the way to 2.5 K. Below 2.5 K, Gd$^{3+}$ moments order according to $g_x$, which transforms according to $\Gamma_5$ (Table 1). This mode mixes with the Fe$^{3+}$ $\Gamma_4$ mode to form a noncentrosymmetric $\Gamma_{45}$ configuration (point group: *m'm'*2), allowing polarization along the crystal *z* axis. This has been observed by Tokunaga *et al.*[188]; exchange striction, that is, the cooperative lattice distortion caused by magnetic coupling, is believed to be the microscopic origin of the polarization.

The Fe$^{3+}$ subsystem in DyFeO$_3$ exhibit an abrupt-type $\Gamma_4 \to \Gamma_1$ SRT (named the Morin transition) at 50 K. Below 4 K, Dy$^{3+}$ ions develop $g_x a_y$-type ordering, consistent with the $\Gamma_5$ mode; the total magnetic phase therefore becomes $\Gamma_{15}$ (magnetic group: 222). This phase belongs to a nonpolar noncentrosymmetric group, which allows a linear magnetoelectric response but forbids polar order. However, polar order and multiferroicity can be switched on by applying a magnetic field along the *z* axis, polarizing the system into the $\Gamma_{45}$ configuration (point group: *m'm'2*). DyFeO$_3$ magnetized along the *z* axis therefore has the same group representation as GdFeO$_3$ at zero field, allowing spontaneous electric polarization along the *z* axis.

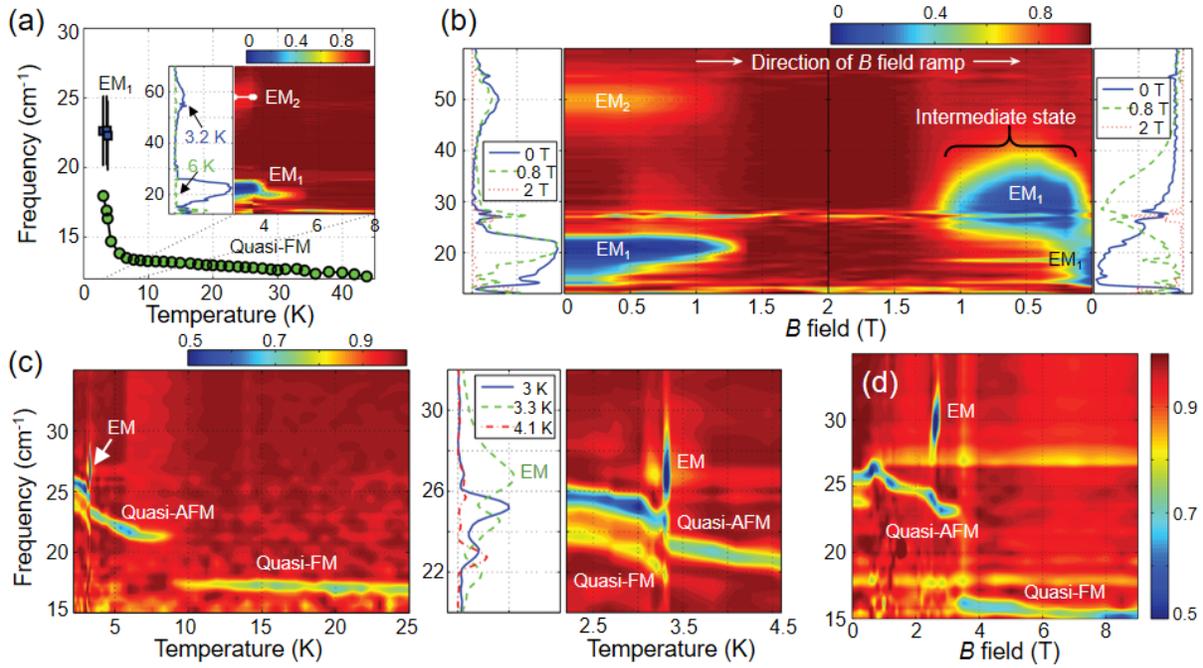

Figure 14. Electromagnons in DyFeO$_3$ and TbFeO$_3$[195,196] measured by far-infrared transmittance spectroscopy. (a) Zero-field mode excitations as a function of temperature in the $\boldsymbol{E}_{THz} \parallel z$, $\boldsymbol{H}_{THz} \parallel y$ geometry for DyFeO$_3$. (b) Magnetic field dependence of DyFeO$_3$ transmittance using the $\boldsymbol{E}_{THz} \parallel z$, $\boldsymbol{H}_{THz} \parallel x$ geometry for $\boldsymbol{B}//y$ at 1.5 K. Up and down sweeps are shown in the left and right panels, respectively. (c) Zero-field temperature-dependent scans using the $\boldsymbol{E}_{THz} \parallel z$, $\boldsymbol{H}_{THz} \parallel x$ geometry for TbFeO$_3$. Right panel zooms in the left panel. (d) Magnetic field dependence of TbFeO$_3$ transmittance for $\boldsymbol{B}//b$ at 1.5 K. Reproduced with permission from [195,196].

Stanislavchuk *et al.*[195] have performed comprehensive far-infrared spectroscopy measurements on DyFeO$_3$, aiming at probing the electromagnon excitations in the low-temperature $\Gamma_{15}$ phase. Figure 14(a) shows the zero-field mode excitations as a function of temperature in the $\boldsymbol{E}_{\text{THz}} \parallel z$, $\boldsymbol{H}_{\text{THz}} \parallel y$ geometry[†]. While the quasi-FM mode shows a prominent blue-shift upon entering the $\Gamma_{15}$ phase, new modes centered at 22 cm$^{-1}$ (0.66 THz) and 58 cm$^{-1}$ (1.74 THz), labeled EM$_1$ and EM$_2$, respectively, are observed. The authors not only ruled out phonons and CFTs as possible origins for these modes but also determined their electric dipole activity, namely, they are only excited for $\boldsymbol{E}_{\text{THz}} \parallel z$. These are consistent with characteristics of electromagnons. Further, the authors demonstrated that EM$_1$ and EM$_2$ can only be observed in the noncentrosymmetric magnetic structure, which corroborates the assignment. Upon applying a static magnetic field $\boldsymbol{B}//y$, at a critical field of 1.3 T, a $\Gamma_{15} \rightarrow \Gamma_{43}$ phase transition occurs, which restores the inversion center (Table 1). EM$_1$ and EM$_2$ abruptly disappear across the field-induced transition [Figure 14(b)]. Interestingly, the eletromagnons exhibit strong hysteresis upon cycling the static magnetic field $\boldsymbol{B}//y$; compare the mode excitations in the up and down sweeps of magnetic field in Figure 14(b). The irreversibility potentially reflects a metastable low-symmetry phase during the $\Gamma_{15} \rightarrow \Gamma_{43}$ phase transition, which has been demonstrated to exhibit intermediate stages.[197]

By using a similar experimental technique, Stanislavchuk et al.[196] have also investigated THz electromagnons in a Tb$^{3+}$-ordered phase in TbFeO$_3$. The low temperature phase diagram of TbFeO$_3$ is more complicated than that of DyFeO$_3$. At $T_{\text{SR1}}$ = 8.5 K, a $\Gamma_4 \rightarrow \Gamma_2$ SRT occurs, which is found in many RFeO$_3$ systems. The Tb$^{3+}$ moment orders according to $a_x g_y f_x c_y$ ($\Gamma_{28}$) at $T_{\text{NTb}}$ = 3.3 K. The $\Gamma_{28}$ phase only holds for a very narrow temperature range, until it transforms into a $\Gamma_{48}$ phase through another SRT at $T_{SR2} = 3.1$ K. Both in the $\Gamma_{28}$ phase (within 3.1 K < $T$ < 3.3 K) and the $\Gamma_{48}$ phase ($T$ < 3.1 K), inversion symmetry is broken, and therefore, potentially electromagnons can be excited.

---

[†] Strictly speaking, the far-infared measurements here are not based on THz time-domain techniques, but since they measure the same quantity, we still denote the electric and magnetic fields of the probe light this way.

In performing zero-field temperature-dependent scans in the $\boldsymbol{E}_{THz} \parallel z$, $\boldsymbol{H}_{THz} \parallel x$ geometry, Stanislavchuk *et al.* found a wealth of mode excitation behavior, shown in Figure 14(c); its right panels show a zoom-in view of data at $T$ < 4 K. While the switching of the polarization selection rule between the quasi-FM and quasi-AFM modes was clearly observed across the $\Gamma_4 \to \Gamma_2$ SRT at $T_{SR1}$, both modes appeared at $T$ < 3.1 K, suggesting a significant relaxation of the polarization selection rule in the $\Gamma_{48}$ phase. In particular, the quasi-FM mode was found to gain electric-dipole activity in this phase. On the other hand, within the narrow temperature range of 3.1 K < $T$ < 3.3 K, a mode at 27 cm$^{-1}$ (0.81 THz) was observed and assigned to an electromagnon (labeled EM, excited only by $\boldsymbol{E}_{THz} \parallel z$). The electromagnon mode also showed delicate appearance within the field-dependent diagram for $\boldsymbol{B}//y$ at 1.5 K [Figure 14(d)]. It was found that $\boldsymbol{B}//y$ leads to multistage phase transitions, and it is only within a narrow field range around 2.5 T can one observe the EM mode. This field-induced phase features a complicated magnetic structure characterized by $\Gamma_{2358}$, which supports spontaneous polarization along the $y$ axis.

## 4  Ultrafast laser manipulation of magnetic order

The ability to manipulate spins in magnetically ordered solids on the shortest possible timescale possesses significant potential for applications in future spintronic and magnetic memory devices, fast-speed computation technology, and quantum information processing.[2,198–200] Among various competing proposals, femtosecond lasers have emerged as the most suitable tool for the control at the fundamental speed limit of quantum spins, namely, within ps or sub-ps time durations. In the dramatic cases where laser pulses drive the material to extreme nonequilibrium, ultrafast magnetic phase transitions occur,[31,49] which has attracted considerable interest over the past two decades.[201–208] Thanks to the rapid pace at which currently laser technology is developing, this research field is expected to gain increasing accessibility, as pulse energies become stronger, pulse durations become shorter, and wavelengths become more tunable.

However, accessibility by no means implies simplicity of the problem of interest. As a matter of fact, ultrafast laser manipulation of magnetic order has been identified as one of the most challenging topics in modern spintronics, encompassing the demand to expand the frontier of knowledge within multiple areas in physics and photonics. Since spins in magnetic solids can never be viewed as an isolated system, but instead, are frequently found to entangle with multiple other fundamental degrees of freedom including charge, lattice, and orbital, one is faced with the demand to understand complex correlated phases and intertwined order of condensed matter.[209] In addition, once laser excitation comes into play, the problem gains an additional level of complexity due to the nonequilibrium nature of the driven system.[39–42,210,211] Besides light–spin interactions, electronic charge transitions and lattice phonons both couple with light strongly and evolve via distinctly different pathways and speeds.[58] Thus, it is necessary to understand how the extremely nonthermal status in multiple coupled reservoirs created by laser excitation evolves in time, and the impact of such evolution on magnetism.

Once a reasonable understanding of the material itself and light–matter interaction is established, what becomes challenging is the way to tailor optical pulses to achieve the desirable type of light–matter interaction. Within the (usually) undiscriminating excitations of multiple reservoirs by light, one identifies the type of excitation that can most efficiently foster the desired modification to spin order. Is it preferable that light interacts only with spins so that charge and lattice excitations should be suppressed? Or, in case spins do not interact with light strong enough, could one strongly excite the lattice, and manipulate spins through lattice–spin interactions? Tailoring light excitations following answers to considerations like these advances the so-called "surgical" approach of laser manipulation of condensed matter.[62] Within the process, one leverages the state-of-the-art laser technology, seeks guidance from theoretical proposals of nonequilibrium quantum matter,[212–215] and creates novel experimental methodologies. All of these activities depend uniquely on the magnetic material of interest, and thus, the amount of opportunities is vast for creating a unique combination of conditions to tailor magnetism on demand.

Historically, magnetic metals,[216,217] semiconductors,[218–222] and insulators[223] have all been used as candidates for laser manipulation of magnetic order; phenomenology and microscopic descriptions are very different, but we will not review all of them at length here. Rather, we focus on the role of $RFeO_3$ along this line of effort. $RFeO_3$ represents the most prototypical Heisenberg-type antiferromagnetic insulator. The magnetic energy hierarchy and temperature-dependent magnetic phase diagram are well understood. THz-frequency magnons, serving as the fingerprints to track distinct magnetic phases, have been well characterized; see Section 3.

Therefore, the stage for the quest for laser manipulation of magnetic order has been set. In the following subsections, we discuss five major pathways to achieve ultrafast laser-induced magnetic phase transitions in RFeO$_3$ that have been reported in the literature.

*4.1 Ultrafast laser-induced heating*

Modification of magnetic order due to laser-induced heating is a widely observed phenomenon and has been proven to be effective. Photons absorbed by the sample (kept in a thermal bath at temperature $T$) through electronic excitations increase the local temperature by $\Delta T$, and the magnetic order evolves towards the status that would be expected for a sample at $T + \Delta T$. Scrutiny is required when heating is supplied by a femtosecond laser pulse. Since the time duration of excitation is shorter than the typical times of electronic decay and electron-lattice thermalization, the way to perceive $\Delta T$ strongly depends on the time delay after pump excitation.

A phenomenological three-temperature model has been frequently applied to describe the time evolution of the thermal properties of the system.[216,224] As shown in Figure 15(a), the model assumes three separate but mutually interacting reservoirs, namely, the electrons, lattice, and spins, which are given respective pump-induced temperature changes – $\Delta T_{electron}$, $\Delta T_{lattice}$, and $\Delta T_{spin}$. At early time delays immediately after the pump excitation of the charge sector, $\Delta T_{electron}$ is the largest. Then $\Delta T_{lattice}$ and $\Delta T_{spin}$ increase with time at the expense of $\Delta T_{electron}$ through energy transfer processes at rates determined by the coupling strengths between the reservoirs [Figure 15(b)]. It is only at time delays when $\Delta T_{electron}$, $\Delta T_{lattice}$, and $\Delta T_{spin}$ have fully converged (quasi-equilibrium condition) can an overall local temperature increase of the system $\Delta T$ be defined, but for the purpose of controlling the magnetic properties, only examining $\Delta T_{spin}$ before reaching quasi-equilibrium is also reasonable.

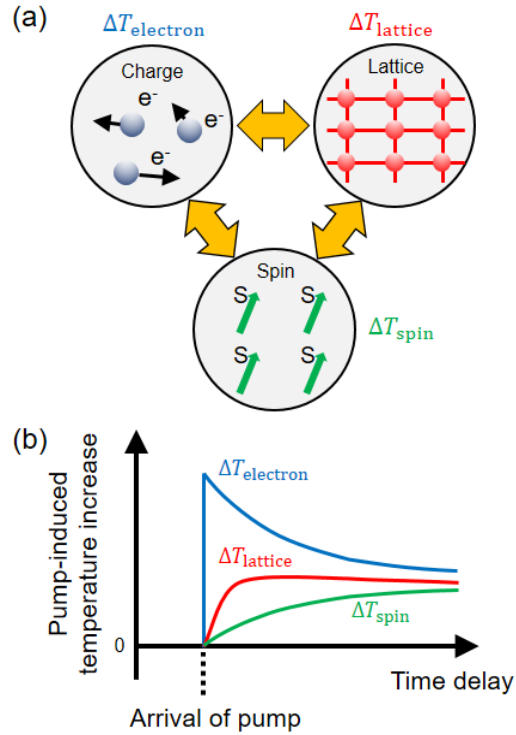

Figure 15. Three-temperature model. (a) Separate but mutually interacting reservoirs. (b) Time dynamics of temperatures of the reservoirs.

Within the RFeO$_3$ class, the temperature range around the Fe$^{3+}$ SRT is ideally suited for ultrafast magnetic phase transitions induced by laser heating. In the spin Hamiltonian, Equation (4), although the isotropic exchange constant and the DM interaction constant are temperature independent, the anisotropy energies depend strongly on the temperature. In TmFeO$_3$, by spectroscopically resolving the quasi-FM and quasi-AFM mode frequencies, the temperature dependence of the $A_x$ and $A_z$ constants has been quantitatively determined. Their difference, $A_x - A_z$, has been found by Zhang *et al.*[179] to change sign across the SRT temperature range (80 – 90 K); see Figure 16. This suggests that a sign change of a similar origin would occur when one imposes ultrafast heating on a sample kept around the SRT temperature as well. The laser-heated sample would then have an easy axis which deviates from that in equilibrium, leading to a spin-structure rotation.

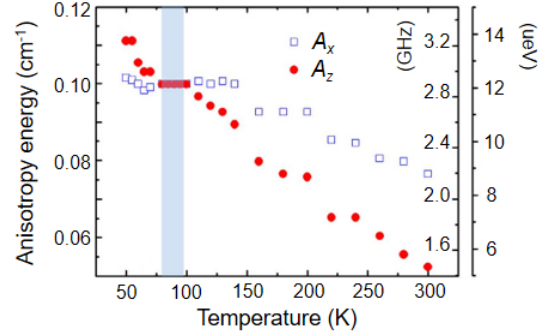

Figure 16. Temperature dependence of the anisotropy constants, $A_x$ and $A_z$, in $TmFeO_3$.[179] Reproduced with permission from [179].

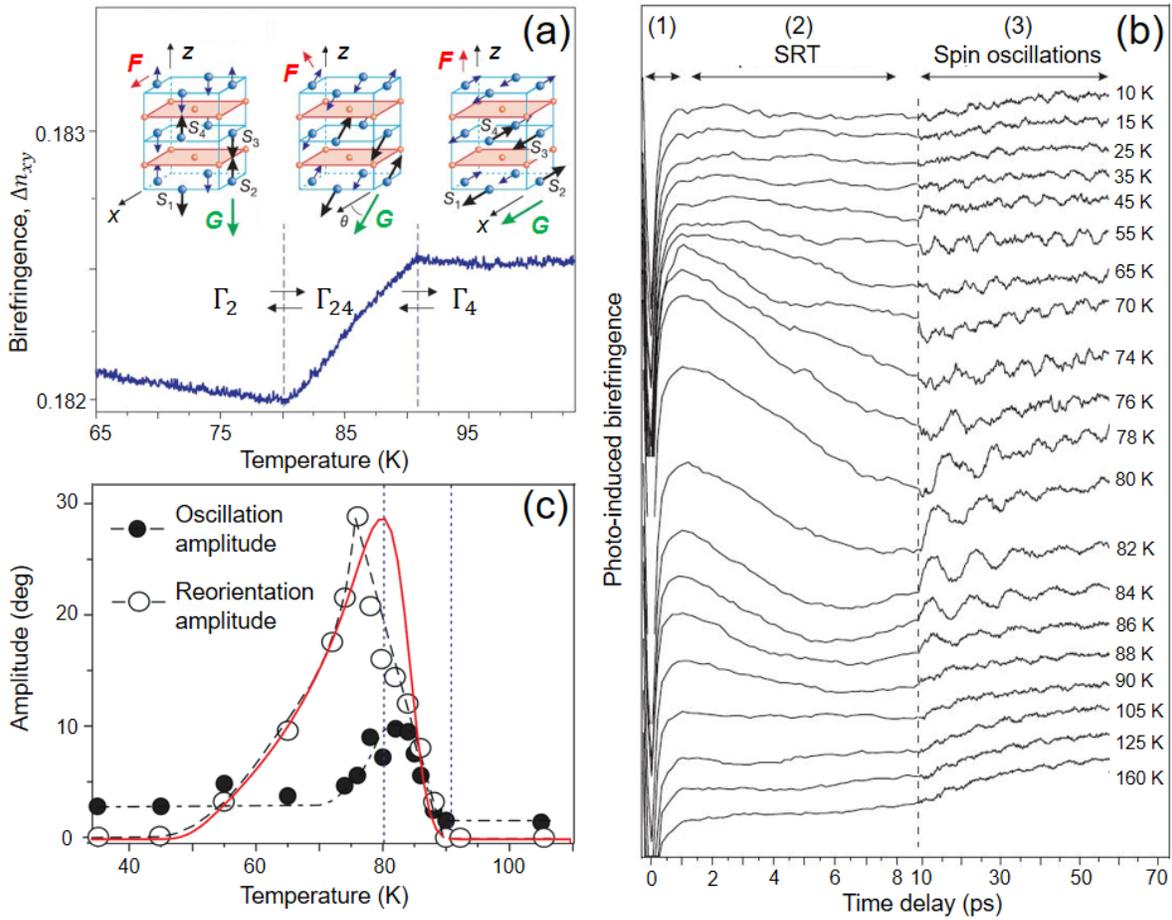

Figure 17. Ultrafast-heating-induced SRT in $TmFeO_3$.[225] (a) Equilibrium linear birefringence as a function of temperature. (b) Time-resolved pump-probe linear birefringence traces at different temperatures. (c) Temperature dependence of the amplitudes of magnon oscillations and reorientation amplitude extracted by fitting to the data in (b). Reproduced with permission from [225].

Kimel et al.[225] have performed time-resolved linear birefringence measurements on a TmFeO$_3$ crystal around the SRT temperature in the presence of ultrafast pump excitation. In equilibrium (with no pump excitation), it was found that linear birefringence is a sensitive indicator of the orientation of the AFM vector $G$ across the $\Gamma_4 \rightarrow \Gamma_{24} \rightarrow \Gamma_2$ transition [Figure 17(a)]. This type of probe mechanism couples to the $G$ vector directly, which is different from the Faraday- or Kerr-effect probes (which couple linearly to the bulk magnetization). As detailed in Sub-subsection 2.2.1, the probe utilizes the symmetric portion of the dielectric permittivity tensor, which contains a term that reads $\varepsilon_{ii} = \beta_{iijk} G_j G_k$; the coupling to $G$ is quadratic and is sensitive to the rotation of $G$ across the SRT. In a z-cut crystal, when the polarization is aligned in-plane between the x and y axes (propagation along z), the birefringence $\Delta n_{xy} = (\varepsilon_{xx} - \varepsilon_{yy})/2n$ was detected by reading the polarization of the transmitted probe pulse.

Upon pump excitation, time-resolved linear birefringence reflects the pump-induced change in $G$. As shown in Figure 17(b), such traces were collected at different sample temperatures. In the time domain, three distinct regimes were identified: Regime (1) is a fast decay due to the charge-phonon thermalization within 1 ps, Regime (2) spans from 1 – 10 ps and makes the signal settle to an offset compared to the equilibrium value, and Regime (3) represents the range with longer time delays showing magnon oscillations. Regime (2) is particularly notable due to the offset structure that is developed in linear birefringence, suggesting the rotation of $G$ within the same time duration.[225] The pump-induced heating mechanism explains the direction of $G$ rotation perfectly. What further corroborates the interpretation is that the pump-induced spin reorientation has the largest amplitude around and slightly below the equilibrium SRT temperature range [Figure 17(c)]. This is expected because at temperatures way below or above the SRT, the laser-induced transient temperature rise is unable to cause a sign change in $A_x - A_z$ (Figure 16), leading to no change to the easy axis. What makes the observation more interesting is that the spin rotation reaches a maximum amplitude of 30 degrees, which is a considerable fraction of the full $\Gamma_4 \rightarrow \Gamma_{24} \rightarrow \Gamma_2$ transition in equilibrium, and completes the rotation within the first 4 ps of Regime (2), which is faster than the spin precession period. The fast and large-amplitude spin control therefore opens up promising applications in magnetic random access memory devices.

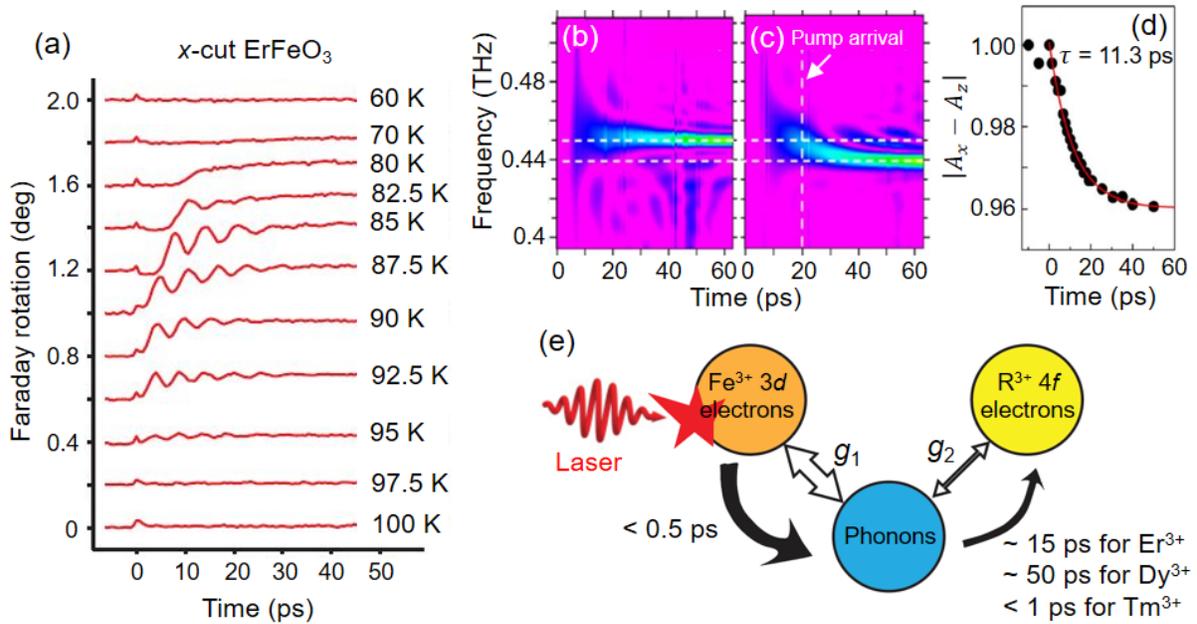

Figure 18. (a) Ultrafast-heating-induced SRT in ErFeO$_3$ probed by time-resolved Faraday rotation.[226] (b) THz pump Faraday rotation probe experiment on ErFeO$_3$.[227] Quasi-FM mode whose frequency is stable with time is observed. (c) Same experiment as (b) except that an additional near-infrared pulse excites the sample at 20 ps to produce ultrafast heating. The temperature was 30 K. (d) Time evolution of the anisotropy energy due to ultrafast heating extracted by fitting the magnon frequency shift in (c). (e) Energy-transfer scheme between three reservoirs during ultrafast heating.[226] 3$d$ electrons and phonon quickly thermalize upon excitation, but the energy transfer rate between 4$f$ ions and lattice depends on the 4$f$ ion species. The model resembles the three-temperature model in essence, but slightly differs in the definition of reservoirs. Reproduced with permission from [226] and [227].

A similar experiment by de Jong et al.[226] performed on ErFeO$_3$ around the SRT temperature exhibited some differences compared to the TmFeO$_3$ experiment.[225] The laser-induced SRT again showed the strongest amplitude in the vicinity of the equilibrium SRT temperature range; see Figure 18(a) for the offsets of the time-resolved Faraday rotation traces on an *x*-cut crystal. However, the rise of the spin rotation took a considerably longer time (~ 40 ps) than in TmFeO$_3$ (~ 4 ps). This raises the question of the microscopic pathway for ultrafast heating. In TmFeO$_3$, the very fast completion of spin rotation suggests that the anisotropy modification gets well-established within 1 ps, while in ErFeO$_3$, the fact that spin rotation is much slower than the precession period implies that the heating process itself, and likewise, the modification of the anisotropy axis, takes a longer time.

Yamaguchi et al.[227] have provided valuable insight into the problem. The quasi-FM mode, whose frequency is sensitive to the anisotropy energy $A_x - A_z$ [Equation (5)], was first launched by a THz pump. Once the oscillation was stabilized [Figure 18(b)], a NIR heating pulse pumped the system again [Figure 18(c)], and the dynamics of the magnon frequency shift following the NIR pulse revealed the time evolution of the anisotropy energy [Figure 18(d)], and thus, the true timescale of heating. For ErFeO$_3$, it was found that heating took ~ 15 ps, which naturally explains why the laser-induced SRT observed by de Jong et al.[226] was slow. This finding also corroborates a model proposed by de Jong et al.[226] that analyzes the microscopic pathway of ultrafast heating in the entire RFeO$_3$ class. As shown in Figure 18(e), when the electronic states of $Fe^{3+}$ are excited by the pump pulse, the energy needs to be passed over to the rare earth ion $R^{3+}$ to modify the anisotropy. Since the $Fe^{3+}$-lattice interaction is strong, the entire pathway is bottlenecked by the lattice-$R^{3+}$ energy transfer; the transfer rate strongly depends on the rare-earth ion species.[228] The knowledge over the ultrafast heating pathway has led to protocols to control the timing and speed of laser-induced SRTs.[226]

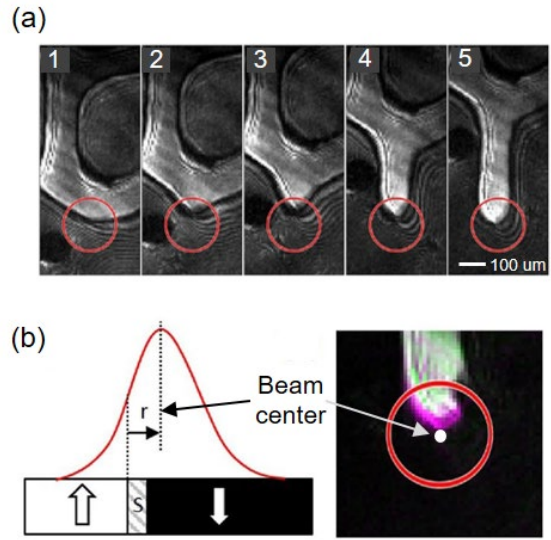

Figure 19. Reconfigurable magnetic domains in ErFeO$_3$ by static heating from a THz free electron laser (FEL). [229] (a) Sequence of Faraday rotation microscopy images showing spin-up domain expansion when the FEL spot (macropulse energy: 8 mJ) scans on the sample surface. Red circle: FEL focal spot. (b) Active area of domain flipping is off from the center of the laser spot. Reproduced with permission from [229].

Kurihara *et al.*[229] have performed another interesting experiment, demonstrating reconfigurable magnetic domain control by laser heating. Unlike the previous examples of ultrafast transient heating, where the effect disappeared when heat diffused outside the pumped area, this work utilized static heating, a time-averaged effect of temperature rise due to laser illumination. An intense THz beam from a free electron laser (FEL) was used. The beam consisted of macropulses with energy up to 40 mJ, spectrally centered around 4 THz, at a repetition rate of 5 Hz. Each macropulse consisted of short micropulses with peak electric fields on the order of several MV/cm (at the focus spot) and a pulse separation of 37 ns. Faraday rotation imaging microscopy on a *z*-cut ErFeO$_3$ crystal at room temperature was performed by a continuous-wave probe laser in the presence of THz FEL illumination. It was observed that the laser heating expands the minority domains (whichever has a smaller volume fraction between spin-up and spin-down domains), when the FEL focal spot scans spatially on the same surface; see Figure 19(a). Upon analysis of an extensive image data set, they found that, as the minority domain expands, the active area of domain flipping does not coincide with the center of the laser spot (where the intensity is highest), but instead, appears at the trailing side of the spot where the spatial gradient of intensity is large. Guided by these observations, the domain reconfiguration was interpreted to result from an entropy force exerted by the laser-induced thermal gradient, assisted by the depinning effect due to static heating.

*4.2 Photomagnetic pathways*

Photomagnetic effects, whose basic properties are overviewed in Sub-section 2.2.1, represent an important nonthermal pathway for achieving laser manipulation of magnetism. The effective magnetic fields generated by IFE and ICME do not rely on the absorption of photons, but rather, depend sensitively on the five property tensors in Table 2 and laser polarization. This suggests a wider range of tuning possibility that can be afforded by this method.

As discussed in Subsection 4.1, de Jong *et al.*[226] observed that the timescale for ultrafast-heating-induced SRT in $ErFeO_3$ is 40 ps. The same study also analyzed the impact of IFE to SRT by pumping with circularly polarized light with opposite helicities and extracted the pump-polarization-dependent response of the Faraday rotation probe. As shown in Figure 20(a), the IFE visibly creates an offset in the Faraday rotation signal around the SRT temperature range in a *z*-cut crystal, indicating a similar type of laser-induced spin rotation as in the case of ultrafast heating. The difference, however, is that the rotation caused by IFE is considerably faster; see the comparison between the 90 K curves shown in Figure 20(a) and (b).

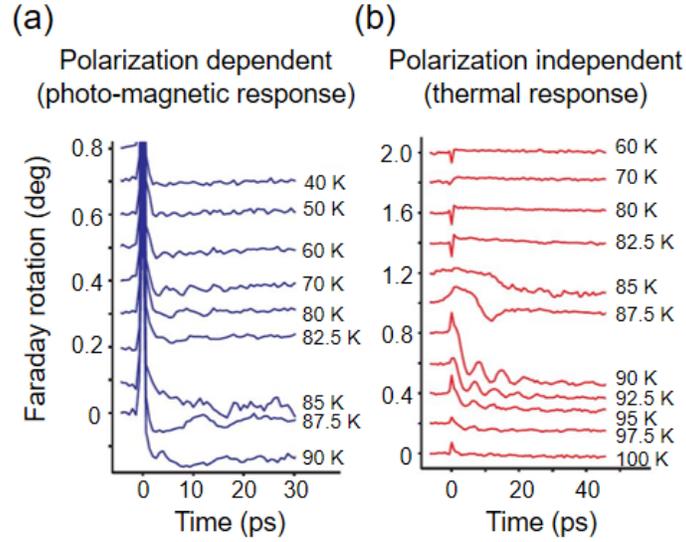

Figure 20. Comparison of timescales between (a) photomagnetic-effect-induced SRT and (b) laser-heating-induced SRT, analyzed by extracting the polarization-dependent and polarization-independent probe responses, respectively.[226] Reproduced with permission from [226].

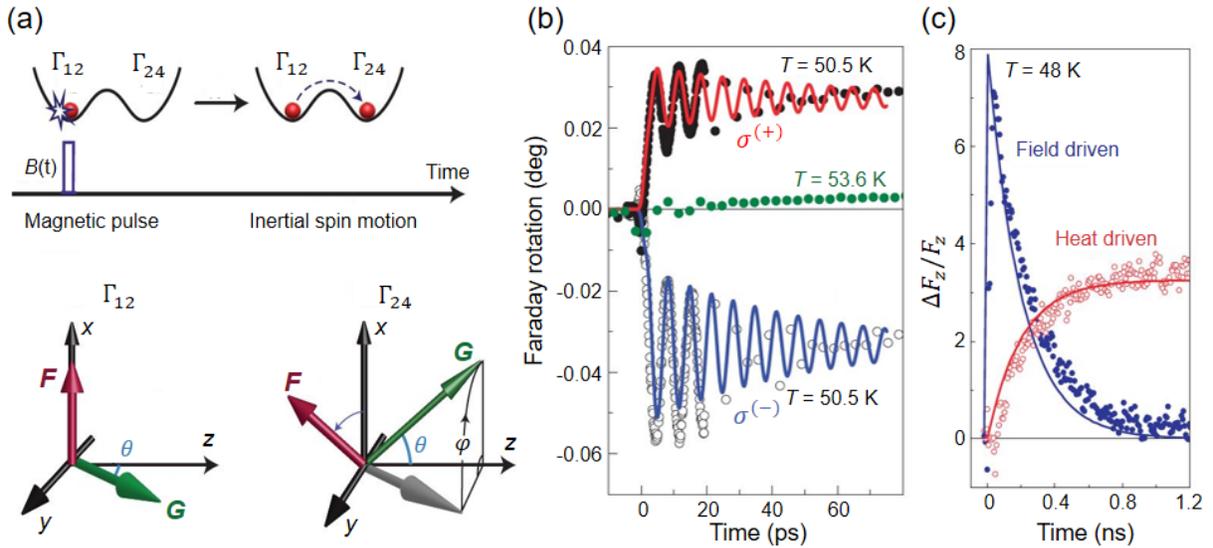

Figure 21. Inertia-driven SRT in HoFeO$_3$.[230] (a) Intense magnetic field pulse generated by IFE supplies enough kinetic energy to enable spin switching to a metastable state. In HoFeO$_3$, the two free energy minima coincide with the $\Gamma_{12}$ and $\Gamma_{24}$ phases, distinguished by an azimuthal angle $\varphi$ of the AFM

vector. (b) Observation of spin switching when the photomagnetic pulse exceeds a critical field strength. (c) Timescale of IFE-induced switching is much faster than the heating timescale. Reproduced with permission from [230].

The potential of fast and impulsive spin switching via photomagnetic effects has been demonstrated by Kimel *et al.*[230] The study attempted to harness the inertia of spins to switch the equilibrium magnetic configuration to a metastable state; this involves overcoming a potential barrier that separates two local minima in the free energy potential, as shown in Figure 21(a). HoFeO$_3$ provides an appropriate double-minima free energy landscape by hosting both $\Gamma_{12}$ and $\Gamma_{24}$ configurations within a particular temperature range [Figure 21(a)]. The "kick" to the spin system was supplied by a short but intense magnetic field pulse generated by IFE.

Upon certain simplifications, the magnetic configuration can be described by a single parameter, that is, the azimuthal angle $\varphi$ of the AFM vector defined in Figure 21(a). For the $\Gamma_{12}$ phase (the initial unperturbed phase), $\varphi = 0$, $\boldsymbol{F}$ is along *x*, and $\boldsymbol{G}$ is in the *y-z* plane. For the $\Gamma_{24}$ phase, there are two possible domains with $\varphi = \pm \pi / 2$, $\boldsymbol{F}$ has a finite *z* component, and $\boldsymbol{G}$ is in the *x-z* plane. The equation of motion is

$$\frac{d^2\varphi}{dt^2} + 2\Gamma \frac{d\varphi}{dt} + \omega_0^2 \frac{dw(\varphi)}{dt} - \frac{\gamma^2 H_D}{\sin\theta} H(t)\cos\varphi = 0. \qquad (15)$$

The first term reflects the spin inertia (second derivative suggests the presence of acceleration), and the second, third, and fourth terms represent the spin damping, restoring force, and driving force, respectively. $\Gamma$ is the damping constant, $\omega_0$ is the magnon frequency, $H_D$ is the DM field, $\theta$ is an angle defined in Figure 21(a) and is assumed to be constant, and $H(t)$ is the IFE-induced magnetic pulse. One can tell that the driving force relies on the presence of DM interaction and is strongest in the $\Gamma_{12}$ phase ($\cos\varphi = 1$).

Experimentally, Kimel et al.[230] then used circularly polarized light to pump a z-cut crystal and probed via Faraday rotation, keeping in mind that $F$ is expected to develop a finite projection onto z when a $\Gamma_{12}$ to $\Gamma_{24}$ switching occurs. Figure 21(b) indeed shows the fast development of a finite offset in the signal, whose sign flips when the pump helicity is reversed; this is the expected behavior for switching into the two possible types of domains, $\varphi = \pm \pi/2$, for the $\Gamma_{24}$ phase. In addition, when the pump intensity is reduced so the IFE-induced field is smaller than the critical field required for the switching (~0.5 T), the offset fails to develop and the system remains in $\Gamma_{12}$ [green curve in (b)]. The pump polarization dependence unambiguously points to the nonthermal nature of the switching, and as shown in Figure 21(c), the switching governs the behavior of the system at early time delays, arising considerably faster than the heating-induced response.

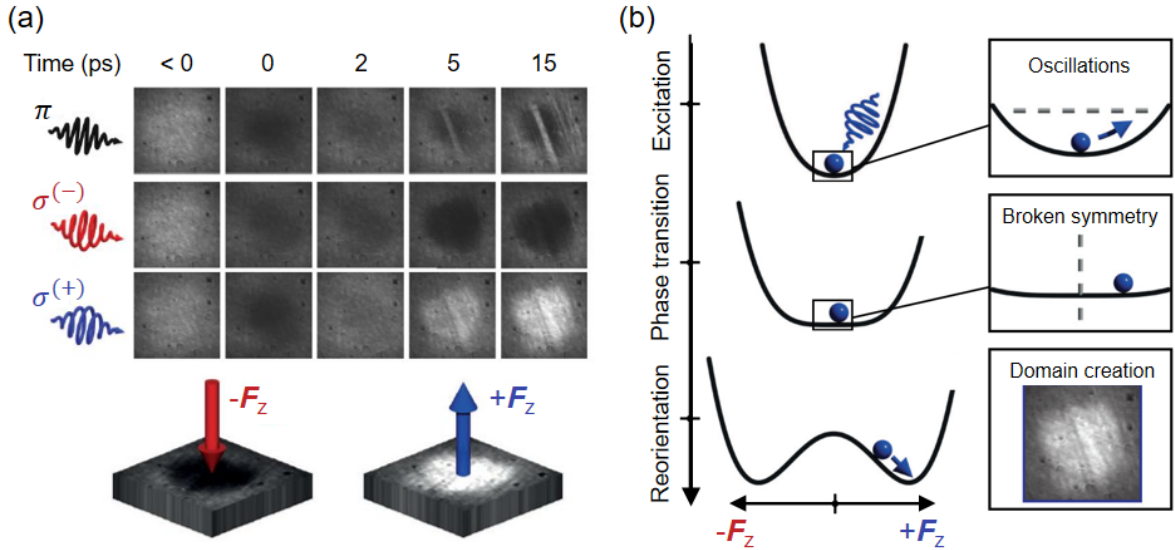

Figure 22. Domain-controllable laser-induced SRT due to the combined effect of IFE and ultrafast heating.[231] (a) Faraday rotation images taken after a single shot of pump pulse by various delay times around the $\Gamma_2 \to \Gamma_4$ SRT temperature range. Grey corresponds to the $\Gamma_2$ phase. Black (white)

corresponds to a spin-down (-up) domain in the $\Gamma_4$ phase. (b) Mechanism for controllable switching. Reproduced with permission from [231].

From the two examples above, one can tell that the thermal effects of laser heating and nonthermal photomagnetic effects usually coexist in an actual experiment, and some pump polarization arguments need to be made to distinguish the two. In fact, their coexistence is not always a drawback, but instead, can be leveraged to realize more powerful control over the routes of laser-induced SRTs. de Jong et al.[231] have established a unique single-shot magneto-optical imaging setup to study this possibility. When a z-cut (SmPr)FeO$_3$ crystal is initially kept in the low-temperature $\Gamma_2$ phase, **F** has zero projection along z, and the Faraday rotation image appears grey [first column in Figure 22(a)]. If a linearly polarized pulse heats up the sample, a transition to $\Gamma_4$ within the laser spot manifests through the appearance of the equal populations of the two energetically degenerate domains with either positive or negative $F_z$ [first row in Figure 22(a)]. However, when the heating pulse is circularly polarized, the domain state that the system transitions into can be selected, and uniformly registered within the spot; see the second and third rows in Figure 22(a). The interpretation for the domain-controllable SRT is shown in Figure 22(b). A spin oscillation whose phase is determined by the pump helicity is initially excited in the single-well potential through IFE. Since the same pulse also induces ultrafast heating, the free energy landscape then evolves according to the temperature-dependent SRT. At the instant where the landscape becomes a double well ($\Gamma_4$), the symmetry of the system is in fact broken due to spin oscillations. The well to which the system is closer at that instant is the more favorable state the system will evolve into eventually. Therefore, by controlling the phase of spin oscillations through IFE, the domain state of the ultrafast-heating-induced SRT can be controlled.

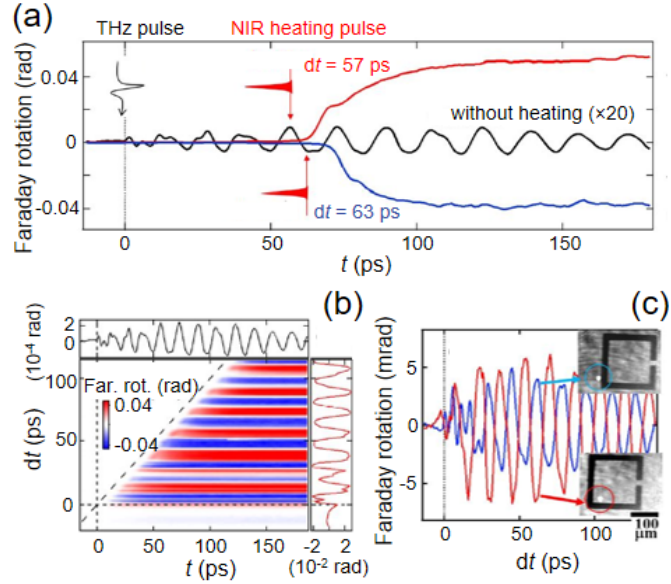

Figure 23. Domain-controllable SRT in ErFeO$_3$ using THz-NIR double pumping.[232] (a) THz pump field (enhanced by a metamaterial) first launches a coherent magnon, and depending on the timing of the NIR heating pulse within the magnon oscillation period, the SRT can be controlled to show a single domain. The sample temperature was 84 K. (b) 2D plot of Faraday rotation as a function of probe time delay $t$ and THz-NIR time interval $dt$. (c) Final-state magnetization as a function of $dt$ has opposite phase when probing inside and outside the metamaterial ring, suggesting the SRT pathway is sensitive to the sign of the THz magnetic field. Reproduced with permission from [232].

This method of utilizing both symmetry breaking by spin precessions and ultrafast heating to achieve domain-controllable SRT has been demonstrated in another notable work by Kurihara *et al.*[232] using a double-pump scheme. Here, the THz magnetic pulse that initiates spin precession is separated from the NIR heating pulse by a time delay of $dt$ [Figure 23(a)]. On a $z$-cut ErFeO$_3$ crystal initially in the $\Gamma_2$ phase, launching the quasi-FM mode induces an oscillation of the out-of-plane magnetization component $F_z$, with the sign of $F_z$ alternating in time. When the NIR heating pulse arrives, the instantaneous sign of $F_z$ determines an SRT into one of the two $\Gamma_4$ domains; the domain with the same sign of $F_z$ would be more favorable. Figure 23(b) clearly shows the alternating final-state magnetization (period synchronized with that of spin precessions) achieved in the $\Gamma_4$ phase by sweeping $dt$. In addition, since a metamaterial ring was used to enhance the THz magnetic field, the authors also compared the final-state magnetization curves between spots inside and outside the ring [Figure 23(c)]. The two curves are out-of-phase, consistent with the fact that the phase of the symmetry-breaking spin precession is determined by the sign of the THz magnetic pulse.

*4.3 Phonon pathway*

A rapidly developing method for nonthermal control of quantum materials is resonant excitation of infrared active phonons. The superexchange interaction, which governs the magnetic properties of transition metal compounds, depends critically on the orbital overlap between transition metal ions and ligands, so it is expected to be effectively modified by lattice distortions induced by phonons. Two particularly notable pathways aligned with this idea are the nonlinear phononics and the phonomagnetic effect within the framework of dynamical multiferroicity.

Nonlinear phononics is a proposal to induce nonequilibrium distortions of a crystal by harnessing the anharmonicity of the lattice free energy potential.[64,233–235] One can consider an anharmonic potential that only includes the cubic term (and neglect all higher orders)

$$V_{\text{anh}} = g_{ijk} Q_i Q_j Q_k, \qquad (16)$$

where $Q$ represents normal coordinates of a phonon, $g_{ijk}$ is the coupling coefficient, and the indices in the sum, $i, j, k$, run over various phonon modes. An important criterion imposed by symmetry is that $V_{\text{anh}}$, being a constituent of the total free energy, must remain invariant under all symmetry operations within the crystallographic point group; this limits the possible combinations of phonons that can couple. The criterion can be expressed as

$$[\Gamma_1 \otimes \Gamma_2] \otimes \Gamma_3 \supset A_g, \tag{17}$$

which means that the direct product of the irreps of all three phonons involved in the coupling in Equation (16) must contain the fully symmetric irrep ($A_g$ for centrosymmetric crystals, which applies to RFeO$_3$).

One widely explored scenario[233,236] is when the first two irreps are from the same infrared (IR) active phonon, while the third is from a Raman (R) phonon, which gives $[\Gamma_{IR} \otimes \Gamma_{IR}] \otimes \Gamma_R \supset A_g$, and $V_{anh} = g(Q_{IR})^2 Q_R$. The equation of motion for the IR and R phonon coordinates under a resonant laser drive of the IR mode can then be derived as[64]

$$\ddot{Q}_{IR} + \gamma_{IR}\dot{Q}_{IR} + (\omega_{IR}^2 + 2gQ_R)Q_{IR} = z_i^* E(t), \tag{18}$$
$$\ddot{Q}_R + \gamma_R \dot{Q}_R + \omega_R^2 Q_R = -g(Q_{IR})^2,$$

where $\gamma_{IR(R)}$ is the IR (R) phonon decay rate, $\omega_{IR(R)}$ is the IR (R) phonon frequency, $E(t)$ is the time-dependent laser driving field, and $z_i^*$ is the Born effective charge of the IR phonon. It is clear that the $Q_{IR}$ is majorly driven by the laser field, but $Q_R$ is driven by $-g(Q_{IR})^2$, which relies entirely on anharmonicity and the $Q_{IR}$ amplitude. Figure 24 plots the time evolution of $Q_{IR}$ and $Q_R$ upon pulsed laser excitation (whose center frequency is resonant with $\omega_{IR}$). The coordinate $Q_{IR}$ oscillates around zero, whose amplitude grows during the drive, and decays after the drive. On the contrary, $Q_R$ develops a rectified response *with a non-zero time average* in the time period during which $Q_{IR}$ is oscillating. This is a significant result since it suggests a distortion of the lattice along the R phonon coordinate lasting on the order of $1/\gamma_{IR}$, which can be much longer than the oscillation periods, $1/\omega_{IR}$ or $1/\omega_R$. Furthermore, the R phonon coordinate is likely to have no equilibrium analog, suggesting access to distorted structures that cannot be achieved by applying static strain fields or pressure. Such a unique lattice control opens up vast possibilities for ultrafast manipulation of magnetic order.

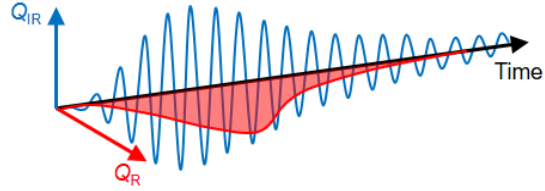

Figure 24. Solution of $Q_{IR}$ and $Q_R$ from Equation (18) upon pulsed laser excitation whose center frequency is resonant with $\omega_{IR}$.[206]

Afanasiev et al.[237] resonantly pumped the highest frequency $B_{1u}$ phonon mode at 85 meV of a (001)-oriented DyFeO$_3$ crystal, and measured time-resolved Faraday rotation signals at various temperatures. In equilibrium, DyFeO$_3$ shows a $\Gamma_1 \to \Gamma_4$ SRT upon warming up across the Morin temperature (51 K). The authors demonstrated the nonequilibrium analog of the same SRT induced by the phonon pumping scheme. As shown in Figure 25(a), it was observed that phonon pumping causes the quasi-AFM mode frequency to red-shift at $T < 51$ K, and blue-shift at $T > 51$ K, and the magnitude of the shifts depends sensitively on the spectral overlap between the pump and the phonon absorption peak [Figure 25(b)]. Since the quasi-AFM frequency is a probe of the local curvature of the magnetic potential, these observations hint at modification of the magnetic potential due to the phonon drive.

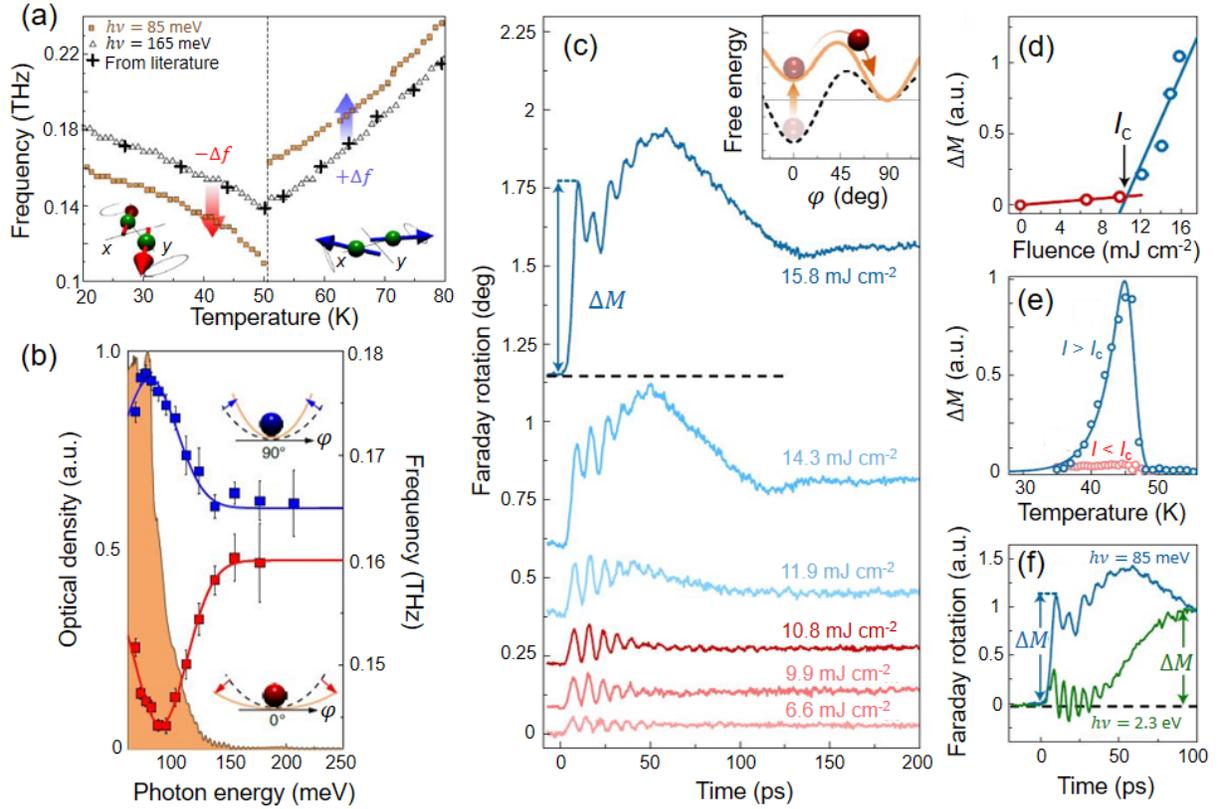

Figure 25. Nonlinear phononic control of magnetic phases in DyFeO$_3$.[237] (a) Frequency shifts (marked by red and blue arrows) of the quasi-AFM mode due to phonon pumping. (b) Frequency shift as a function of pump photon energy. Shaded curve shows the phonon absorption spectrum. (c) Faraday rotation transients with increasing pump fluence at 45 K. (d) Offset versus fluence. (e) Offset versus temperature for fluences below (red) and above (blue) the threshold. $I_c$ denotes the fluence threshold. (e) Phonon pumping versus charge pumping (ultrafast heating). Reproduced with permission from [237].

Figure 25(c) shows that, when the crystal in the $\Gamma_1$ phase is pumped with increasingly strong fluences, an offset $\Delta M$ develops in the Faraday rotation transients; this indicates a $\Gamma_1 \rightarrow \Gamma_4$ SRT because the $\Gamma_4$ phase possesses a weak magnetization along $z$. The offset $\Delta M$ exhibits a threshold behavior as a function of fluence [Figure 25(d)], showing the largest amplitude slightly below the Morin temperature [Figure 25(e)]. Through density functional theory (DFT) calculations, Afanasiev et al.[237] argued that nonlinear phononic coupling with the form $V_{\text{anh}} = g(Q_{B_{1u}})^2 Q_{A_g}$ is the cause of the SRT. By driving the $B_{1u}$ IR phonon to large amplitude, the crystal distorts along a fully symmetric $A_g$ phonon coordinate which involves antipolar motions of $Dy^{3+}$ ions. The $Fe^{3+}$-$Dy^{3+}$ exchange interaction is thus renormalized by a few μeV, which modifies the magnetic potential strong enough [inset to Figure 25(c)] to induce the SRT. The ultrafast heating mechanism was ruled out through a control experiment where the pump photon energy was tuned to be resonant with the charge gap. As shown in Figure 25(f), starting from the same initial condition, charge excitation (which is known to generate ultrafast heating) is considerably slower in creating the $\Delta M$ offset, compared with the phonon driving scenario. This points again to the usefulness of the dynamical lattice control approach. One interesting detail is that the switched magnetic phase persists for hundreds of ps, even after $B_{1u}$ has stopped ringing and thus $A_g$ distortion is expected to be absent. Similar phenomenology of magnetic dynamics extending beyond the rectified lattice response has also been observed in another experiment.[206] Further work is needed to clarify the underlying mechanism.

The quadratic-linear anharmonic coupling of the type $V_{\text{anh}} \propto (Q_{B_{1u}})^2 Q_{A_g}$ employed by Afanasiev et al.[237] creates a distortion along an $A_g$ coordinate, which does not break the symmetry of the crystal. More dramatic dynamical effects are expected if the rectified distortion has an irrep that breaks some symmetries. Radaelli[238] has pointed out that the only scenario that this can happen for quadratic-linear coupling $V_{\text{anh}} \propto (Q_{\text{IR}})^2 Q_R$ is when the IR active phonon is a degenerate phonon. On the other hand, Juraschek et al.[239] have proposed symmetry-breaking nonlinear phononics through the trilinear anharmonic coupling, $V_{\text{anh}} \propto Q_{\text{IR}_1} Q_{\text{IR}_2} Q_R$, where $IR_1$ and $IR_2$ are two distinct phonons driven by the laser. This does not necessarily require two laser beams in an experiment, especially for orthorhombic systems such as $RFeO_3$.

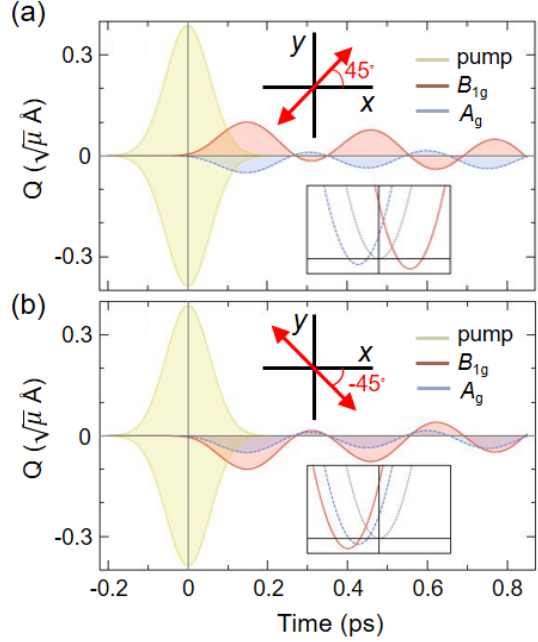

Figure 26. Evolution of the amplitude of the two Raman modes with (a) 45 deg and (b) -45 deg pump polarizations. Notice the sign change of the $B_{1g}$ mode.[239] Insets show shifts of the free energy minima of the Raman modes. Reproduced with permission from [239].

Juraschek et al.[239] have pointed out that since the lattice constants along the $x$ and $y$ axes in RFeO$_3$ are similar, the in-plane IR phonons of $B_{3u}$ (polarization along $x$) and $B_{2u}$ (polarization along $y$) symmetries are nearly degenerate. This enables driving both the $B_{3u}$ and $B_{2u}$ phonons to large amplitudes by aligning the pump polarization in between the $x$ and $y$ axes. The criterion in Equation (17) can be satisfied with $V_{anh} \propto Q_{B_{2u}} Q_{B_{3u}} Q_{B_{1g}}$. Notably, the R distortion now has $B_{1g}$ symmetry, which breaks the $x$-$z$ and $y$-$z$ mirror planes of the crystal. What makes the proposal more interesting is that the direction of the rectified $B_{1g}$ distortion can be manipulated by the relative phase between $B_{3u}$ and $B_{2u}$ oscillations; this is a feature not afforded by the quadratic-linear anharmonic coupling. Using ErFeO$_3$ as an example, the first-principles calculations in Figure 26 show phonon dynamics under laser driving with different in-plane polarizations. While the $A_g$ distortion (resulting from the quadratic-linear coupling) has a fixed sign for 45 deg and –45 deg polarizations, the $B_{1g}$ distortion (from trilinear coupling) clearly switches sign. The possibility of directional control of ultrafast symmetry-breaking lattice distortions has thus been demonstrated.

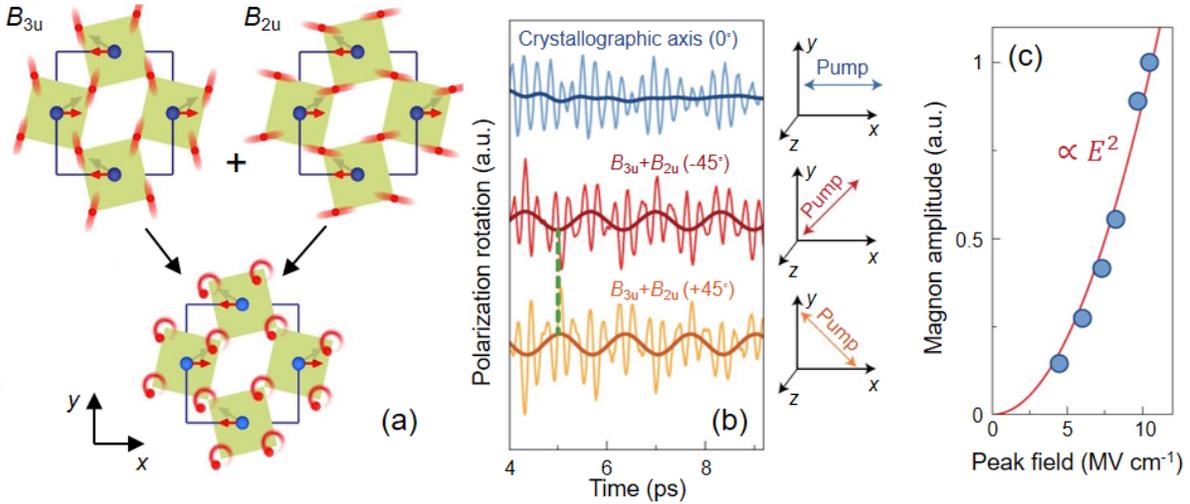

Figure 27. Phonon IFE in ErFeO$_3$.[240] (a) Eigenvectors of in-plane IR phonons, and their superposition. (b) Time-resolved Faraday rotation transients under different THz pump polarizations. Oscillations contain fast Raman phonons and the slow quasi-AFM magnon (outlined by thick solid curves). No magnon is observed when the pump polarization aligns with a crystal axis. (c) Quadratic dependence of magnon amplitude on the pump field. Reproduced with permission from [240].

In fact, an experiment that attempts to simultaneously pump the $B_{3u}$ and $B_{2u}$ phonons in ErFeO$_3$ has already been performed by Nova *et al.*[240] However, instead of observing prominent effects of nonlinear phononics, they discovered a new phonomagnetic effect that has later been interpreted within the framework of dynamical multiferroicity.[241] As shown in Figure 27(a), intense THz fields resonantly drive the two $B_u$ phonons with orthogonal polarizations in a z-cut ErFeO$_3$ in the $\Gamma_4$ phase. With appropriate choices of amplitudes and phases, a superposition of the two phonons can provide circular trajectories for the oxygen atoms. Time-reversal symmetry is broken by the circularly polarized phonons, perturbing the magnetic degree of freedom. Coherent quasi-AFM magnon oscillations are excited, whose amplitude is proportional to the product of the $B_{3u}$ and $B_{2u}$ phonon amplitudes, while no magnon is observed when the pump polarization aligns with one of the crystal axes (driving only one of the two phonons); see Figure 27(b). Considering the selection rule for the quasi-AFM mode, the magnon excitation can be understood as a result of the generation of a transient effective magnetic field along the $z$ axis from the circularly polarized phonons in the $x$-$y$ plane. The effective magnetic field, written as $H_z^{\text{eff}} = -i\alpha_{xyz}Q_{ux}Q_{uy}$, resembles the form of IFE in Table 2, with the only difference that the phonon coordinates replace the laser electric fields in the original expression. The phenomenon is therefore termed the phonon IFE.[242]

Juraschek et al.[241] have later placed the phonon IFE in a general framework called dynamical multiferroicity. They pointed out that, in a way similar to how symmetry arguments predict a spatially varying magnetization $M$ to produce polarization $P$ as $P \sim M \times (\nabla \times M)$, the reciprocal effect, where $P$ produces $M$, should also be allowed. The authors provided

$$M \sim P \times (\partial_t P). \tag{19}$$

Therefore, a time-varying polarization should be able to generate a finite magnetization.

Applying this expression to the experiment by Nova et al.,[240] one can write the in-plane polarizations induced by in-plane phonons as

$$P = \begin{pmatrix} P_1(t) \\ P_2(t) \end{pmatrix} = \begin{pmatrix} A_1 \sin(\omega_1 t) \\ A_2 \sin(\omega_2 t) \end{pmatrix}, \tag{20}$$

where $A_1$ ($A_2$) and $\omega_1$ ($\omega_2$) are the amplitude and frequency of the $B_{3u}$ ($B_{2u}$) phonon, respectively. From Equation (19), this gives

$$M(t) \sim [\frac{\omega_+}{2}\sin(\omega_- t) - \frac{\omega_-}{2}\sin(\omega_+ t)]A_1 A_2 \hat{z}, \tag{21}$$

where there are two oscillation components. The one at the difference frequency $\omega_- = \omega_1 - \omega_2$ oscillates with a larger amplitude ($\propto \omega_+$), while the other at the sum frequency $\omega_+ = \omega_1 + \omega_2$ oscillates with a smaller amplitude ($\propto \omega_-$). In ErFeO$_3$, $\omega_1/2\pi = 17$ THz, $\omega_2/2\pi = 16.2$ THz, giving the large-amplitude oscillation of $M(t)$ to be at $\omega_-/2\pi = 0.8$ THz. Juraschek et al.[241] mentioned that the frequency of $M(t)$ satisfies the resonant condition with the quasi-AFM mode (0.75 THz) of ErFeO$_3$, which explains the observation of magnons by Nova et al.[240]. Furthermore, the fact that $M(t)$ in Equation (21) is along the crystal $z$ axis and is proportional to the product $A_1 A_2$ is consistent with the phonon IFE picture and the observation of the quadratic dependence of the magnon amplitude on the pump field [Figure 27(c)].

Phonomagnetic effects therefore present themselves as a powerful alternative for achieving laser manipulation of magnetic order, and a variety of phenomena are expected, since every photomagnetic effect in Table 2 can have a phonon analog. Furthermore, realistic calculations have shown that for comparable pulse energies, phonomagnetic effects can sometimes generate considerably larger effective fields compared with photomagnetic effects.[242] This further widens the applicability of this technique to dynamical material engineering.

*4.4 Rare-earth pathways*

An alternative approach for inducing SRTs in RFeO$_3$ is to resonantly pump a certain resonant transition of the R$^{3+}$ ions. The magnetic anisotropy of RFeO$_3$ is determined by the magnetic coupling between the R$^{3+}$ and Fe$^{3+}$ moments. While the spin of Fe$^{3+}$ can be viewed as a constant, the moment of $R^{3+}$ is a sensitive function of temperature, owing to the thermal population of the ions within its crystal-field-split levels (separated by ~meV, which is the thermal energy scale of a few tens of Kelvin). The temperature dependence thus enters the $R^{3+}$–$Fe^{3+}$ coupling, and thereby causes the strong temperature dependence of magnetic anisotropy. An idea then arises, asking whether magnetic anisotropy can be transiently manipulated by a nonthermal distribution of $R^{3+}$ in its crystal-field-split levels driven by laser pulses.

Baierl *et al.*[243] have verified this proposal in $TmFeO_3$. The $^3H_6$ ground state of the $Tm^{3+}$ ions in $TmFeO_3$ is fully split by the low-symmetry crystal-field into singlets with energy separations on the order of 1-10 meV; see Subsection 3.2. When an intense THz pulse resonantly excites the electric-dipole-active transitions between the crystal-field-split levels, the $Tm^{3+}$ ions would be driven into an excited state with a nonthermal population distribution; see Figure 28(a). This affects the strength of $Tm^{3+}$-$Fe^{3+}$ coupling, leading to a modification to the magnetic anisotropy and setting a new easy axis for the $Fe^{3+}$ spins. Since the new easy axis deviates from that in equilibrium, magnon oscillations are launched, as shown by the Faraday rotation traces in Figure 28(b). Notably, as the intensity of the pump THz field increases (up to 0.3 T in peak magnetic field), the time-domain magnon waveform clearly develops nonlinearity, with an increasingly prominent low-frequency component emerging from a fast-oscillating background. A frequency-domain analysis of the magnon modes shows that it is the quasi-FM mode that follows a nonlinear amplitude scaling relation with the pump field [Figure 28(c) and (d)]. The nonlinear excitation of magnons falls outside the description of the Zeeman-torque-type excitation in the weak THz field limit, and therefore, is a unique manifestation of THz-field-induced magnetic anisotropy change. What further corroborates this interpretation is the temperature dependence of the nonlinearity in exciting the quasi-FM mode. Since the static anisotropy is close to zero in the SRT temperature range, laser-induced anisotropy modification is expected to be dominant in the $\Gamma_{24}$ phase; this is indeed observed [Figure 28(e)].

Laser-induced SRTs through the rare-earth pathway and the ultrafast heating pathway both rely on modification of the magnetic anisotropy. However, the advantage of the rare-earth pathway is that it is more direct, and the onset of anisotropy modification is expected to be instantaneous. On the contrary, in the case of ultrafast heating, one would need to rely on the $R^{3+}$-lattice coupling [Figure 18(e)], whose speed is sometimes bottlenecked in certain $R^{3+}$ species.

Recently, Fitzky et al.[244] have compared the rise times of anisotropy modification for the rare-earth pathway and the ultrafast heating pathway. As shown in Figure 29(a), in a z-cut $Sm_{0.7}Er_{0.3}FeO_3$ crystal, midinfrared pump fields centered at 25 THz ad 33 THz were used to pump optical phonons and $Sm^{3+}$ $^6H_{5/2} \to {}^6H_{7/2}$ transitions, respectively. The temperature was tuned to the SRT range to cause the effect of laser-induced magnetic anisotropy change to stand out. Within the SRT range (310–330 K), the orientation of the easy axis for spins can be directly correlated with the anisotropy; this scenario is similar to that proposed by Baierl et al.[243] The offset structure developed in Faraday rotation signal probes laser-induced spin rotation, which, on the scale of a 100 ps window, looks similar for 25 THz pumping and 33 THz pumping [Figure 29(b)]. However, a closer comparison of the time dynamics in Figure 29(c) shows that the onset of spin rotation is much quicker for the 33 THz pump than for the 25 THz pump. Such dynamics of spin rotation directly reveal the timescale for setting the transient magnetic anisotropy. As shown in Figure 29(d), the rise time of anisotropy modification is nearly zero (instantaneous) for the 33 THz pump but can be as long as 20 ps for the 25 THz pump for certain temperatures. The 20 ps rise time is exactly the time that is needed for the lattice heat generated by the 25 THz phonon pumping to be transferred to the $Sm^{3+}$ subsystem [Figure 29(a), dashed arrow], as expected for the ultrafast heating scenario. This observation therefore shows that rare-earth pumping is a more direct pathway of manipulating magnetic anisotropy and realizing a laser-induced SRT.

Laser-induced SRTs represent just one of the many examples where $R^{3+}$–$Fe^{3+}$ interaction can be utilized to achieve interesting physics in $RFeO_3$. In a broader context, it points to a variety of possibilities towards engineering magnetic properties of compounds that harbor both a rare-earth subsystem and a transition-metal spin system; examples include relaxation of optical selection rules of electromagnons,[245] cooperative quantum coupling[169] that we will cover in depth in Section 6, and coherent manipulation of spin wave amplitudes.[246] The $R^{3+}$-$Fe^{3+}$ interaction is therefore a property that one should always keep in mind while developing protocols for laser control of $RFeO_3$.

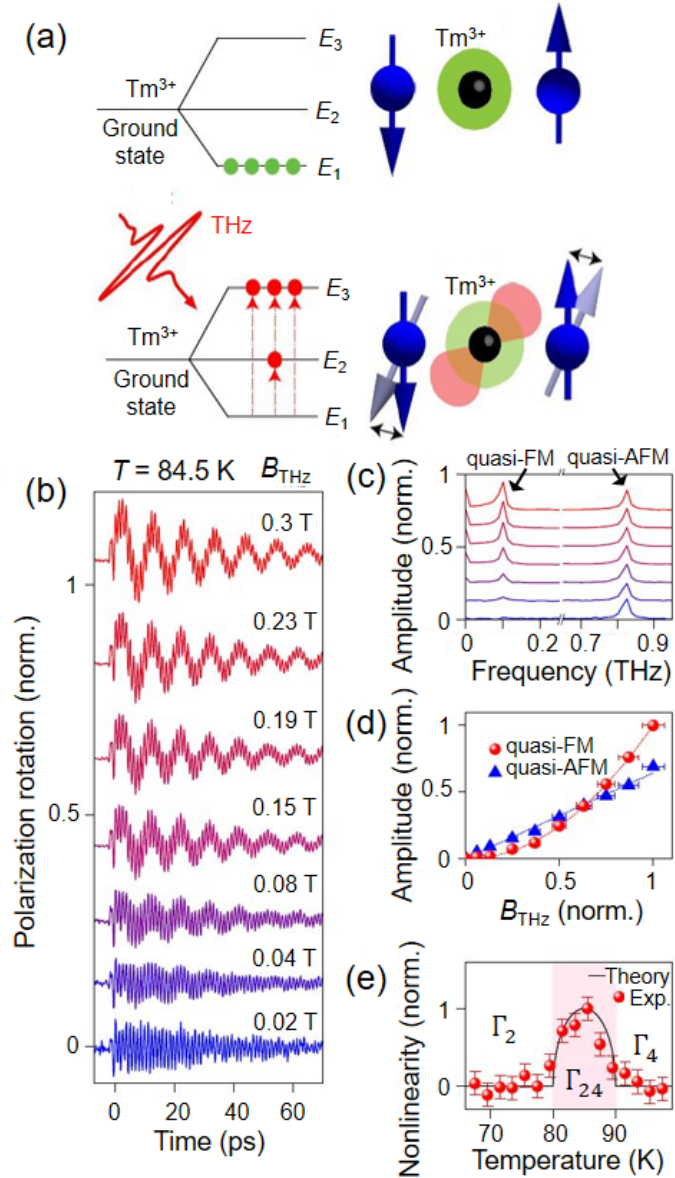

Figure 28. Nonlinear excitation of magnons by pumping rare-earth crystal-field transitions.[243] (a) Intense THz pump repopulates $Tm^{3+}$ ions within its crystal-field levels, which in turn modifies the magnetic anisotropy. (b) Faraday rotation transients under increasing THz pump fields. (c) Fourier transform of (b). (d) Spectral weights of the quasi-FM and quasi-AFM modes extracted from (c). (e) Nonlinearity of quasi-FM excitation versus temperature. Reproduced with permission from [243].

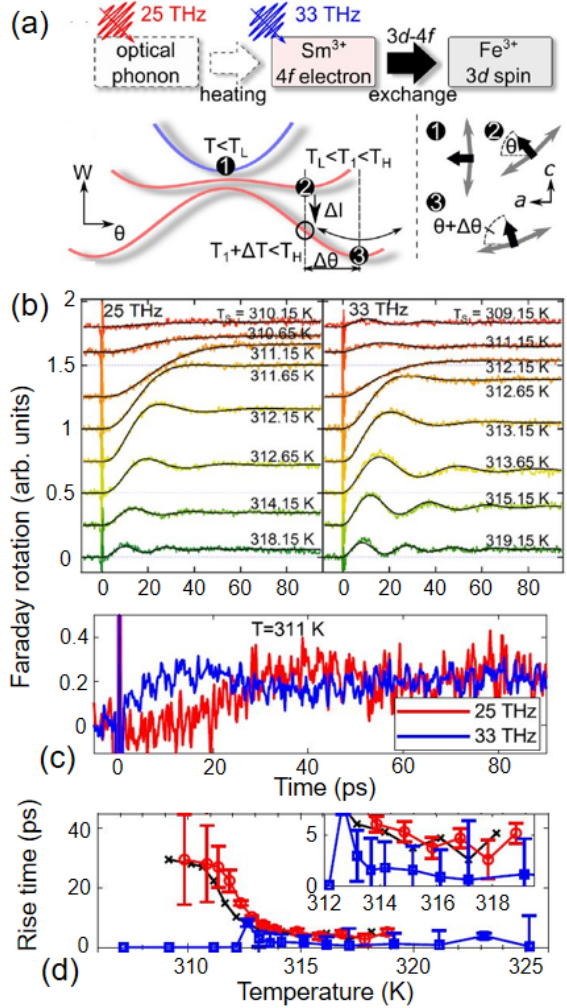

Figure 29. Comparing the onset of magnetic anisotropy due to rare-earth pumping and phonon pumping.[244] (a) The pathway that leads to anisotropy modification. 25 THz pump drives optical phonons, while 33 THz pump drives $Sm^{3+}$ atomic transitions. Lower: anisotropy energy landscape within the SRT temperature range. Orientation of the easy axis is parameterized by $\theta$. Under laser illumination, the easy axis switches to a new direction of $\theta + \Delta\theta$. (b) Faraday rotation transients at various temperatures within the SRT range for a 25 THz pump and a 33 THz pump. (c) Detailed comparison of the rise time of the offset signal. (d) Rise time for establishing anisotropy for the two types of pumps. Blue: 33 THz pump. Red: 25 THz pump. Inset zooms into the 312 K - 320 K range. Reproduced with permission from [244].

*4.5 Floquet pathways*

In recent years Floquet engineering has emerged as a promising tool for controlling quantum many-body systems. The idea is to utilize time-periodic driving conditions to modulate orbital states and dynamically renormalize microscopic interaction parameters in quantum materials, thereby achieving active control of their properties or eliciting novel functionalities that are inaccessible in equilibrium. Since the oscillating electromagnetic fields of radiation provide an ideal source of time-periodic potentials, Floquet engineering of quantum materials in the context of light–matter interaction has invoked a variety of novel ideas and concepts, including band structure engineering,[29,247–251] electronic topology,[252–254] magnetic correlations,[213,215,255–258] and discrete time crystals.[259] Floquet engineering of magnetism is arguably the most exciting proposal with practical applications.

Here we will first briefly introduce the basics of Floquet theory, following Ref. [260] and [261]. Given a time-periodic Hamiltonian $H(t) = H(t+T)$, where $T$ is the period, the evolution of the system can be expanded in the basis of Floquet states $|\psi_n(t)\rangle$. Since the Schrödinger equation of $H(t)$ can be viewed as a time-domain analog to the problem of solving a spatially periodic Hamiltonian (which gives the Bloch theorem $\varphi_k(x) = e^{-ikx}u(x)$ for crystals), the solution to $H(t)$ is similar in format to the Bloch's theorem. The ansatz reads

$$|\psi_n(t)\rangle = e^{-i\varepsilon_n t/\hbar}|\phi_n(t)\rangle, \tag{22}$$

where $\varepsilon_n$ is the quasienergy, and $|\phi_n(t)\rangle$ is a periodic function $|\phi_n(t)\rangle = |\phi_n(t+T)\rangle$ that plays the same role as $u(x)$ in the Bloch theorem. We then write $|\phi_n(t)\rangle$ in Fourier series in terms of harmonics of the driving frequency, $\omega = 2\pi/T$, as

$$|\phi_n(t)\rangle = \sum_m e^{-im\omega t} |\phi_n^{(m)}\rangle, \tag{23}$$

where $|\phi_n^{(m)}\rangle$ is the coefficient of the *m*-th harmonic. By using Equations (22) and (23), the time-dependent Schrödinger equation becomes

$$(\varepsilon_n + m\hbar\omega)|\phi_n^{(m)}\rangle = \sum_{m'} H^{(m-m')}|\phi_n^{(m')}\rangle, \tag{24}$$

where $H^{(m)} = (1/T)\int_0^T dt\, e^{im\omega t} H(t)$ are the Fourier components of the Hamiltonian. Derivation of Equation (24) suggests that the time-dependent problem has been converted into a time-independent one in an enlarged Floquet space.

The solutions for the quasienergy in Equation (24) can be obtained by diagonalizing the matrix equation[261]

$$\mathcal{H}\varphi_n = \varepsilon_n \varphi_n, \tag{25}$$

where

$$\mathcal{H} = \begin{pmatrix} \ddots & H^{(-1)} & H^{(-2)} & & \\ H^{(1)} & H^{(0)} - m\hbar\omega & H^{(-1)} & H^{(-2)} & \\ H^{(2)} & H^{(1)} & H^{(0)} - (m+1)\hbar\omega & H^{(-1)} & \\ & H^{(2)} & H^{(1)} & \ddots & \end{pmatrix}, \quad \varphi_n = \begin{pmatrix} \vdots \\ |\phi_n^{(m)}\rangle \\ |\phi_n^{(m+1)}\rangle \\ \vdots \end{pmatrix}. \tag{26}$$

The enlarged basis vector $\varphi_n$ is constructed by stacking up $|\phi_n^{(m)}\rangle$ for various orders. Furthermore, each entry of $\mathcal{H}$ in Equation (26) represents a matrix of dimension $d$, which is the dimension of the original Hilbert space specified by $H(t)$. The dimension of $\mathcal{H}$ should be infinite due to the infinite number of Floquet sectors, but in order to perform diagonalization, the matrix needs to be truncated up to the $N$-th Floquet sector (while making sure that $N$ is large enough to make the results converge), so the dimension of the actual matrix to diagonalize would be $N \times d$.

Using the method above, Mentink et al.[255] have theoretically proposed Floquet engineering of exchange interactions between spins in Mott insulators. The minimal Hubbard model for a Mott insulator reads

$$H = -t_0 \sum_{\langle ij \rangle \sigma} c^\dagger_{i\sigma} c_{j\sigma} + U \sum_j n_{j\uparrow} n_{j\downarrow}, \qquad (27)$$

where $t_0$ is the nearest-neighbor hopping, $c^\dagger$ ($c$) is the electron creation (annihilation) operator, $\sigma = \uparrow, \downarrow$ is the spin index, $U$ is the onsite Coulomb repulsion, and $n$ is the density operator. The authors considered a two-site cluster model filled with two electrons (half filling), which is exactly solvable, and later found that the simple model is able to capture all the essential Floquet physics that would be expected for a full lattice.[255] Therefore, we will only consider the two-site model here.

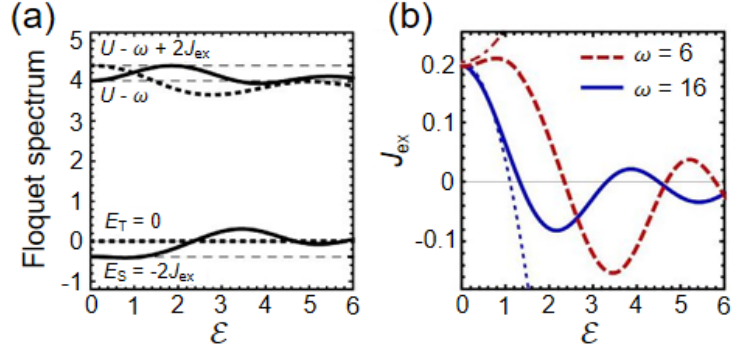

Figure 30. Floquet spectrum and exchange interaction energy of a two-site cluster Hubbard model.[255] (a) Energy level structure versus the Floquet parameter $\mathcal{E}$. (b) Exchange interaction versus $\mathcal{E}$ for two pumping frequencies. The unit is the hopping amplitude $t_0$. Onsite repulsion is assumed to be $U = 10$. The red dash dotted line and blue dashed line are obtained with perturbation theory (valid for $\mathcal{E} \ll 1$). Reproduced with permission from [255].

In the atomic limit ($t_0 = 0$), the electronic states are $|\uparrow,\downarrow\rangle$ and $|\downarrow,\uparrow\rangle$ at $E_1 = 0$, and $|\uparrow\downarrow,0\rangle$ and $|0,\uparrow\downarrow\rangle$ at $E_2 = U$. Turning on the hopping term ($t_0 \neq 0$) lifts the degeneracy, producing the singlet state at $E_S = -4t_0^2/U$ and the triplet state at $E_T = 0$ as the lowest-energy states; their energy separation $E_T - E_S$ is twice the exchange interaction, $2J_{ex}$. The periodic light field drive is considered by using a Peierls substitution $t_0 \rightarrow t_0 \exp[ieaA_{ij}(t)]$, where $e$ is the electron charge, $a$ is the lattice constant, and $A$ is the vector potential of light. The new time-dependent Hamiltonian can be used to derive the $m$-th Fourier component $H^{(m)}$, which can then be plugged in Equation (26) for a full solution of the quasienergies (among which $E_T$ and $E_S$ are most important).

Figure 30(a) and (b) show how $E_T$, $E_S$, and $J_{ex} = (E_T - E_S)/2$ depend on the Floquet parameter $\mathcal{E} = eaE/\hbar\omega$ ($E$ being the light field amplitude), which is nothing but the ratio of the Bloch frequency to the driving frequency. Depending on the pump frequency $\omega$, and in the $\mathcal{E}<1$ range, the Floquet drive can lead to either an increase or decrease in $J_{ex}$, representing strengthening or weakening of the AFM correlation. For certain regimes of $\mathcal{E}>1$, the sign of $J_{ex}$ is reversed upon increasing field strength, suggesting a transition into a FM-type correlation. Since dynamical modification of $J_{ex}$ is expected to cause dramatic impacts on the magnetic order, this set of calculations,[255] derived from the simple cluster model and light-induced hopping renormalization, has profound implications to future experimental attempts.

On the other hand, Chaudhary et al.[262] have taken the orbital degree of freedom into account and proposed the concept of orbital Floquet engineering of magnetism. For most Mott insulating crystals, say, transition metal (TM) compounds, the spin exchange between TM spins is mediated by ligand ions. In addition, TM ions, being placed in strong crystal-field environments themselves, can have distinct orbital ground states and energy level configurations. Therefore, supplying Floquet drives to either ligand orbitals or TM orbitals is expected to cause orbital inter-mixing, and in turn, modify the exchange interaction $J_{ex}$. Using fourth-order perturbation theory, Chaudhary et al.[262] have given analytical expressions of $\Delta J_{ex}/J_{ex}$, the percentage change in $J_{ex}$, due to various types of laser drives. After taking realistic parameters into account, they found that both schemes (pumping ligands or pumping TMs) are challenging for experiments, either due to insufficient transition dipole moments or the requirement to use unusually high pumping frequencies, which, in most experimental situations, cause significant heating effects. However, they found that orbital hybridization caused by *vibrational motions* arising from certain coherent Raman phonons can produce large $\Delta J_{ex}/J_{ex}$; the required pumping conditions are also realistic. Since the time periodicity of the Floquet drive in this case is supplied by coherent phonons rather than laser fields, the proposal represents a unique example of "phonon Floquet physics".

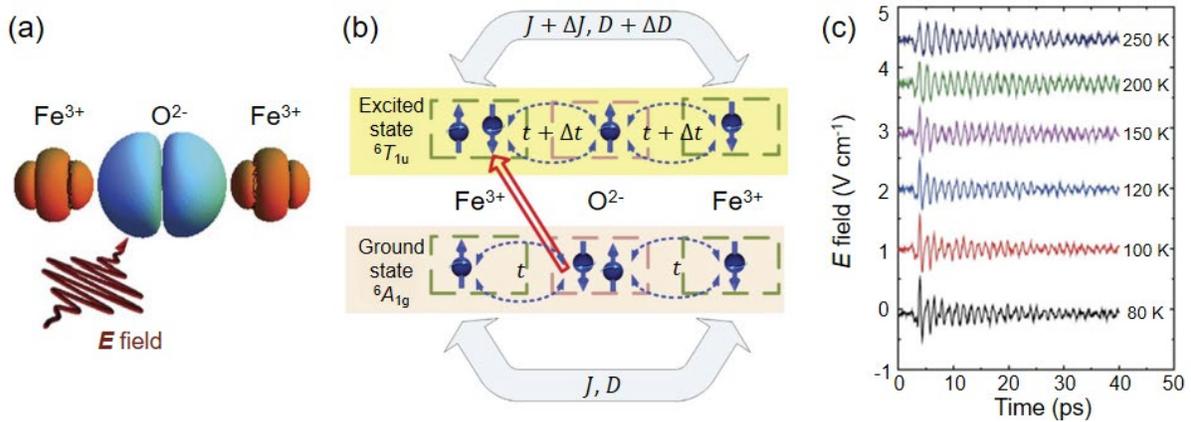

Figure 31. Floquet modification of exchange interaction in $RFeO_3$.[123] (a) Infrared pump pulses (centered at 1.55 eV) nonresonantly drive a Fe-O-Fe cluster. (b) Ultrafast modification of exchange interaction is due to virtual charge transfer arising from light-induced orbital state mixing. (c) Temperature-dependent THz emission signal for $ErFeO_3$. Reproduced with permission from [123].

To date there has been limited success in experimental demonstrations of Floquet engineering of magnetism. Using RFeO$_3$ as a material platform, Mikhaylovskiy et al.[123] have reported THz emission signatures of Floquet modification of exchange interactions. As shown in Figure 31(a), the superexchange interaction between Fe$^{3+}$ spins in RFeO$_3$ are mediated by O$^{2-}$ ligands. A nonresonant laser drive creates *virtual* excitations to the excited level, that is, the charge-transferred orbital configuration of the Fe-O-Fe cluster [Figure 31(b)]. This leads to Rabi splittings and a coherent mixing of the ground state and the excited state of the cluster, thereby providing an instantaneous modification to the exchange interaction. Mikhaylovskiy et al.[123] proposed that the coherent light-spin interaction should obey the interaction Hamiltonian

$$H_I = I_{opt}\alpha \sum_{i,j}(\boldsymbol{S}_i \cdot \boldsymbol{S}_j) + 2I_{opt}\boldsymbol{\beta} \cdot \sum_{i,j}(\boldsymbol{S}_i \times \boldsymbol{S}_j), \qquad (28)$$

where $I_{opt}$ is light intensity, $\alpha$ and $\boldsymbol{\beta}$ are scalar and vector coefficients, respectively. When $H_I$ is added to the equilibrium spin Hamiltonian of RFeO$_3$, the first (second) term in $H_I$ leads to an effective change to the symmetric (antisymmetric) exchange integral, $\Delta J = \alpha I_{opt}$ ($\Delta \boldsymbol{D} = \boldsymbol{\beta} I_{opt}$). As the ratio $D/J$ determines the canting angle of Fe$^{3+}$ spins, the Floquet engineered ratio, $(D+\Delta D)/(J+\Delta J)$, would deviate from the equilibrium value ($D/J$), providing an impulsive torque to change the canting angle, and thereby, launching the quasi-AFM mode. As shown in Figure 31(c), the quasi-AFM mode is indeed observed in THz emission measurements performed on ErFeO$_3$. The mode amplitude does not show prominent dependence on temperature, even across the SRT; this is in contrast to quasi-FM magnons typically launched by ultrafast heating, where the magnon amplitude is largest around the SRT temperature. Furthermore, a unique feature of the interaction Hamiltonian in Equation (28) is that the interaction is isotropic, meaning that it only depends on the light intensity, but remains insensitive to the pump light polarization and propagation direction. This is also found to be consistent with experiments,[123] and distinguishes the Floquet mechanism from photomagnetic effects such as the IFE and ICME, which rely sensitively on the helicity and polarization of the pump light.

Floquet engineering of magnetism is a rapidly developing field. With theoretical models that take more material details into account, say, through first-principles calculations,[215] the proposed novel phenomena would be closer to experimental demonstrations. On the experimental side, in addition to ultrafast optics, a variety of time-resolved scattering techniques[116,257,263] may have the capability to detect the coherent manipulation of magnetic order, fostering more ways for experiment-theory comparisons. This field is therefore expected to see way more interesting results and fruitful collaborations in the future.

# 5 Antiferromagnetic spintronics and coherent magnonic control

Spintronics refers to a field of study that aims to harness electron spins as the information carrier in computational devices.[2,3,264] Spintronic devices and systems are expected to bring revolutionary developments to multiple technological applications related to magnetism, including magnetic recording, sensing, and logical applications, and are believed to hold the potential to complement the existing semiconductor-based microelectronic industry in the next-generation information technology. In the past, the choice of material platforms for performing specific spintronic tasks has predominately been solid-state crystals or films with FM ordering.[265–268] However, the past ten years has witnessed a surging interest in promoting AFM materials to take this central role, establishing the subfield of AFM spintronics.[4,5,67–69,269,270]

Several notable advantages have been identified for antiferromagnet-based spintronics compared to their ferromagnet counterparts. Their imperviousness to external magnetic field fluctuations improves the robustness of information storage. They are more naturally abundant, meaning a wider variety of materials with distinct properties are within the library to choose from. Further, their intrinsically faster response times, set by the magnon frequency and the exchange energy scale, are expected to increase the operation speeds of devices by up to three orders of magnitude.

Experimental efforts to date have been devoted to finding methods for efficiently reading and writing AFM states, in which AFM insulators and metals play different roles. For AFM metals, significant attention has been paid to using electrical current as a direct switch of AFM states. Breakthrough discoveries for efficient switching protocols include spin-transfer torque and spin-orbit torque,[271,272] both of which can generate staggered local effective fields that are commensurate with the periodicity of the AFM sublattices, and thereby, incur much more efficient switching than external uniform fields. On the other hand, AFM insulators are particularly suitable for magnonics applications.[4,7] This subfield of spintronic is concerned with what comes after a spin perturbation (in the form of magnons) is made to an AFM material, and how the magnon excitation can be utilized to the largest extent to transport and process information, especially under the general framework of wave-based computing technologies. Insulators are well suited for magnonics due to their capability of maintaining spin coherence and minimizing energy dissipation due to Ohmic losses. In addition, AFM insulators also find more frequent applications in interfacing optical systems with magnonic devices, since magnons can be selectively excited by THz radiation while accidental excitation of charge carriers is suppressed.

$RFeO_3$ clearly falls under the category of AFM insulators that possess huge application potential in THz-frequency magnonic devices. Therefore, the purpose of this Section is to review various tools discovered so far for exciting, propagating, and manipulating coherent magnons in $RFeO_3$, with special emphasis on how they can facilitate realistic applications. Topics for this review are restricted to $RFeO_3$ systems and include coherent magnon amplitude control through double-pulse interference, magnon-polaritons and magnon propagation effects, $RFeO_3$/FM metal heterostructuring as a means for efficient magnon excitation and SRT control, and finally, nonlinear magnonics. Most of these topics are still at the fundamental physics level, meaning that magnonic devices made from $RFeO_3$ will not enter the market any time soon, but they provide important clues for physicists and engineers to take the most important steps towards the goal.

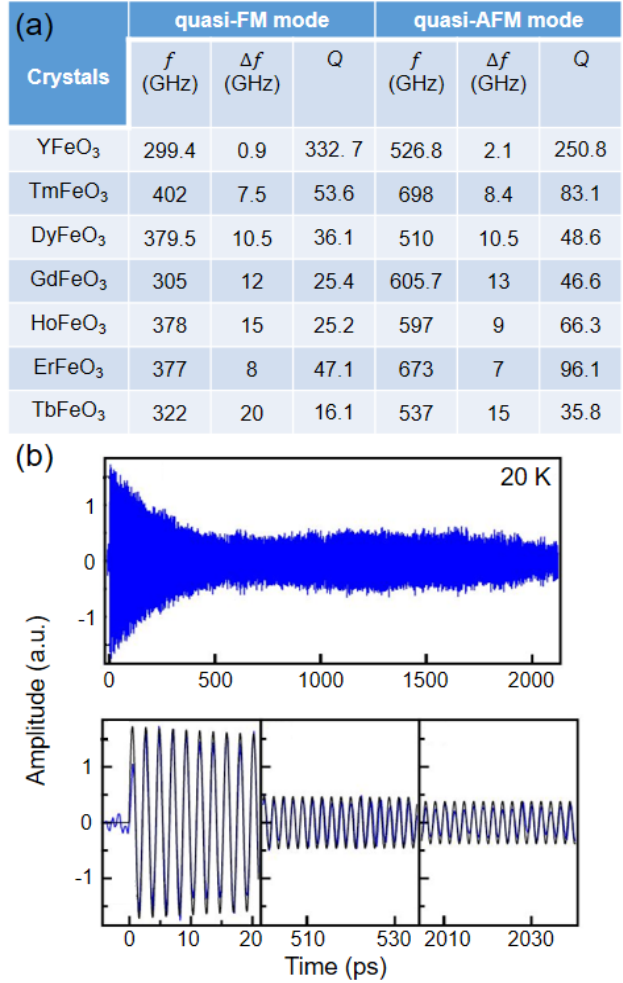

Figure 32. High $Q$ factors of THz magnons in RFeO$_3$. (a) Table summarized using data from Ref. [140], which lists frequency $f$, linewidth $\Delta f$ (in FWHM), and $Q$ factor for quasi-FM and quasi-AFM magnons in various RFeO$_3$ crystals at room temperature. (b) Long-lived temporal oscillations due to quasi-FM magnons in ErFeO$_3$ at 20 K.[273] The bottom row shows zoom-in views of -4 – 21 ps, 500 – 535 ps, and 2005-2040 ps. Reproduced with permission from [273].

Before going into the detailed discussion, we first demonstrate the high quality-factors ($Q$-factors) of the RFeO$_3$ magnons. This is the most important fact because none of the proposals in magnonics that follow would be possible without a demonstrated record of ultrahigh $Q$ factors of magnons in the host materials. Figure 32(a) is a table that summarizes magnon frequencies $f$ and linewidths $\Delta f$ [in full width half maximum (FWHM)], and the corresponding $Q$ factors $Q = f / \Delta f$ for various RFeO$_3$ crystals at room temperature. The data was obtained in the last century by a frequency-domain far-infrared spectroscopy instrument with an outstanding frequency resolution, and was summarized by Balbashov et al.[140] YFeO$_3$ stands out by showing $Q > 300$ for the quasi-FM mode and $Q > 250$ for the quasi-AFM mode. More recently, an even higher value of $Q > 1000$ has been demonstrated by Watanabe et al.[273] in a high-quality ErFeO$_3$ crystal cooled to 20 K; see Figure 32(b) for time-domain quasi-FM oscillation signal that lasts for at least 2 ns. Such a high $Q$-factor readily exceeds typical $Q$-factors (of 10-100) for other model AFM insulators such as oxides NiO/MnO[274,275] and difluorides MnF$_2$/FeF$_2$/CoF$_2$[276,277], and even approaches that of yttrium iron garnet,[278–280] a material widely applied in microwave magnonics due to its exceptionally low damping rates. The high $Q$-factor ensures good spatial and temporal coherence, thereby setting the stage for RFeO$_3$ to play a central role in THz-frequency magnonic devices in the future.

*5.1 Magnon amplitude control through double-pulse interference*

We first describe previous work demonstrating control of magnon amplitude by interference effects. To observe this phenomenon directly in the time domain using ultrashort laser pulses, it is important to first clarify a concept called the free induction decay (FID), which we briefly mentioned in Section 3 but did not elaborate. FID signals appear in the context of interaction between an ensemble of two-level systems (TLSs) with a pulsed light field whose frequency is tuned near resonant with the TLS level spacing. For spin systems (since they are TLSs, with spin up and down being the two levels), the status of the spin ensemble can be represented by a Bloch vector, which develops a tilt away from its equilibrium orientation after the impulsive laser-spin interaction. The Bloch vector then precesses around the equilibrium orientation, leading to emission of a light field that persists until the Bloch vector decays back to the equilibrium state. The FID is defined as the subsequent emission field by the oscillating Bloch vector after the impulsive laser drive.

Coherent control of magnons refers to coherent control of FID signals arising from magnon oscillations, which is useful in magnonic logic gates compatible with wave-based computation protocols.[6–10] When the binary data is encoded into the magnon amplitude, a Mach-Zender-type interferometer setup depicted in Figure 33(a) can be used to achieve NOT and XNOR gates.[7,281] The idea is to split a magnon wave into two branches, each complemented by a phase shifter, and combine the phase-shifted signals from the two branches into one at the output. Such an interferometer setup can be effectively demonstrated by an optical experiment with double-pulse pumping capabilities, as we discuss below.

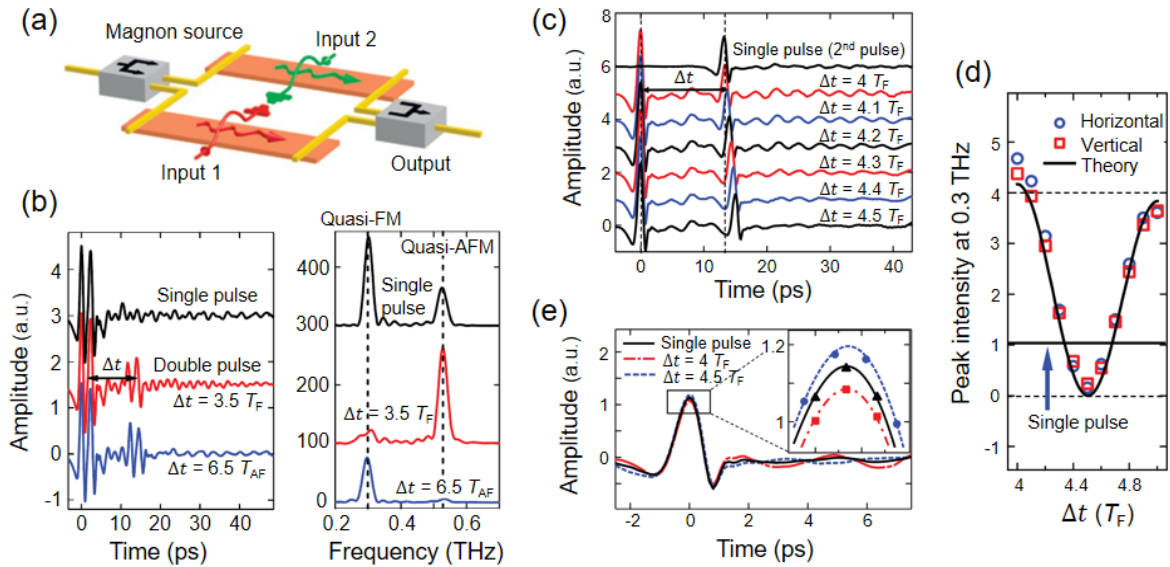

Figure 33. Double-pulse coherent control of magnons in $YFeO_3$.[282] (a) Mach-Zender-type interferometer setup used to achieve NOT and XNOR gates.[7,281] (b) Double-pulse pumping experiment on an *x*-cut

crystal. Left: time-domain signals with single-pulse and double-pulse excitations. Right: Fourier transform of the FID signal. (c) Coherent control of the quasi-FM mode in a z-cut crystal. THz electric field is parallel along the y axis. (d) Fourier transform of the FID signal in (c) versus inter-pulse time delay. (e) Energy lost in the magnon sector due to destructive interference is restored back to the second pulse. Reproduced with permission from [7] and [282].

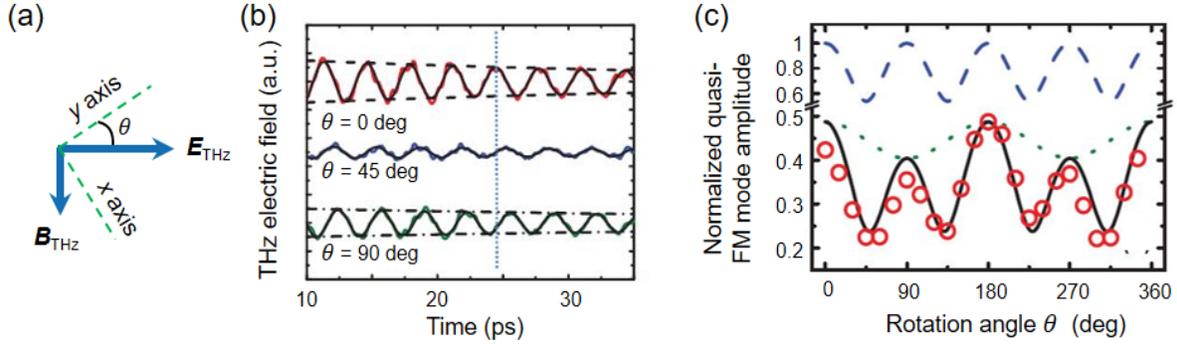

Figure 34. Single-pulse coherent control of magnons in YFeO$_3$.[283] (a) Polarization configuration. (b) FID signal versus the polarization angle $\theta$. (c) FID signal amplitude (derived from Fourier transform, shown in red circles) versus $\theta$. Black solid line is model calculation using Equation (31). Blue dashed and green dotted lines consider either the birefringence or the dichroism (but not both). Reproduced with permission from [283].

Yamaguchi et al.[282] have demonstrated coherent control of magnon FID signals through interference in a double-pulse pumping experiment on YFeO$_3$. Two THz pulses with controllable delay excite two sets of FID oscillations with a controllable phase shift, exactly mimicking the two branches of magnons that propagate within a Mach-Zender-type interferometer. The detector reads a superposition of the two FID signals in the time domain. The pump polarization is first aligned along a direction of the crystal to enable simultaneous excitation of both the quasi-FM and quasi-AFM modes by a single pulse. As shown in Figure 33(b), when double-pulse pumping is used, one of the two modes can be selectively suppressed by destructive interference, which is achieved by setting the inter-pulse time delay to be half-integer multiples of the magnon oscillation period. The authors then switched to a new configuration where only the quasi-FM mode can be excited, and the coherent control of FID amplitude through constructive or destructive interference is again demonstrated; see Figure 33(c) and (d) for how time- and frequency-domain FID amplitudes evolve with inter-pulse delay. In addition, energy restoration back to the radiation field due to destructive interference of magnons is also demonstrated. As shown in Figure 33(e), when the second pulse is examined, its integrated energy (represented by the peak) is higher for the double-pulse pumping scenario in which the magnon launched by the first pulse is quenched by the second pulse due to destructive interference. This suggests that the energy lost in the spin sector is transferred back to the radiation field, manifesting as an energy gain by the second pulse.

Jin et al.[283] have subsequently developed another protocol that enables a double-pulse coherent control experiment by using only a single THz pulse. The key idea is to utilize the crystal anisotropy to split the pulse into two with perpendicular polarizations and mismatched velocities. As shown in Figure 34(a), a THz pulse incident onto a $z$-cut YFeO$_3$ crystal is polarized in the $x$-$y$ plane; the angle formed by the THz electric vector and the $y$ axis is $\theta$. In this configuration, only the quasi-FM mode can be excited, and tuning $\theta$ leads to a branching of power according to

$$E_x(0) = E_{\text{THz}}^{\text{in}} \sin(\theta),$$
$$E_y(0) = E_{\text{THz}}^{\text{in}} \cos(\theta),$$
(29)

where $E_x(0)$ and $E_y(0)$ represent the electric field amplitude along the $x$ and $y$ axes at the incident surface, respectively, and $E_{\text{THz}}^{\text{in}}$ is the incident field amplitude. The fields $E_x$ and $E_y$ then propagate with distinct speeds in the crystal with thickness $d$, so that at the exit

$$E_x(d) = E_{\text{THz}}^{\text{in}} \sin(\theta) e^{-\alpha_x d/2} e^{i\omega n_x d/c}, \tag{30}$$
$$E_y(d) = E_{\text{THz}}^{\text{in}} \cos(\theta) e^{-\alpha_y d/2} e^{i\omega n_y d/c},$$

where $\alpha_{x(y)}$ and $n_{x(y)}$ are the absorption coefficient and refractive index along the $x$ ($y$) axis, respectively, $\omega$ is the angular frequency, and $c$ is the speed of light. The output THz signal at the exit can be derived as follows:

$$E_{\text{THz}}^{\text{out}} = E_x(d)\sin(\theta) + E_y(d)\cos(\theta) = E_{\text{THz}}^{\text{in}} e^{-\alpha_y d/2}[\cos^2(\theta) e^{i\omega(n_x - n_y)d/c} e^{-(\alpha_x - \alpha_y)d} + \sin^2(\theta)]. \tag{31}$$

This equation provides an analytical expression useful for realizing coherent control. Figure 34(b) and (c) show how the FID signal varies with $\theta$. The on/off ratio of the FID amplitude reaches 3 and can be explained well by Equation (31) by using realistic parameters for the refractive index and absorption coefficient.

Double-pulse coherent control experiments using ideas similar to the two examples above have been demonstrated with various types of excitations, including magnetic resonances,[274] phonons,[284–286] and charge excitations.[287] All these studies have helped establish a complete picture of wave-based logical gates that operate based on the linear superposition principle.

## 5.2 Magnon-polaritons and magnon propagation

Another key subject of research in magnonics is the transport of information carried by magnons from one place to another.[288] This can be done through two methods. One is to first attempt converting the energy stored in magnons into photons, and then to propagate the photons to transfer the information. The method requires fabrication of a hybrid magnon-photon coupled device to form the so-called magnon-polaritons.[289–292] The second method is to utilize the intrinsic propagation effect of magnons themselves to transfer information. Therefore, this Subsection aims to review magnon-polaritons and magnon propagation effects in RFeO$_3$. The reason for arranging these two topics together is because they are not fundamentally distinct effects from the level of physical principles. Some magnon-polariton devices do not offer confinement in all spatial dimensions, and therefore, allow polariton propagation along certain directions. Furthermore, for magnons propagating in a magnetic crystal, their magnetic dipoles are spontaneously coupled to a light field even in free space, which makes polaritonic effects appear. We will discuss below how polaritonic effects cross over to propagation effects as the spatial confinement is released step by step. These unique observations have been well established in RFeO$_3$ thanks to the high spatial and temporal coherence of magnon excitations afforded by the material system.

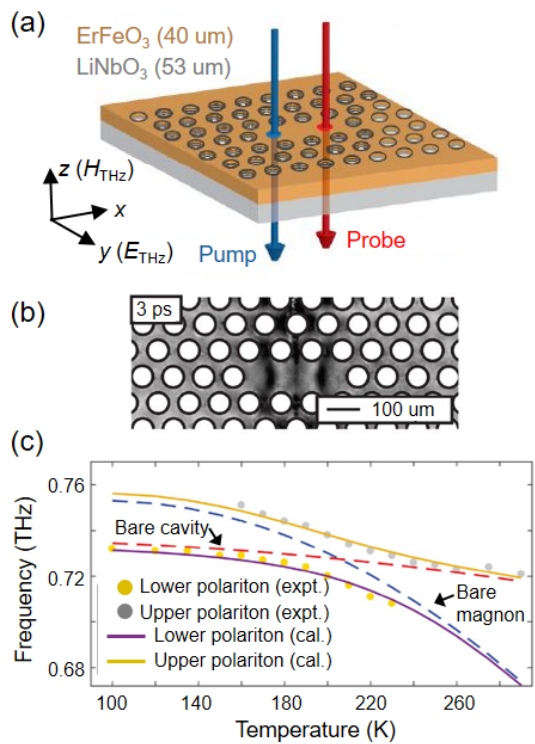

Figure 35. Magnon-phonon-polaritons in a photonic crystal cavity.[293] (a) Experimental configuration. (b) Electro-optic sampling imaging of the cavity mode 3 ps after pump excitation. (c) Anticrossing branches of the magnon-phonon-polariton. Temperature is adjusted to detune the magnon and the phonon-polariton

frequencies. Grey markers: data. Solid lines: polariton branches. Dashed lines: mode frequencies assuming no coupling. Reproduced with permission from [293].

Sivarajah *et al.*[293] have realized magnon-phonon-polaritons in a LiNbO$_3$-ErFeO$_3$ cavity. The device, depicted in Figure 35(a), consisted of a two-layer structure with a polished ErFeO$_3$ crystal stacked on top of a LiNbO$_3$ wafer. A hexagonal lattice of through holes were drilled on the structure by femtosecond laser machining; three sites were intentionally left undrilled to create a defect in the photonic crystal, supporting a cavity mode. Through impulsive stimulated Raman scattering (ISRS), a 3 eV pump laser beam directed into the cavity excited coherent phonon-polaritons in LiNbO$_3$, whose magnetic field component aligns parallel to the canted moment in the adjacent ErFeO$_3$ layer and therefore couples to its quasi-FM magnon mode. The spatial and temporal profiles of the magnon-phonon-polariton were mapped out by electro-optic sampling microscopy [Figure 35(b)]. Utilizing the different temperature dependences of the magnon mode in ErFeO$_3$ and the phonon-polariton mode in LiNbO$_3$, the authors performed a temperature scan to achieve resonance frequency detuning, as shown in Figure 35(c). Around the temperature range where the magnon mode (red dashed) and the phonon-polariton mode (blue dashed) are supposed to intersect in frequency, two coupled modes appear with a frequency splitting of 16 GHz, suggesting the formation of strongly coupled magnon-phonon-polaritons.

Cavity magnon-polaritons based on RFeO$_3$ have also been demonstrated by Bialek *et al.*[294] They performed normal-incidence THz transmission experiments on a polycrystalline DyFeO$_3$ disk sample. The front and back surfaces of the sample provided the cavity confinement and the Fabry–Pérot cavity modes. As shown in Figure 36, the scan was performed in the high temperature range around the Fe$^{3+}$ ordering temperature, where both the quasi-FM and quasi-AFM modes soften significantly according to $f \propto (1-T/T_\mathrm{N})^\beta$, with $\beta \sim 1/3$. The THz setup was a continuous-wave spectrometer based on frequency extensions to a vector network analyzer, and the measurement quantity was the temperature-differential $S$ parameter $\partial S_{21}/\partial T$. Both the experimental data [Figure 36(a)] and numerical simulations of the phase of $\partial S_{21}/\partial T$ [Figure 36(b)] show clear avoided crossing features in the eigenmodes of the sample; these are hallmarks of polariton modes formed by strong coupling between the magnons and the Fabry–Pérot cavity modes.

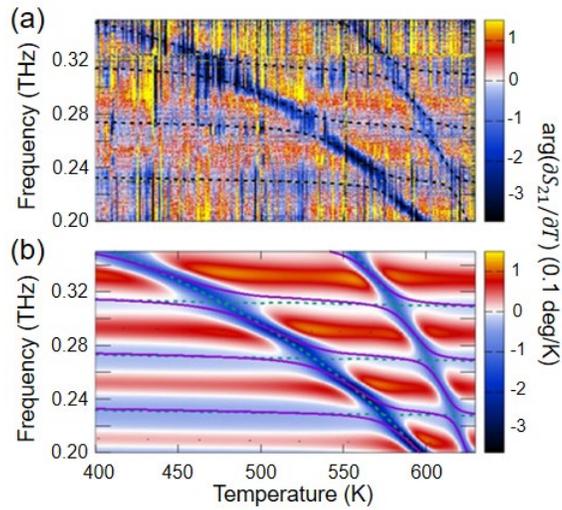

Figure 36. Magnon-polaritons in a Fabry–Pérot cavity.[294] (a) Experimental phase shift of $\partial S_{21}/\partial T$. Dashed lines mark the polariton modes. (b) Model simulations. Dashed green lines: uncoupled modes. Modes with strong (weak) temperature dependence are magnon (cavity) modes. Purple solid lines: polariton modes. Reproduced with permission from [294].

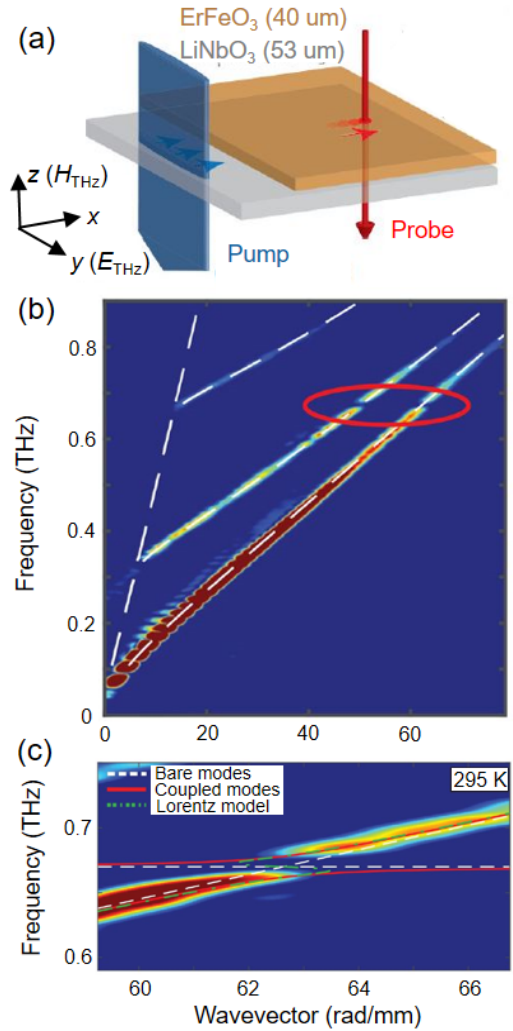

Figure 37. Magnon-phonon-polaritons in a hybrid waveguide.[293] (a) Experimental configuration. (b) Dispersion relation of transverse-electric phonon-polariton modes. (c) Zoom-in view of the red-oval-enclosed region, where polariton branches form an anticrossing pattern. Reproduced with permission from [293].

The two examples above both represent magnon-polaritons in fully confined cavities. However, magnon-polaritons can also exist in a structure that supports propagation, such as in a waveguide. This has also been demonstrated by Sivarajah et al.,[293] in a LiNbO$_3$–ErFeO$_3$ device which had a similar layered structure to that shown in Figure 35(a), only that all the air holes were removed to lift the spatial confinement in the slab plane; see Figure 37(a). Such a waveguide device therefore allowed propagation (the wavevector can be continuously varied) within the sample plane, but still confined the field in the out-of-plane direction, due to the index mismatch of the slab with the surrounding environment. The pump was shaped into a line, which, when directed onto the bare $LiNbO_3$ region, excited phonon-polaritons through ISRS. The phonon-polaritons then propagated into the $ErFeO_3$-$LiNbO_3$ region and coupled to the quasi-AFM magnons in $ErFeO_3$. A 2D Fourier transformation directly gave the dispersion relation of the polariton [Figure 37(b)] when the authors used the probe to sample the polariton field spatially and temporally. The dispersion relation showed two prominent near-linear dispersive lines outside the light cone that were identified as the first two transverse-electric modes of the polariton waveguide. Signature of strong coupling to magnons was again found by zooming into the frequency window around the quasi-AFM frequency [red oval enclosed region in Figure 37(b)] and observing the mode anticrossings [Figure 37(c)]. The coupling strength of the magnon-phonon-polariton in this waveguide structure reached 20 GHz.

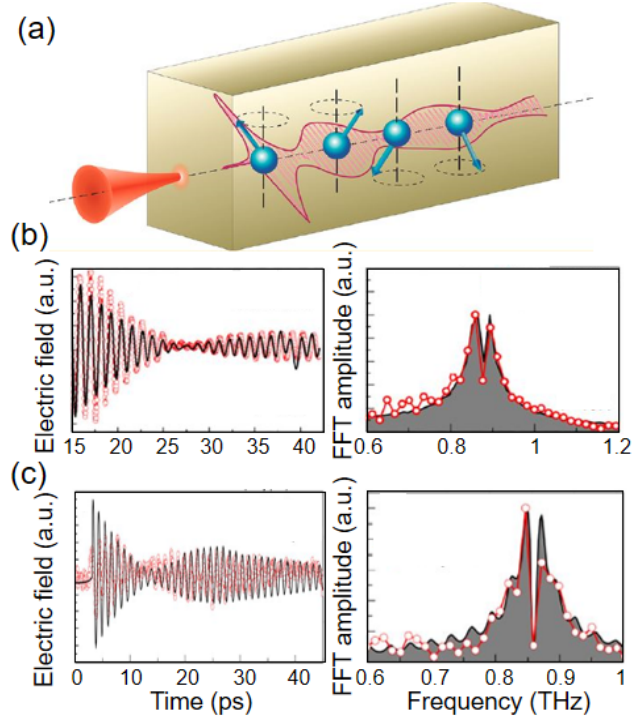

Figure 38. Signatures of magnon-polaritons in free space.[295] (a) As optically excited magnons propagate in a TmFeO$_3$ crystal, they couple with a free-space light field and form polariton modes. (b) FID signal in transmission experiments (left) and its Fourier transform (right). (c) Emission fields in THz emission experiment (left) and its Fourier transform (right). Simulation results (black lines and grey shades) are overlaid on data (red hollow circles). Temperature: 40 K. Reproduced with permission from [295].

Having observed waveguide-based magnon-polaritons, one natural question that arises is whether polaritonic effects can still be seen within a simple setup where magnons propagate freely in a magnetic crystal, with all photonic confinements removed. Grishunin *et al.*[295] investigated this problem by performing THz transmission and emission experiments in a TmFeO$_3$ crystal [Figure 38(a)]. By using an appropriate crystal cut and polarization configuration, they selectively excited the quasi-AFM magnon. Surprisingly, both the FID signal (in the transmission experiment) and the emitted THz field (in the emission experiment) developed beating patterns when the measurement time window became longer than 50 ps [Figure 38(b) and (c)]. The time-domain beating suggests a frequency splitting of the quasi-AFM mode, which can be interpreted as the upper and lower branches of magnon-polaritons in the dispersion relation, where the light line is expected to intersect the magnon line; the interpretation is supported by a match between experiment and a classical electromagnetic simulation without any adjustable parameter [Figure 38(b) and (c)]. Interestingly, the polariton branches could be observed only for crystals that are thick enough, indicating that it arises from a delicate energy exchange process between magnons and photons as magnons are propagating through the crystal. It sets the requirement that both a long interaction length and long interaction time between magnons and photons are needed for polaritonic effects to stand out in free-space magnonic devices.

One may notice that the magnon propagation effect Grishunin *et al.*[295] pointed out is exclusively associated with the phase velocity. The group velocity is assumed to be near zero because optically excited magnons are expected to have very small momentum. This statement holds true for most systems employed so far where coherent magnons are launched through optical methods. A true demonstration of the transport of an optically excited magnon wavepacket at a high group velocity therefore would be of high scientific value. This task has recently been achieved by Hortensius *et al.*[296] using a $DyFeO_3$ crystal.

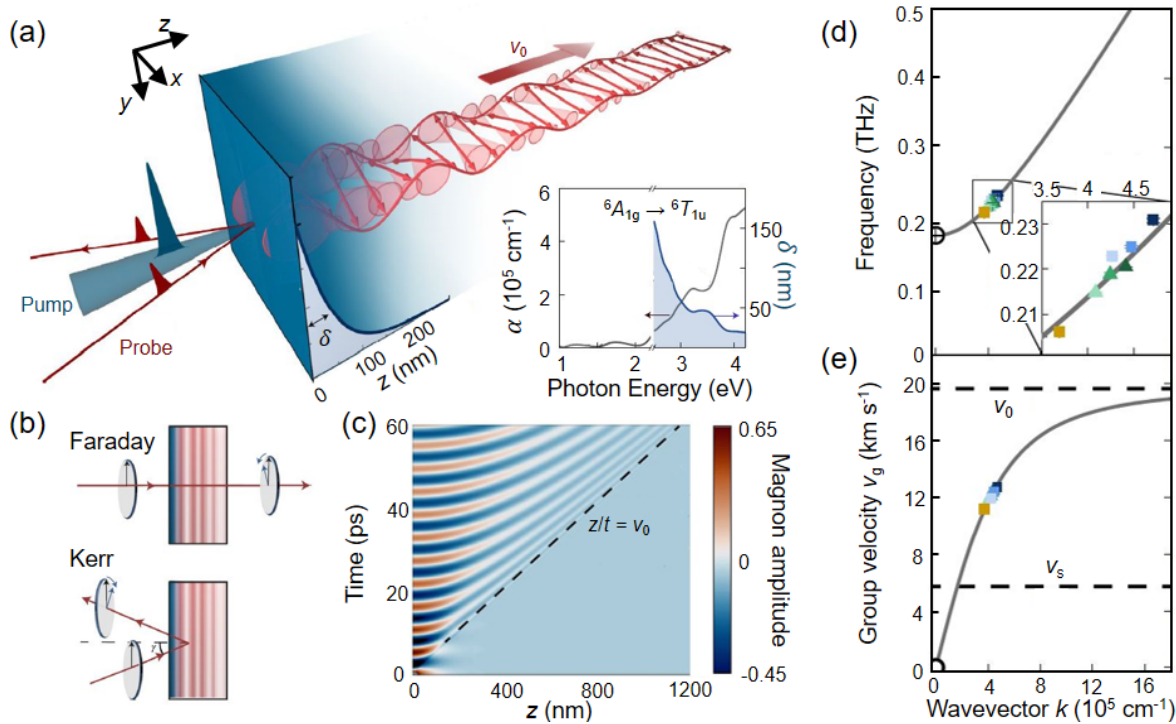

Figure 39. Demonstration of magnon propagation at a supersonic group velocity.[296] (a) High-$k$ magnons can be excited if pump excitation is confined in a very small penetration depth. Inset: penetration depth and absorption coefficient versus photon energy. (b) Within the Faraday and Kerr measurement geometries, the Kerr geometry is sensitive to high-$k$ modes. (c) Simulated chirping of the magnon wavepacket. (d) Reconstruction of the magnon dispersion relation by probe-wavelength-dependent measurements using the Kerr geometry. (e) Group velocity plot obtained from the dispersion. Colored markers: experimental data. Black dashed lines: sound velocity $v_s$ and limiting group velocity $v_0$. Reproduced with permission from [296].

As shown in Figure 39(a), the key idea of Hortensius *et al.*[296] that enabled launching high-$k$ ($k$: wavevector) magnons optically was to tune the pump photon energy up to the strongly absorbing charge-transfer peak for which the penetration depth is very small (~ 50 nm). This resulted in confining the pump excitation (including pump-excited magnons) within a very thin layer of the sample, promoting the upper limit of wavevector that becomes accessible. The way the authors discerned the high-$k$ magnons was through the peculiar observation that the quasi-AFM magnon frequency measured by time-resolved Kerr rotation was consistently higher than that obtained from time-resolved Faraday rotation [Figure 39(b)]. This result was interpreted by considering the magnon dispersion

$$\omega_k = \sqrt{\omega_0^2 + (v_0 k)^2}, \tag{32}$$

where $\omega_k$ ($\omega_0$) is the magnon frequency at finite $k$ (zone center) probed in the Kerr (Faraday) geometry, and $v_0 \approx 20$ km/s$^{-1}$ is the limiting group velocity. What makes the Kerr geometry sensitive to finite-$k$ modes is the Bragg condition that relates the probe wavelength $\lambda_0$ with the bright wavevector $k_m$ through

$$k_m = 2k_0 n(\lambda_0) \cos\gamma, \tag{33}$$

where $k_0 \cos\gamma$ is the surface-normal component of the wavevector of the probe beam, and $n(\lambda_0)$ is the index of refraction at $\lambda_0$. This means that, although the initial magnon excitation contains a broad spectrum of $k$ [see Figure 39(c) for how the wavepacket gets quickly chirped as time elapses], for a given $\lambda_0$, only one $k_m$ component is detected (and therefore "bright"), giving $\omega_{k_m} = \sqrt{\omega_0^2 + (v_0 k_m)^2}$. The authors performed a series of probe-wavelength- and angle-dependent measurements on the magnon frequency to map out the dispersion within a certain range of $k$ [Figure 39(d)], and the slope of the dispersion around these data points determined the group velocity defined as

$$v_g = (\partial \omega_k / \partial k)|_{k=k_m}. \tag{34}$$

Figure 39(e) shows $v_g$ versus $k$, and quantitatively gives the supersonic $v_g$ that is achieved by the high-$k$ magnon modes observed in this experiment.

The rapidly developing research on magnon-polaritons and magnon propagation effects in RFeO$_3$ is expected to open a new avenue for versatile design and control of future magnonic devices that operate in the THz frequency range.

*5.3 Heterostructuring and exchange-biased interfaces*

A useful strategy in constructing functional spintronic devices is to place two magnetic materials in proximity, typically by depositing two layers of films together to form a heterostructure.[297–300] The exchange coupling between spins at the interface changes the spin states in both layers, and therefore, provides a tuning knob for magnetic control. The most common usage of antiferromagnetic materials in magnetic heterostructures is to form an exchange bias device.[301] In a heterostructure where an AFM layer is adjacent to a soft FM layer, the exchange coupling causes a shift in the magnetic hysteresis loop of the FM layer. Since the AFM state cannot be easily perturbed by external magnetic fields, once the AFM spin state is set, the direction and magnitude of the loop shift would be determined, causing a robust pinning of the FM layer magnetization at zero magnetic field. In this Subsection, we review efforts to study FM/RFeO$_3$ heterostructures and the unique properties they provide for magnonic applications and SRT control.

Tang *et al.*[302] have discovered unusually efficient magnon excitation by optical pulses in a Fe/RFeO$_3$ (R = Er or Dy) heterostructure. One of their sample structures consisted of a 3.5-nm-thick FM Fe film deposited on top of a polished *x*-cut crystal of ErFeO$_3$, and another 2-nm-thick nonmagnetic Cu layer coated on Fe as the protective layer. Static characterization of the soft Fe film using the longitudinal MOKE revealed a clear exchange bias effect; see the shift of the hysteresis loop in Figure 40(a) from zero under a weak magnetic field applied along the *z*-axis of the ErFeO$_3$ crystal. Figure 40(b) shows results of time-resolved MOKE measurements on the same sample at room temperature (ErFeO$_3$ in $\Gamma_4$ phase). Oscillations are observed, whose amplitude is considerably larger than that from a bare ErFeO$_3$ substrate (without the Fe film coating) under the same experimental conditions. Through a Fourier-domain analysis, they found that the oscillations arise from a superposition of multiple mode excitations including the quasi-FM magnon mode, an impurity mode, and a low-frequency coherent phonon mode; measurements on a heterostructure fabricated on a *z*-cut crystal demonstrated efficient excitation of the quasi-AFM mode as well (quasi-AFM only observable in *z*-cut samples due to a selection rule). The extraordinarily efficient magnon excitation in Fe/RFeO$_3$ heterostructures opens up opportunities in magnonic applications. Its microscopic mechanism should be distinct from ultrafast laser-induced heating, because otherwise it would only be expected around and slightly below the SRT temperature (Subsection 4.1). In the mechanism that the authors proposed [Figure 40(c)], interfacial antiferromagnetic exchange interactions cant the macroscopic moments of RFeO$_3$ out of the sample plane. Pump excitation quenches the interfacial exchange transiently, making the moments experience a restoring force back to the in-plane *z*-axis, which is the orientation expected for a bare RFeO$_3$ substrate without the Fe capping layer. This mechanism enables launching magnon modes efficiently across the entire temperature range.

Joly *et al.*[303] have explored the possibility of triggering an SRT in the FM layer through coupling to the SRT in RFeO$_3$ in a FM/RFeO$_3$ heterostructure. As shown in Figure 41(a), the sample consisted of a *y*-cut SmFeO$_3$, on top of which a FM 2-nm-thick Co film and a 1-nm-thick Pt protective layer were deposited. SmFeO$_3$ undergoes a $\Gamma_4 \to \Gamma_2$ SRT between 460 K and 470 K, so a 90-degree in-plane spin rotation is expected for a *y*-cut crystal. A circularly polarized x-ray beam, whose energy is resonant with the Co L$_3$-edge, impinged on the sample with grazing incidence (16 degrees). The x-ray magnetic circular dichroism (XMCD) directly reveals spin alignment in the Co film, with XMCD being finite only when spins have finite projections to the propagation direction of the x-ray beam. Figure 41(b) shows a set of temperature-dependent XMCD images taken in two incident geometries. When the x-ray propagates parallel to the *x*-axis (*z*-axis), the image shows high (low) contrast at 300 K but low (high) contrast at 485 K. Since a high image contrast indicates the coexistence of strong XMCD signals with opposite signs, it directly proves that (1) spins are along the current x-ray propagation direction, and (2) magnetic domains are forms in the Co film. Figure 41(b) therefore shows an SRT with spins parallel to the *x*-axis at 300 K and to the *z*-axis at 485 K in the Co film. This coincides with the SRT in $SmFeO_3$, and therefore, provides the proof for spin interlocking at the interface.

Using the XMCD image contrast as an indicator of spin orientation of the Co film, Joly *et al.*[303] also attempted to induce a correlated SRT through laser-induced heating. Figure 41(c) shows analogous maps to Figure 41(b), only that the heating source was a picosecond laser beam; the temperature was set to 425 K, which is slightly below the SRT in $SmFeO_3$. An SRT in the Co film that occurs in tandem with the SRT in $SmFeO_3$ upon static laser heating is again observed. Furthermore, Le Guyader *et al.*[304] have later extended this work to the dynamical realm by providing time resolution to the experiment. They adjusted the sample temperature to be initially within the SRT range, used an optical pulse to provide transient heating, and collected the XMCD images by a time-delayed x-ray pulse. Figure 41(d) shows images before and after time zero, and their subtractions, for the two incident geometries. The subtraction highlights the pump-induced change, and one can tell that pumping causes a decrease (increase) in the image contrast for x-ray propagating along the *x*-axis (*z*-axis). Although the change is small, it reflects a spin rotation by a few degrees from the *x*-axis to the *z*-axis caused by the optical pump. The rotation builds up within sub-100 ps, and a more accurate measurement of the rotation time is needed with a better instrumental resolution. Investigating an SRT in a FM film exchange-locked to an AFM material provides opportunities to interrogate the ultimate speed limit of spin switching in FM materials. The experiments carried out by this series of work here is expected to reveal much insight towards this goal.

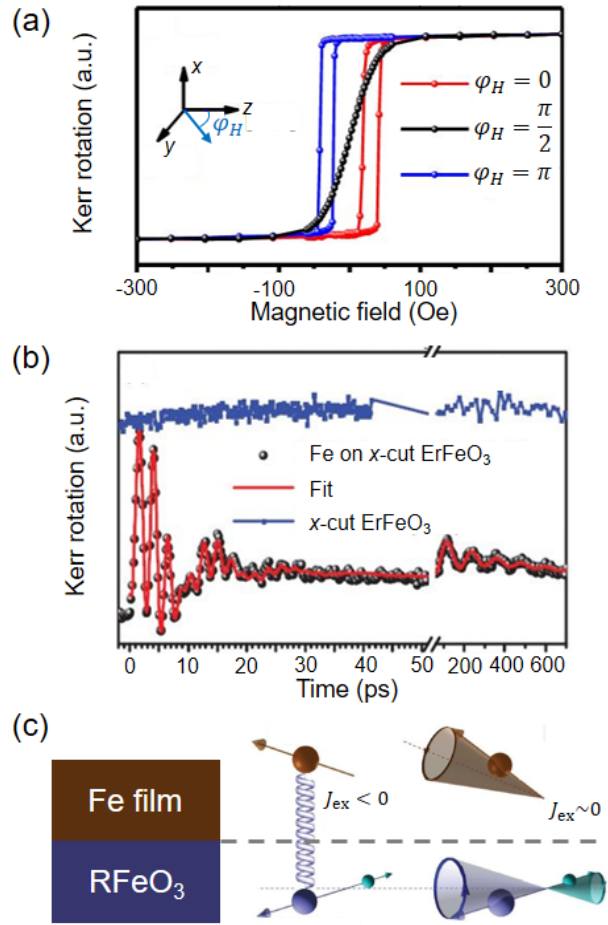

Figure 40. Efficient magnon excitation in a Fe/RFeO$_3$ heterostructure.[302] (a) Magnetic hysteresis loops of the Fe/ErFeO$_3$(100) sample for various field orientations. Fields never exceed the coercive field of RFeO$_3$. (b) Time-resolved MOKE transients for Fe/ErFeO$_3$(100) and the bare ErFeO$_3$(100) substrate under comparable pump fluences. (c) The proposed microscopic mechanism that explains efficient magnon excitation across a wide temperature range. Spins are canted initially due to interfacial exchange (left).

Quenching of exchange by optical pulses leads to a restoring force and thereby launches magnons (right). Reproduced with permission from [302].

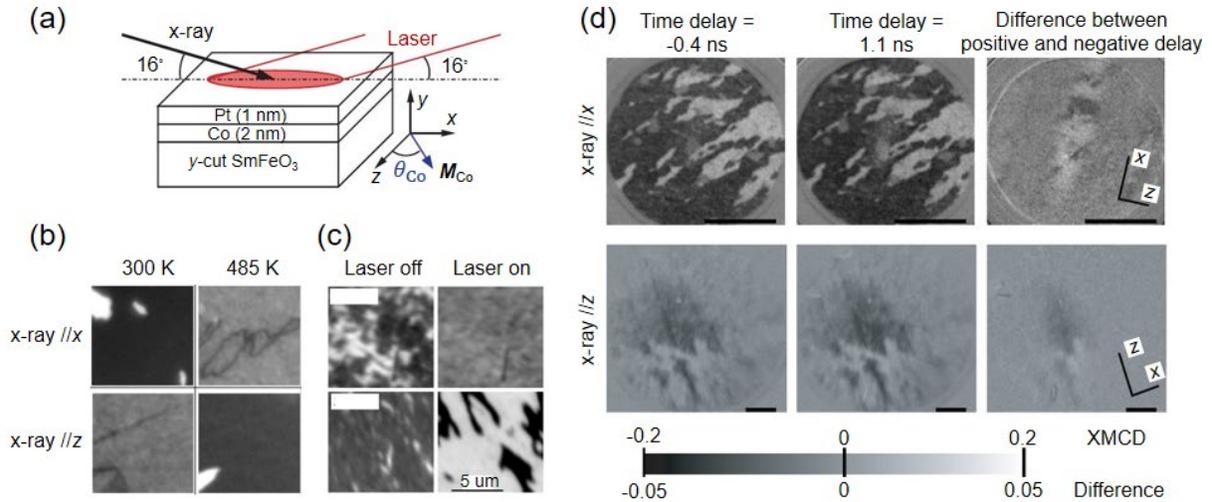

Figure 41. Correlated SRT in a Co/SmFeO$_3$ heterostructure.[303,304] (a) Experimental configuration. (b) Heating-induced SRT in the Co film probed by XMCD imaging. (c) Analogous maps to (b) except that heating is supplied by a laser beam. (d) Time-resolved XMCD images before and after a pulsed optical pump, and the pump-induced change obtained by a subtraction. Reproduced with permission from [303,304].

## 5.4 Nonlinear magnonics

In this Subsection, we examine novel phenomena that arise when the spin oscillation amplitude becomes so large that the assumption of a harmonic magnetic potential is no longer valid. In these cases, anharmonicity sets in, and in the second quantization formulism, it is equivalent to considering the magnon-magnon scattering as the number of magnons increases.[305] On the application side, nonlinear magnonics is crucial because it enables the control of a magnon wave by another magnon wave, which is the fundamental working principle to construct all-magnon data processing units such as magnon transistors.[306] In addition, nonlinearity during magnon propagation also facilitates the generation of magnon solitons and bullets,[307,308] opening up the possibility of long-distance magnon-based data transfer without distortion.[7]

In the past, exciting large-amplitude THz-frequency magnons in AFM materials such as $RFeO_3$ was challenging. However, the recent advances of various types of techniques for generating intense THz radiation have opened up new possibilities to perform an in-depth study of the subject; see Sub-subsection 2.2.2. We provide two examples below that demonstrate nonlinear magnonics in $RFeO_3$.

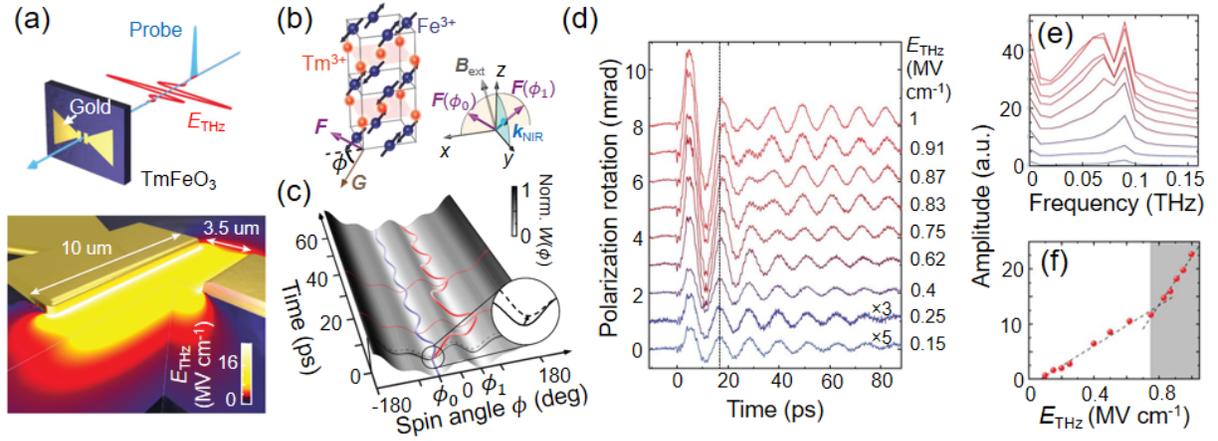

Figure 42. Nonlinearity of magnons as evidence for all-coherent spin switching.[309] (a) Experimental configuration. (b) Spin orientations in the $\Gamma_{24}$ phase and the two potential energy minima centered at $\phi_0$ and $\phi_1$. (c) Spin trajectories with (red curve) and without (blue curve) spin switching. Inset: modification to the potential due to an impulsive THz drive. (d) Fluence dependence of Faraday rotation transients. Dashed line highlights the phase shift. (e) Fourier transform of signals in (d). (f) Fluence dependence of the long-lived offset signal. Grey area marks the fluence range higher than the switching threshold. Reproduced with permission from [309].

One type of anharmonic potential is a potential with two local minima separated by a shallow barrier. Assuming that the equilibrated system is initially in a valley, increasing the magnon oscillation amplitude will lead to further and further excursions of the system from the local minimum, until the spins gain large enough energy to overcome the barrier and switch to the other valley. This suggests transitioning of the magnetic configuration to a metastable state, a phenomenon in which both spin switching effect and nonlinear magnonics manifest.

Schlauderer et al.[309] have demonstrated all-coherent spin switching in a double-well potential system in $TmFeO_3$, providing evidence for extreme nonlinearity of magnons. Being aware that the magnetic field component of THz pulses is too weak to excite large-amplitude magnons, the authors adopted the scheme discovered by Baierl et al.[243] (reviewed in Subsection 4.4), where THz-field driving of the electric-dipole-active crystal-field transitions of $Tm^{3+}$ supplies an anisotropy torque to $Fe^{3+}$ spins to generate large-amplitude quasi-FM magnons. A gold bowtie antenna was fabricated on the crystal to provide further electric field enhancement within the feed gap [Figure 42(a)]. The crystal was set at a temperature in the middle of the SRT (within the $\Gamma_{24}$ phase), where the equilibrium spin orientation should be such that the AFM vector is at an intermediate angle between the $x$-axis and the $z$-axis [Figure 42(b)]. There are four energy-equivalent domain types consistent with the configuration, but two were selected to be more favorable energetically as a magnetic field was applied, forming the two local minima of the desired double-well potential. With the angle between the AFM vector and the $x$-axis denoted by $\phi$, the two minima are expected to be centered at $\phi_0$ with $-\pi/2 < \phi_0 < 0$, and $\phi_1$ with $0 < \phi_1 < \pi/2$ [Figure 42(c)].

As Schlauderer et al.[309] increased the THz pump field strength and probed the magnon dynamics within the feed gap using time-resolved Faraday rotation measurements, they observed a beating pattern developing on the oscillation signal, accompanied by a phase flip [Figure 42(d)]. Fourier transform of the oscillations revealed an asymmetric splitting of the quasi-FM magnon mode, proving the existence of magnon nonlinearity. At the onset of peak splitting, the Faraday rotation signal also gained a long-lived offset, suggesting a spin-switching to a new metastable state. All these observations, being consistent with a microscopic model, are signatures of spin switching from $\phi_0$ to $\phi_1$ as the quasi-FM magnon amplitude surpasses a threshold at which the excursion of the system from equilibrium is large enough to overcome the barrier. They also engineered the height of the potential barrier through temperature or magnetic field tuning and observed behaviors consistent with the interpretation. Their demonstration of ballistic switching provides a scalable approach to switching magnetic bits with the least amount of energy dissipation.

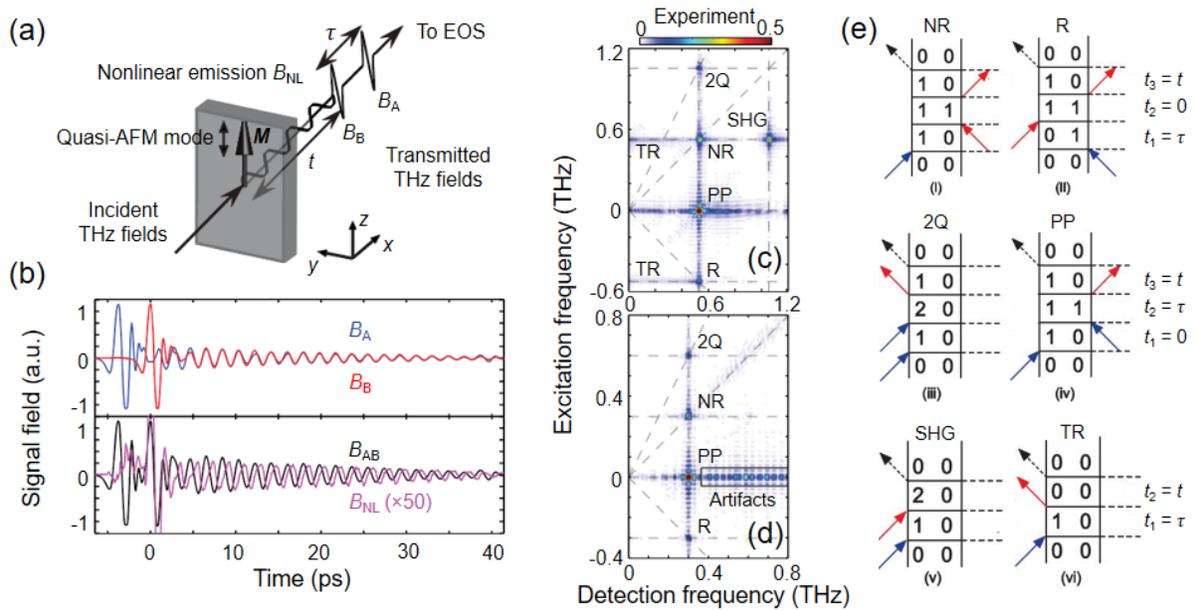

Figure 43. 2D coherent THz spectroscopy.[310] (a) Experimental configuration. (b) Various signal fields with either a single pulse or double pulses. The nonlinear signal $B_{NL}(t,\tau)$ is obtained by a subtraction process achieved in the experiment by a differential chopping detection method. (c) and (d) 2D magnitude spectra of $B_{AB}(f,v)$ for measurements focusing on the quasi-FM and quasi-AFM frequency ranges. (e) Double-sided Feynman diagrams showing the excitation pathways of various nonlinear processes marked in (c) and (d). Blue and red arrows represent interactions with pulses A and B, respectively. Black dashed arrow represents the measured emission field. Reproduced with permission from [310].

For systems that appear distant from a phase transition (and therefore the double-well model does not apply), magnon nonlinearity can also set in when spins sample a large enough phase space on the Bloch sphere during magnon oscillations. Two-dimensional (2D) coherent spectroscopy, which is a technique frequently used on bio-chemical molecular systems to analyze lattice anharmonicity and vibronic couplings, is an ideal tool to be applied here for studying spin anharmonicity. The difference, however, is that this method needs to operate in the THz frequency range, which is much less common than 2D infrared spectroscopy performed in the mid-infrared range.

Lu et al.[310] have used 2D THz coherent spectroscopy to reveal magnon nonlinearity in YFeO$_3$. As shown in Figure 43(a), the experiment was carried out by directing two intense THz pulses, labeled A and B, onto a YFeO$_3$ crystal and measuring the transmitted fields. If we denote the interpulse time delay by $\tau$, and the measurement time delay after the second pulse (pulse B) by $t$, the 2D time-domain nonlinear signal field $B_{NL}(t,\tau)$ is written as

$$B_{NL}(t,\tau) = B_{AB}(t,\tau) - B_{A}(t,\tau) - B_{B}(t,\tau), \tag{35}$$

where $B_{AB}(t,\tau)$ is the total field in the presence of both pulses, and $B_{A}(t,\tau)$ ($B_{B}(t,\tau)$) is the field with the pulse A (B) only. The subtraction in Equation (35) therefore cancels all signals of linear order and those which do not depend simultaneously on both pulses [Figure 43(b)]. A 2D Fourier transform of $B_{AB}(t,\tau)$ gives $B_{AB}(f,\nu)$, which is the standard 2D spectrum as a function of excitation frequency $\nu$ and detection frequency $f$. The magnitude spectra of $B_{AB}(f,\nu)$ for measurements focusing on the quasi-FM and quasi-AFM frequency ranges are shown in Figure 43(c) and (d), respectively.

A total of six types of peaks are resolved,[310] which can be ascribed to distinct nonlinear processes. The rephasing (R), nonrephasing (NR), pump-probe (PP), and two-quantum (2Q) peaks all arise from $\chi^{(3)}$ processes. A general 2D spectroscopy experiment probing $\chi^{(3)}$ processes requires three pump pulses. Since only two are used in this experiment, one needs to view one of the pulses here as the superposition of two (each with partial amplitude, and zero delay) to complete the analysis of the various possibilities of three-pulse interaction. In addition, $\chi^{(2)}$ processes that can be ascribed to second-harmonic generation (SHG) and THz rectification (TR) are also observed. All these nonlinear processes are graphically presented in terms of double-sided Feynman diagrams in Figure 43(e), and we will not describe each of them at length here since the diagrams contain all the essential information about the field-spin interaction sequence, and standard textbooks provide procedures to interpret them.[311]

Being able to resolve six types of nonlinear field-spin interactions is the unique capability afforded by 2D THz spectroscopy,[312–318] and multidimensional spectroscopy is sometimes viewed as "the ultimate ultrafast spectroscopy experiment." As more intense THz sources and more elegant ways to manipulate THz beams are being developed, this general technique is expected to see further development.[319] The traditional power of 2D spectroscopy in revealing anharmonic potentials and anharmonic couplings between resonances is broadly useful for a large array of exotic quantum magnets[320,321] to provide valuable insights on the fundamental physics of magnetism.

# 6   Ultrastrong cooperative magnetic coupling and the Dicke phase transition

In Subsection 5.2, we discussed a few experimental attempts to observe and analyze magnon-polaritons based on $RFeO_3$ systems. These are hybrid excitations formed by the strong coupling of THz magnons in $RFeO_3$ crystals with photons from the surrounding environment (cavity/waveguide/free space). When describing the light-matter coupling as being "strong," we referred to the criterion that the frequency splitting of the polariton modes are large enough to be resolved from the polariton linewidths,[73,322] so that at least one full cycle of light–matter energy exchange that arises from Rabi flopping can stand out from damping.

In this Section, we focus on another series of THz-frequency phenomena in $RFeO_3$ systems that are again closely related to both magnetism and light–matter coupling. There are two important aspects, however, that distinguish these phenomena from those in Subsection 5.2, and make them extraordinary. The first is that they reside in the *ultrastrong* coupling (USC) regime of light–matter interaction.[73,74] This is a regime where the coupling strength needs to not only overcome the decay rate but also become a sizable fraction of the natural frequencies of the uncoupled light and matter.[323–329] The quantitative argument translates to the fact that light and matter mix to an extreme degree, and unusual phenomena that are crucial for quantum applications are expected to arise.[330–333] Secondly, although the phenomena we are about to discuss reveal unprecedented information on light–matter coupled systems, they do not really require a standard polaritonic setup, in which a carefully designed photonic cavity should be fabricated to enclose a crystal supporting matter excitations. Instead, they are matter–matter coupled systems that are capable of simulating light–matter coupled Hamiltonians. The role cavity photons play in a standard polaritonic system is taken by a collective bosonic matter excitation, which ultrastrongly couples with another matter excitation in the same material.

These condensed-matter quantum simulators provide opportunities to reveal novel phenomena that are predicted for light–matter coupling Hamiltonians but have so far remained difficult to access.

To illustrate why RFeO3 can be such quantum simulators of light–matter coupling Hamiltonians in the USC regime, and what exact problems they address, we arrange our presentation below according to a designed sequence. We will first provide an introduction to the fundamentals of light–matter coupling models, and then describe the predictions related to the Dicke superradiant phase transition (SRPT) in the USC regime. A discussion will then follow addressing why a Dicke SRPT in equilibrium is difficult to achieve in genuine quantum-optical systems, and how $Er^{3+}$–$Fe^{3+}$ interactions in ErFeO3 is an ideal platform to simulate a magnonic SRPT in equilibrium. Finally, we will show that even a RFeO3 system without $R^{3+}$–$Fe^{3+}$ magnetic coupling, such as YFeO3, can simulate the anisotropic Hopfield model by forcing the quasi-FM and quasi-AFM modes to couple through a properly oriented external magnetic field. The unusual aspects associated with the counter-rotating term within the simulated Hamiltonian cause large-amplitude ground-state squeezing, which is expected to be useful in decoherence-free quantum applications.

*6.1 Ultrastrong light-matter coupling*

To describe the unusual physical phenomena in the USC regime of light–matter interaction, we first introduce the model that describes $N$ identical two-level atoms (with ground state $|g\rangle$ and excited state $|e\rangle$) interacting with a single-mode light field in a photonic cavity[334] [Figure 44 (a) and (b)]. The first-quantized Hamiltonian that describes the atomic ensemble reads $H_{at} = \sum_{j=1}^{N} \frac{\hat{p}_j^2}{2m} + \hat{U}_j$, where $\hat{p}_j$ is the electron momentum of the $j$-th atom, $m$ is the electron mass, and $\hat{U}_j$ is the potential energy. Coupling of the ensemble with the light field can be described by the Peierls substitution $\hat{p}_j \to \hat{p}_j - e\hat{A}(r_j)$, where $\hat{A}(r_j)$ is the vector potential at the atomic position $r_j$. This gives

$$H_{at} \to H_{at} + H_{int} + H_{A^2}. \tag{36}$$

The light–matter interaction term $H_{int}$ and the $A^2$-term $H_{A^2}$ read

$$H_{int} = -\sum_{j=1}^{N} \frac{e}{m} p_j \cdot A_0 (a^\dagger + a), \tag{37}$$

$$H_{A^2} = \sum_{j=1}^{N} \frac{e^2}{2m} |A_0|^2 (a^\dagger + a)^2,$$

where $A_0$ is the amplitude of $\hat{A}(r_j)$, and $a(a^\dagger)$ is the photon annihilation (creation) operator. These two terms can be rewritten in the second-quantized form

$$H_{int} = -i\hbar\Omega_0 (a^\dagger + a) b^\dagger + h.c., \tag{38}$$

$$H_{A^2} = \hbar D (a^\dagger + a)^2,$$

where the collective atomic excitation operator $b^\dagger$ is defined as

$$b^\dagger = \frac{1}{\sqrt{N}} \sum_{j=1}^{N} (|e\rangle\langle g|)_j, \tag{39}$$

the Rabi frequency as an indicator of the light-matter coupling strength is

$$\Omega_0 = \frac{\omega_{eg}}{\hbar} d_{eg} \cdot A_0 \sqrt{N}, \tag{40}$$

and the coefficient for the $A^2$-term is

$$D = (\Omega_0)^2 / \omega_{eg}. \tag{41}$$

Here $\omega_{eg}$ ($d_{eg}$) represents the frequency difference (transition dipole) between the ground state and the excited state of the two-level atoms.

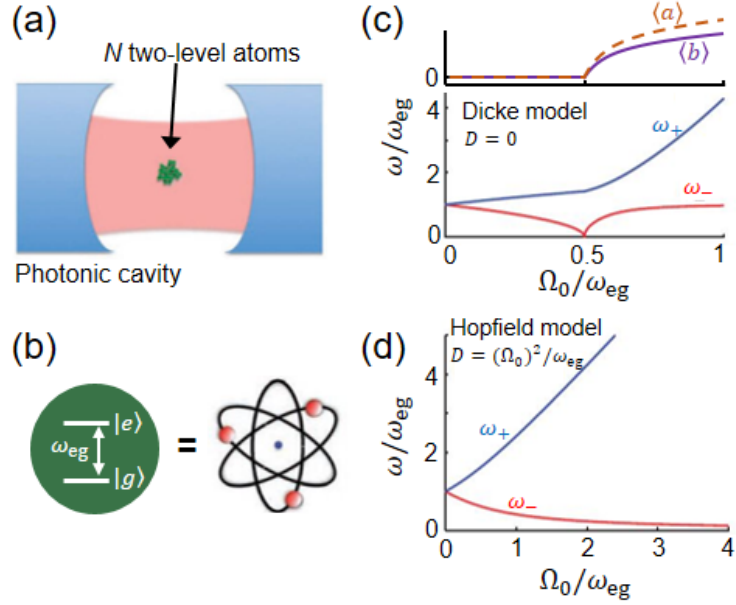

Figure 44. Light–matter interaction setup for studying the Dicke phase transition in the USC regime.[334] (a) $N$ atoms in a photonic cavity. (b) Diagram of an individual atom. (c) Static cavity and matter fields (upper)[335] and polariton frequencies versus coupling strength (lower) calculated from the Dicke model. (d) Same as the lower panel of (c), but for the Hopfield model (full Hamiltonian retaining the $A^2$-term). Reproduced with permission from [334].

By writing the atomic Hamiltonian as $H_{at} = \hbar\omega_{eg} b^\dagger b$ and the photonic Hamiltonian as $H_{cav} = \hbar\omega_{cav} a^\dagger a$ ($\omega_{cav}$ is the frequency of the photonic mode), and using the expressions in Equation (38), we obtain the total Hamiltonian as the sum

$$H = H_{cav} + H_{at} + H_{int} + H_{A^2}. \tag{42}$$

There are a few important aspects about this Hamiltonian. The first is the appearance of terms proportional to $(ab^\dagger - a^\dagger b)$ and $(a^\dagger b^\dagger - ab)$ in the interaction Hamiltonian $H_{int}$; the former are known as the co-rotating terms and the latter as the counter-rotating terms. Co-rotating terms correspond to resonant absorption ($ab^\dagger$) and emission ($a^\dagger b$) processes, while the counter-rotating terms either create ($a^\dagger b^\dagger$) or annihilate ($ab$) two excitations simultaneously, and therefore, correspond to nonresonant processes that do not conserve the excitation number. Second, when the light–matter coupling strength is much smaller than the energy spacing of the two-level systems ($\Omega_0 / \omega_{eg} \ll 1$), it is legitimate to neglect the counter-rotating terms in $H_{int}$ (known as the rotating-wave approximation, RWA) and the entire $A^2$-term $H_{A^2}$ (since it appears higher order in terms of $\Omega_0 / \omega_{eg}$), making the remainder of the Hamiltonian fully diagonalizable. These approximations are not valid in the USC regime, in which $\Omega_0$ becomes so large that $\Omega_0 / \omega_{eg} \sim 1$. Third, the $\Omega_0 / \omega_{eg} \sim 1$ criterion for the USC regime is usually considered to be demanding to realize experimentally, but the appearance of the $\sqrt{N}$ enhancement factor in Equation (40) provides an opportunity to boost $\Omega_0$ simply by increasing the number of atoms in the cavity. The $\sqrt{N}$ enhancement points to the collective nature of $\Omega_0$, since it suggests that $N$ atoms cooperatively contribute to the coupling so that the ensemble can be viewed as one giant combined oscillator whose dipole is $\sqrt{N}$ times stronger than individual dipoles.[289,324,336] This effect is reminiscent of the collective radiative decay of atoms predicted by Dicke's model of superradiance,[337] and thus, is named Dicke cooperativity. Finally, based on Dicke cooperativity, a number of solid-state systems which host numerous numbers of oscillators (large $\sqrt{N}$) with huge dipole moments (large $d_{eg}$) have been used to reach the USC regime;[290,324,336] comprehensive reviews on this topic exist.[73,74]

One can then proceed to analyze the total Hamiltonian in Equation (42) to discover the unusual phenomena associated with the USC regime. In 1973, Hepp and Lieb[338] as well as Wang and Hioe[339] studied the thermodynamic properties of the Dicke model, using a simplified Hamiltonian adopting the RWA and dropping the $A^2$-term. They discovered a second-order phase transition at large enough $\Omega_0$ and low enough temperature, across which a static photonic field and an atomic polarization appear simultaneously. This is a significant result known as the Dicke superradiant phase transition (SRPT). Despite its original prediction as a finite-temperature classical transition, the model was shown in the zero-temperature limit to host a quantum phase transition as one increases $\Omega_0$; see the zero-temperature expectation values of photon and matter operators plotted versus $\Omega_0$ in Figure 44(c) and the order parameter onset at $\Omega_0 / \omega_{eg} = 0.5$.

The SRPT can be understood as the outcome of critical mode softening. Starting from the total Hamiltonian in Equation (42), eigenfrequencies of the model can be obtained by a Hopfield-Bogoliubov transformation[334] to convert the Hamiltonian into the diagonalized form

$$H = \hbar \sum_{i=\pm} \omega_i P_i^\dagger P_i + \text{Const.}, \tag{43}$$

where $P_i$ are polaritonic operators ($i = \pm$, representing upper or lower polariton) given by

$$P_i = u_i^{\text{ph}} a + u_i^{\text{el}} b + v_i^{\text{ph}} a^\dagger + v_i^{\text{el}} b^\dagger, \tag{44}$$

satisfying the bosonic commutation relations $[P_+, P_+^\dagger] = [P_-, P_-^\dagger] = 1$ and $[P_+, P_-^\dagger] = [P_-, P_+^\dagger] = 1$.

The coefficient vectors $(u_i^{ph}, u_i^{el}, v_i^{ph}, v_i^{el})^T$ and the polaritonic frequencies $\omega_i$ are, respectively, the eigenvectors and eigenvalues of the matrix

$$M = \begin{pmatrix} \omega_{cav} + i2D & -i\Omega_0 & -2D & -i\Omega_0 \\ i\Omega_0 & \omega_{eg} & -i\Omega_0 & 0 \\ 2D & -i\Omega_0 & -(\omega_{cav} + i2D) & -i\Omega_0 \\ -i\Omega_0 & 0 & i\Omega_0 & -\omega_{eg} \end{pmatrix}. \quad (45)$$

Polariton frequencies calculated by using this method assuming $D = 0$ (neglecting $A^2$-term) and $\omega_{cav} = \omega_{eg}$ (zero detuning) are shown in the lower panel of Figure 44(c). As $\Omega_0$ increases, the coupling makes the polariton branches to move apart from each other. The energy gap is linear versus $\Omega_0$ initially, but the softening of the lower polariton branch $\omega_-$ accelerates upon further increasing $\Omega_0$, until it softens to zero frequency at $\Omega_0 / \omega_{eg} = 0.5$; this is the exact point where a quantum phase transition into the SRPT is achieved.

However, the conspicuous flaw in the Hamiltonian assumed by the early proposal of the SRPT was the RWA and the neglect of the $A^2$-term, since they, as we have mentioned, are not simplifications that can be made when $\Omega_0 / \omega_{eg} \sim 0.5$, even though such a large coupling strength is needed for the transition to occur. In particular, when the $A^2$-term $H_{A^2}$ is included, calculations show that the lower polariton frequency would only approach but never reach zero [Figure 44(d)], no matter how large $\Omega_0$ becomes. This means that the SRPT should be forbidden for the full model; the underlying reason can be traced back to the oscillator strength sum rule and the gauge invariance.[340] Later, the assumed conditions for proving the inaccessibility of an SRPT are clarified, which are reduced to the statement that a SRPT in thermal equilibrium can never be achieved for minimally coupled (Peierls substituted) systems under the dipole approximation.[341] This is known as the no-go theorem for the SRPT, which appears in the discussion of later works.[342–344]

Because the unique ground state afforded by the SRPT is appealing, a question then arises asking whether it is possible to circumvent the no-go theorem. A variety of proposals have been put forward, aiming to breach at least one of the assumptions used to derive the theorem. A nonequilibrium scheme of laser-driven atoms to mimic the Dicke model (without the $A^2$-term) was proposed and later demonstrated experimentally.[345,346] Early proposals recommended inclusion of the spin degree of freedom in the light–matter interaction.[347] A few studies have searched for certain superconducting circuit diagrams for which the oscillator strength sum rule does not apply.[334,348] More recent studies aim to harness spatially-varying electric fields in multimode cavities to achieve light–matter coupling beyond the dipole approximation.[342,349]

*6.2 Magnonic superradiant phase transition in ErFeO$_3$*

Despite the large variety of proposals of carefully designed light–matter coupled systems aiming at invalidating the no-go theorem, an experimental demonstration of the SRPT in thermal equilibrium has remained elusive. This, on the one hand, calls for more quantum-controllable experimental systems that allow easier detection of potential superradiant ground states,[350–352] but on the other hand, suggests the necessity of a more overturning strategy that falls outside the scope of existing proposals.

Since the essence of the Dicke model is the coupling of an ensemble of two-level systems with a boson field, the boson field being a photonic mode confined in a cavity, it would be interesting to think whether the model can be simulated by a near-equivalent quantum system where another collective boson field in matter, such as magnons, plays the role of photons, to couple to the two-level systems. The advantage of such a mapping would be two-fold. The first is that the matter boson field (that plays the role of photons) and the two-level systems can both reside in one material system, and a photonic cavity is not required. This significantly improves the experimental accessibility for simultaneous probing of order parameters in both sectors across a potential SRPT. Second, when magnons are used to replace photons, one could use electron spins as two-level systems, and the resulting spin-magnon coupling is ensured to neither be minimally coupled (since the charge degree of freedom is not involved) nor be describable in any sense by the dipole approximation (since the dominant interaction is the exchange interaction). This implies that both of the two necessary conditions for the no-go theorem are likely to be violated in a spin–magnon interaction system and therefore opens up a unique possibility to achieve a SRPT in thermal equilibrium.

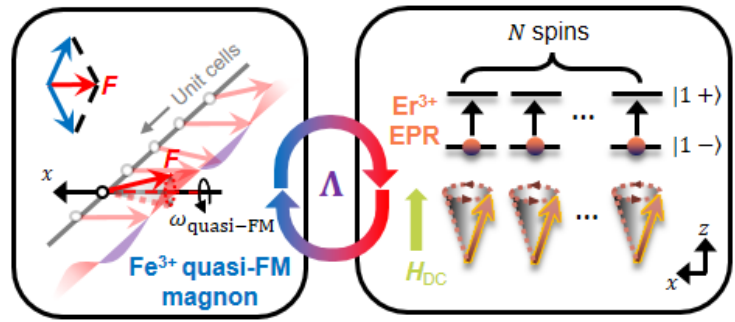

Figure 45. Spin–magnon interaction in ErFeO$_3$.[169] Quasi-FM oscillations of Fe$^{3+}$ are resonantly coupled with the EPR of $N$ Er$^{3+}$ spins at a rate of $\Lambda$. Reproduced with permission from [169].

RFeO$_3$ crystals are ideal spin–magnon interacting systems that can be used as the desired quantum simulator of the Dicke Hamiltonian. Taking ErFeO$_3$ as an example, the situation is depicted in Figure 45. The quasi-FM magnon mode of Fe$^{3+}$ plays the role of the cavity photonic mode in the Dicke model, while Er$^{3+}$ ions have the appropriate energy level structure to be treated as two-level systems. Within a low-symmetry crystal field, spin-orbit coupled states of $Er^{3+}$ ions ($4f^{11}$) split into Kramers doublets (Subsection 3.2). In the low temperature limit, only the ground state doublet $|1-\rangle$ and $|1+\rangle$ would be relevant as thermal excitation cannot populate the higher doublets; these states mimic a two-level atom with $|g\rangle = |1-\rangle$ and $|e\rangle = |1+\rangle$. The transition between $|1-\rangle$ and $|1+\rangle$ is the electron paramagnetic resonance (EPR) of $Er^{3+}$, whose frequency is tunable by an external magnetic field. Because $Fe^{3+}$ and $Er^{3+}$ are known to be coupled by exchange interactions,[80] $Fe^{3+}$ magnons and $Er^{3+}$ EPR will couple; the coupling strength, denoted by $\Lambda$ here, is equivalent to $\Omega_0$ in Equation (38).

To ensure that the spin–magnon interaction in ErFeO3 is a faithful simulator of the Dicke model, Li et al.[169] have performed an experiment to search for Dicke cooperativity, a key trait in the original Dicke Hamiltonian that enables a $\sqrt{N}$-fold scale-up of $\Omega_0$. Figure 46(a) shows THz absorption spectra for a z-cut ErFeO3 crystal at 45 K ($\Gamma_2$ phase) at various external magnetic fields ($H_{DC} \parallel z$). The constant-frequency line at 0.39 THz is assigned to be the $Fe^{3+}$ quasi-FM and the line that increases linearly with $H_{DC}$ is the $Er^{3+}$ EPR. At lower temperatures, anticrossings begin to develop around the zero-detuning magnetic fields where the $Fe^{3+}$ magnon and $Er^{3+}$ EPR are close in frequency [Figure 46(b)]; the frequency splitting at zero detuning is quantitatively proportional to $\Lambda$, the spin–magnon coupling strength. With decreasing temperature, it is observed that anticrossings become more prominent, suggesting that $\Lambda$ increases [Figure 46(b)-(e)]. Furthermore, at a constant temperature, doping the crystal with nonmagnetic $Y^{3+}$ ($Er_xY_{1-x}FeO_3$) is also able to tune $\Lambda$, with $\Lambda$ decreasing as the $Er^{3+}$ concentration $x$ decreases [compare Figure 46(c) with (f) and (i), (d) with (g) and (j), (e) with (h) and (k)].

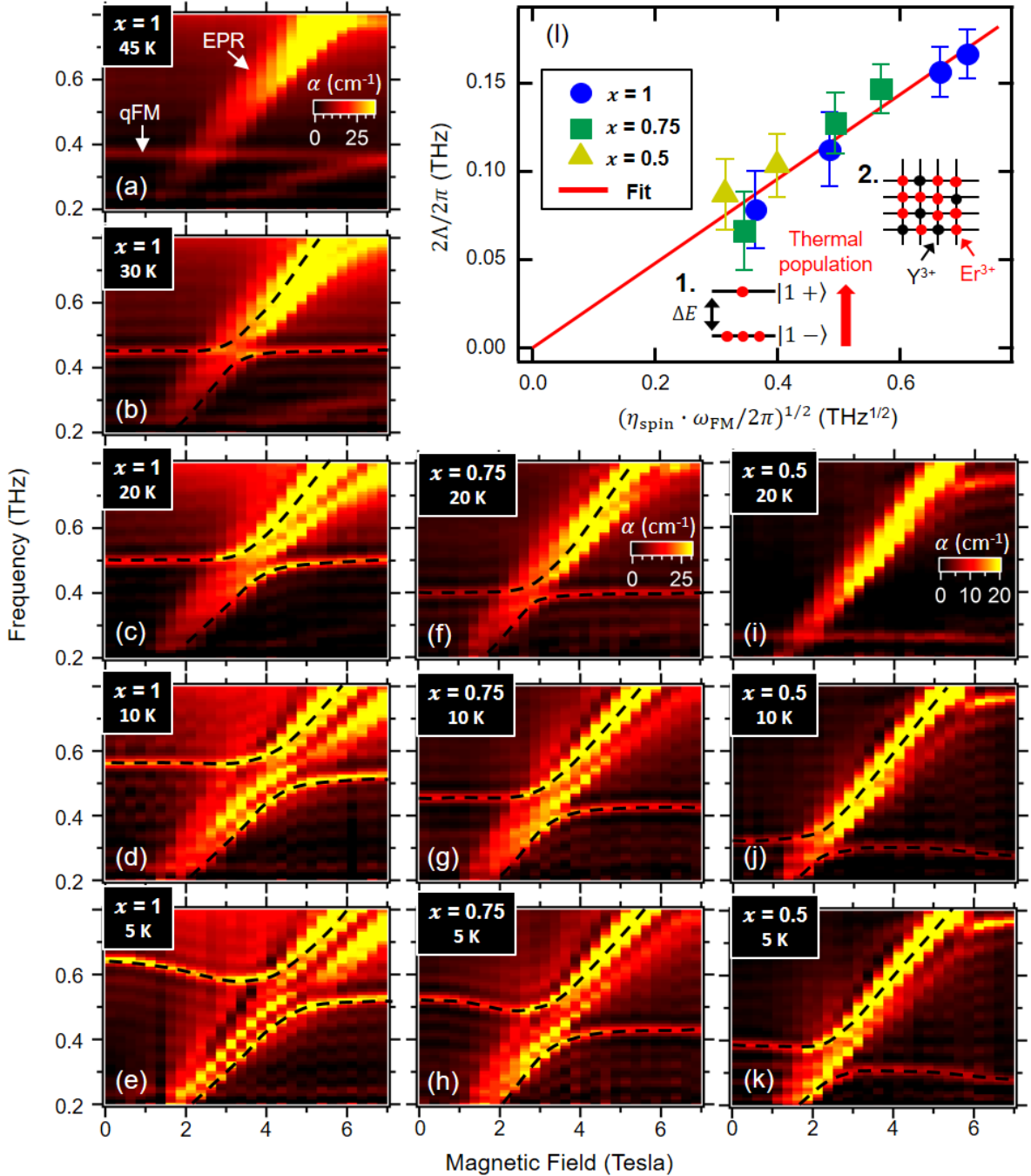

Figure 46. Evidence for Dicke cooperativity in magnetic interactions.[169] (a) to (k) THz absorption spectra of ErFeO$_3$ crystal under a $z$-oriented external magnetic field at various temperatures and Y$^{3+}$ doping levels. (l) The coupling rate shows proportionality with the square root of Er$^{3+}$ density. Inset shows two types of mechanisms that tunes the effective spin density of Er$^{3+}$. Reproduced with permission from [169].

Li et al.[169] interpreted the tunability of $\Lambda$ both by temperature and by $Y^{3+}$ doping as an indicator of Dicke cooperativity, because, as discussed below, both experimental control knobs are effectively tuning the $Er^{3+}$ spin density $n_{spin}$. As shown in the inset of Figure 46(l), temperature tuning changes the thermal distribution of $Er^{3+}$ within its $|1-\rangle$ and $|1+\rangle$ states; the lower the temperature, the more spins populate the lower $|1-\rangle$ state, which effectively increases the number of excitable $Er^{3+}$ spins according to $\tanh(-\Delta E / k_B T)$, where $\Delta E$ is the energy separation and $k_B T$ is the thermal energy. Second, in experiments using $Er_xY_{1-x}FeO_3$ ($x$ being $Er^{3+}$ concentration) crystals, nonmagnetic $Y^{3+}$ doping simply reduces the $Er^{3+}$ density through dilution without changing the crystal and magnetic structure of the sample. Putting these facts together, the $Er^{3+}$ spin density that can contribute to the spin–magnon coupling is subject to a correction factor $\eta_{spin} = x \cdot \tanh(-\Delta E / k_B T)$, which contains both temperature and doping information. Indeed, when one plots $\Lambda$ versus $\sqrt{\eta_{spin}}$, the data points from all temperature- and doping-dependent measurements fall onto a single line that passes through the origin, evidencing the $\Omega_0 \propto \sqrt{N}$ relation of Dicke cooperativity; see Figure 46(l), and note the additional $\sqrt{\omega_{FM}}$ factor, where $\omega_{FM}$ is the quasi-FM frequency, that comes from the vacuum field amplitude of magnons.

The observation of Dicke cooperativity suggests that $Er^{3+}$ spins as two-level systems are cooperatively coupled with the $Fe^{3+}$ magnons in $ErFeO_3$. Furthermore, the extracted $\Lambda$ from Figure 46(k) occupies a considerable fraction of the magnon/EPR resonance frequency, proving that the spin–magnon coupling is in the USC regime. In addition, as discussed before, the model is known to violate a few key premises used to derive the no-go theorem. All of these add favorably to the necessary conditions to achieve an SRPT in thermal equilibrium.

What would an SRPT for a spin–magnon coupled system look like then? Spontaneous appearance of a static light field and an atomic polarization in the light–matter coupled superradiant ground state straightforwardly translates to spin ordering in a spin–magnon coupled system. And just like how ordering sets in both sectors that participate in the coupling in the Dicke model, a SRPT in $ErFeO_3$ would be a cooperative spin ordering involving both the $Er^{3+}$ and $Fe^{3+}$ subsystems. Within the well-known temperature-dependent magnetic phase diagram of $ErFeO_3$ (Figure 3), one indeed can identify a phase transition ($\Gamma_2 \rightarrow \Gamma_{12}$) at 4 K, known as the low-temperature phase transition (LTPT), across which $Er^{3+}$ spins develop AFM order along the $z$ axis and the AFM vector of the $Fe^{3+}$ spins rotates in the $y$-$z$ plane; see Figure 47(a) and (b).

Bamba et al.[353] have theoretically demonstrated that the LTPT in $ErFeO_3$ is in fact a SRPT for a spin–magnon coupled system. The term "magnonic SRPT" was given to the phenomenon to distinguish it from the standard light–matter coupled SRPT. A spin model Hamiltonian taking into account the $Fe^{3+}$ subsystem, the $Er^{3+}$ subsystem, and the $Fe^{3+}$–$Er^{3+}$ spin interactions was established:

$$H = H_{Fe} + H_{Er} + H_{Fe\text{-}Er}. \tag{46}$$

The $Fe^{3+}$ Hamiltonian, written as

$$H_{Fe} = \sum_{s=A,B} \sum_{i=1}^{N_0} \mu_B \hat{\boldsymbol{S}}_i^s \cdot \boldsymbol{g}^{Fe} \cdot \boldsymbol{B}^{DC} + J_{Fe} \sum_{n.n} \hat{\boldsymbol{S}}_i^A \cdot \hat{\boldsymbol{S}}_{i'}^B - D_y^{Fe} \sum_{n.n} (\hat{S}_{i,z}^A \hat{S}_{i',x}^B - \hat{S}_{i',z}^B \hat{S}_{i,x}^A) \tag{47}$$
$$- \sum_{s=A,B} \sum_{i=1}^{N_0} [A_x (\hat{S}_{i,x}^s)^2 + A_z (\hat{S}_{i,z}^s)^2 + A_{xz} \hat{S}_{i,x}^s \hat{S}_{i,z}^s],$$

includes (in the order given in the equation) the field–spin Zeeman interaction, the isotropic exchange interaction summed over the nearest neighbors (n.n), the antisymmetric DM interaction summed over the n.n, and the single-ion anisotropy. The treatment follows the two-sublattice ($s = A$ and $B$ within a unit cell) description first developed by Herrmann[107] (same as Equation (4), but added by the Zeeman term), where $\hat{S}_i^s$ represents the $Fe^{3+}$ spin operator for sublattice $s$ within the $i$-th unit cell, a total of $N_0$ unit cells are considered, $\mu_B$ is the Bohr magneton, $g^{Fe}$ is the g-factor tensor, $\boldsymbol{B}^{DC}$ is the external static magnetic field, $J_{Fe}$ ($D_y^{Fe}$) is the symmetric (DM) exchange constant, and $A_x$, $A_z$, and $A_{xz}$ are anisotropy energies. The $Er^{3+}$ Hamiltonian

$$H_{Er} = \sum_{s=A,B} \sum_{i=1}^{N_0} \frac{1}{2} \mu_B \hat{\boldsymbol{R}}_i^s \cdot g^{Er} \cdot \boldsymbol{B}^{DC} + J_{Er} \sum_{n.n} \hat{\boldsymbol{R}}_i^A \cdot \hat{\boldsymbol{R}}_{i'}^B, \tag{48}$$

consists of field–spin Zeeman interaction and the isotropic $Er^{3+}$–$Er^{3+}$ interactions. Here, the $Er^{3+}$ spins are modeled to have two sublattices in each unit cell ($s = A$ and $B$) described using vectors of Pauli operators $\hat{\boldsymbol{R}}_i^s = \hat{\boldsymbol{\sigma}}_i^s \equiv (\hat{\sigma}_{i,x}^s, \hat{\sigma}_{i,y}^s, \hat{\sigma}_{i,z}^s)^t$, $g^{Er}$ is the $Er^{3+}$ g-factor tensor, $J_{Er}$ is the $Er^{3+}$–$Er^{3+}$ exchange constant. Finally, the $Fe^{3+}$–$Er^{3+}$ interaction term, which is responsible for the spin–magnon interaction in the Dicke-like Hamiltonian, reads

$$H_{Er\text{-}Fe} = \sum_{s,s'=A,B} \sum_{i=1}^{N_0} [J \hat{\boldsymbol{R}}_i^s \cdot \hat{\boldsymbol{S}}_i^{s'} + \boldsymbol{D}^{s,s'} \cdot (\hat{\boldsymbol{R}}_i^s \times \hat{\boldsymbol{S}}_i^{s'})], \tag{49}$$

where the interaction is considered to be closed in each unit cell, and $J$ and $\boldsymbol{D}^{s,s'}$ are the isotropic and antisymmetric $Er^{3+}$–$Fe^{3+}$ exchange constants, respectively. Symmetry properties of the crystal reduces $\boldsymbol{D}^{s,s'}$ so they can be described by only two scalars $D_x$ and $D_y$ as

$$\begin{aligned}
\boldsymbol{D}^{A,A} &= (D_x, D_y, 0)^t, \\
\boldsymbol{D}^{A,B} &= (-D_x, -D_y, 0)^t, \\
\boldsymbol{D}^{B,A} &= (-D_x, D_y, 0)^t, \\
\boldsymbol{D}^{B,B} &= (D_x, -D_y, 0)^t.
\end{aligned} \tag{50}$$

By using Equations (46)-(50), one is able to obtain a complete picture of the thermodynamic properties of the system. Mean-field calculations correctly predict the order parameter onsets of both the $Er^{3+}$ and $Fe^{3+}$ subsystems across the LTPT; see Figure 47(c) and (d) for the thermal averaged values of $Er^{3+}$ spins $\bar{\sigma}^{A/B}$ and $Fe^{3+}$ spins $\bar{S}^{A/B}$ versus temperature at zero magnetic field. The fact that $\bar{\sigma}_z = \bar{\sigma}_z^A = -\bar{\sigma}_z^B$ becomes finite suggests the development of collinear AFM order in the $Er^{3+}$ subsystem along the $z$ axis, while the exchange of weight between $\bar{S}_z = -\bar{S}_z^A = \bar{S}_z^B$ and $\bar{S}_y = \bar{S}_y^A = -\bar{S}_y^B$ suggests rotation of the $Fe^{3+}$ AFM vector within the $y$-$z$ plane; these match the experimental phenomenology of the LTPT. Furthermore, mean-field calculations with $\boldsymbol{B}^{DC} \parallel x$ give the phase diagram shown in Figure 47(e), in which the color intensity encodes the AFM order parameter within the $Er^{3+}$ sector. Application of a magnetic field destabilizes the low-temperature $\Gamma_{12}$ phase by restoring the system back to the $\Gamma_2$ phase; this again agrees with previous experimental findings.[91,354]

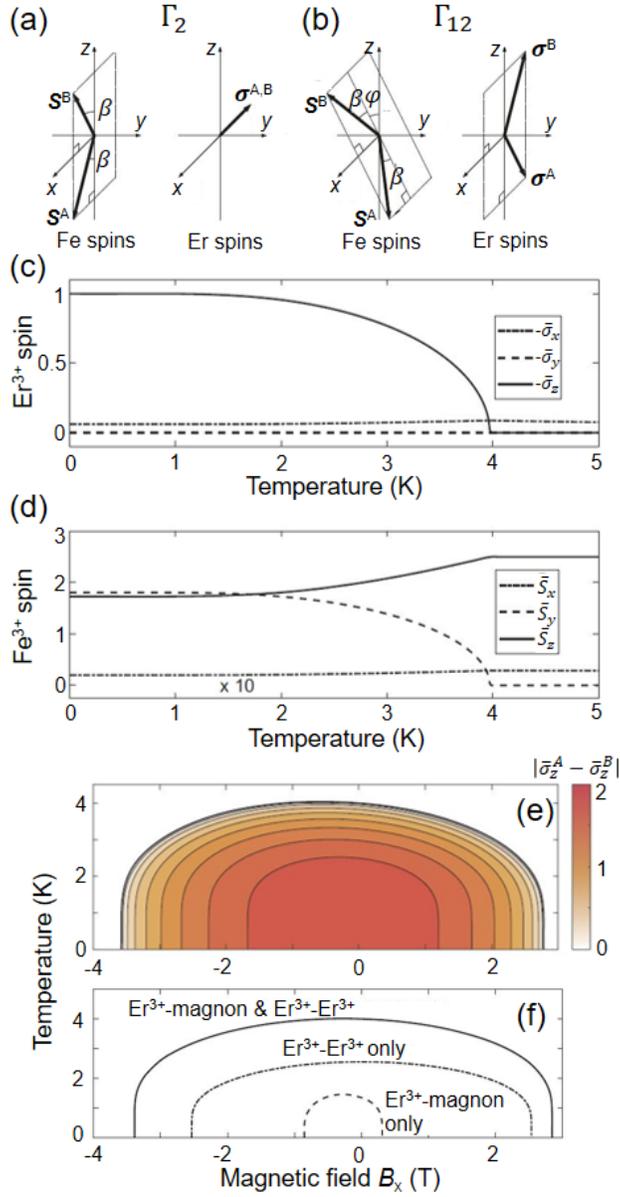

Figure 47. Magnonic superradiant phase transition in $ErFeO_3$.[353] (a) and (b) The $\Gamma_2 \to \Gamma_{12}$ LTPT, across which $Er^{3+}$ develops AFM order and $Fe^{3+}$ spins rotate in the *b-c* plane. Averaged spin components of (c) $Er^{3+}$ and (d) $Fe^{3+}$ versus temperature calculated via the mean-field method. (e) Temperature–field phase diagram for $\boldsymbol{B}^{DC} \parallel x$, highlighting the phase boundary of LTPT, calculated by the mean-field method. (f) Phase boundaries calculated from the extended Dicke Hamiltonian, by selectively including certain terms in the Hamiltonian. Solid: both the $Er^{3+}$–magnon and $Er^{3+}$–$Er^{3+}$ interactions are included. Dash-dotted: only the $Er^{3+}$–$Er^{3+}$ interaction is included. Dashed: only the $Er^{3+}$–magnon interaction is included. Reproduced with permission from [353].

To establish the relation of the spin model of $ErFeO_3$ with the Dicke model in the context of quantum optics and SRPT, Bamba et al.[353] rewrote the overall spin Hamiltonian Equations (46)-(50) into the second quantized form:

$$H = \sum_{K=0,\pi} \hbar\omega_K \hat{a}_K^\dagger \hat{a}_K + E_x\hat{\Sigma}_x^+ + \sum_{\xi=x,y,z} g_\xi^{Er} \mu_B B_\xi^{DC} \hat{\Sigma}_\xi^+ + \frac{4z_{Er}J_{Er}}{N_0} \hat{\Sigma}^A \cdot \hat{\Sigma}^B \qquad (51)$$

$$+ \frac{\sqrt{2}\hbar g_x}{\sqrt{N_0}}(a_\pi^\dagger + a_\pi)\hat{\Sigma}_x^+ + \frac{i\sqrt{2}\hbar g_y}{\sqrt{N_0}}(a_0^\dagger - a_0)\hat{\Sigma}_y^+ + \frac{\sqrt{2}\hbar g_{y'}}{\sqrt{N_0}}(a_\pi^\dagger + a_\pi)\hat{\Sigma}_y^-$$

$$+ \frac{i\sqrt{2}\hbar g_z}{\sqrt{N_0}}(a_\pi^\dagger - a_\pi)\hat{\Sigma}_z^- + \frac{\sqrt{2}\hbar g_{z'}}{\sqrt{N_0}}(a_0^\dagger + a_0)\hat{\Sigma}_z^+.$$

The two $Fe^{3+}$ magnons with frequency $\omega_K$ are written as creation ($\hat{a}_K^\dagger$) and annihilation ($\hat{a}_K$) operators, with $K = 0, \pi$ denoting the FM and AFM modes, respectively. $E_x = h \times 0.023\,THz$ is the zero-field splitting of the $Er^{3+}$ EPR mode. $\hat{\Sigma}^{A/B} \equiv \frac{1}{2}\sum_{i=1}^{N_0} \hat{R}_i^{A/B}$ is the large-spin operator for $Er^{3+}$, with $\hat{\Sigma}^\pm \equiv \hat{\Sigma}^A \pm \hat{\Sigma}^B$. $g_x$, $g_y$, $g_{y'}$, $g_z$, and $g_{z'}$ are coefficients for terms that couple one of the $Fe^{3+}$ magnon fields with a $Er^{3+}$ large-spin component, and can be calculated by a material parameter set already known through prior experiments. Estimate of energy scales enables certain coupling terms to be dropped, but the coupling term led by the $g_z$ coefficient, the first term on line 3 of Equation (51), is the most important because it relates the $Fe^{3+}$ quasi-AFM magnon ($a_\pi^\dagger - a_\pi$) with the z-axis AFM vector ($\hat{\Sigma}_z^-$) of the $Er^{3+}$ spins, which develop spontaneous order across an LTPT. Equation (51) formally resembles the Dicke Hamiltonian in quantum optics, but it extends the quantum optical model in complexity because it involves multiple magnon modes (as opposed to the single-mode description in the Dicke model) and the magnon–spin interactions are anisotropic; Bamba et al.[353] named it an *extended, multi-mode, and anisotropic Dicke model*. Furthermore, the key feature of Equation (51) is that the $A^2$-term does not appear for spin–spin exchange interactions, making the model possess significant potential to host a magnonic SRPT.

To investigate whether the LTPT is a magnonic SRPT, Bamba et al.[353] adopted a semiclassical method with the extended, multi-mode, and anisotropic Dicke model, which fully reproduced the thermodynamic properties predicted by the mean-field calculations. Furthermore, they switched on and off certain terms in the Hamiltonian to distinguish impacts from different terms on the phase boundary of the LTPT. As shown in Figure 47(f), the LTPT temperature remains finite even when the $Er^{3+}$–$Er^{3+}$ exchange interaction is eliminated, suggesting that the $Er^{3+}$–$Fe^{3+}$ interaction alone is able to cause the LTPT. While the $Er^{3+}$–$Er^{3+}$ interaction energy scale ($J_{Er}$) is identified to play a significant role in describing the $Er^{3+}$ AFM phase boundary, it is only by including the $Er^{3+}$-$Fe^{3+}$ interaction can one describe simultaneous ordering of both $Er^{3+}$ and $Fe^{3+}$ across the LTPT. These results determine the nature of the LTPT to be magnonic SRPT, representing the first-time demonstration of the long-sought SRPT in thermal equilibrium.

Building upon the extended, multi-mode, and anisotropic Dicke Hamiltonian [Equation (51)], Marquez Peraca et al.[355] have discovered an additional field-induced phase emerging from the low-temperature phase diagram when an external magnetic field is applied along the z axis of the $ErFeO_3$ crystal. The $\Gamma_{12}$ phase, described as the magnonic superradiant phase, is denoted as the "S" phase, while the normal $\Gamma_2$ phase is denoted as the "N" phase. As shown in Figure 48, for temperatures above 2.5 K, an increase of the magnetic field simply shrinks the S phase boundary in a similar fashion to the $\boldsymbol{B}^{DC} \parallel x$ case shown in Figure 47(e). However, when the system is cooled down below 2.5 K, an intermediate phase, labeled "I", emerges. This phase exhibits a strong $Er^{3+}$ AFM order parameter $\bar{\sigma}_x = \bar{\sigma}_x^A - \bar{\sigma}_x^B$ along the x axis of the crystal, as opposed to the strong $\bar{\sigma}_z$ component in the S phase. Computation of the free energy landscape suggests that the S-to-I phase transition is a first-order spin-flop transition where the ordered $Er^{3+}$ spins abruptly switch their easy axis from z to x as the magnetic field increases. If the field increases further, an I-to-N transition occurs, and the transition is shown to be second-order. Low-temperature magneto-spectroscopy data was found to be in agreement with the theoretical phase diagram. The appearance of the I phase from the extended Dicke Hamiltonian highlights the richness of the magnon SRPT model that goes beyond the standard Dicke model in quantum optics.

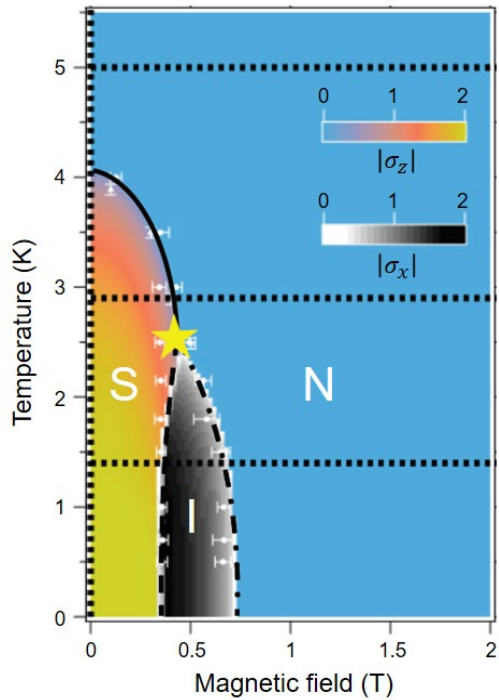

Figure 48. Temperature-field phase diagram for $\boldsymbol{B}^{DC} \parallel z$.[355] Between the normal "N" phase and the superradiant "S" phase, a new intermediate "I" phase emerges. The S-to-I transition is first-order. The S-to-N and I-to-N transitions are both second-order.

*6.3 Quantum simulation of the anisotropic Hopfield model and two-mode vacuum squeezing*

In addition to the magnon SRPT, RFeO$_3$ systems have been applied in a broader context as quantum simulators of cooperative light–matter Hamiltonians. Recently, Makihara *et al.*[356] have used YFeO$_3$ to demonstrate the unusual properties of magnon squeezing that arises from an anisotropic Hopfield Hamiltonian. Here, the word "anisotropic" has a different meaning from the anisotropic Dicke model of ErFeO$_3$ mentioned in Subsection 6.2, which describes the direction-dependent coupling of Er$^{3+}$ moments with Fe$^{3+}$ magnons. In YFeO$_3$, Y$^{3+}$ is nonmagnetic, so it would not couple to Fe$^{3+}$ magnons. Instead, the "anisotropy" here refers to the fact that the coupling strengths characterizing the co-rotating terms and the counter-rotating terms in the interaction Hamiltonian are different. Following this definition, the standard Dicke Hamiltonian [Equation (38)] is isotropic because the co-rotating terms $(ab^\dagger - a^\dagger b)$ and the counter-rotating terms $(a^\dagger b^\dagger - ab)$ share the same coupling coefficient $\Omega_0$.

Makihara *et al.*[356] achieved the anisotropic Hopfield Hamiltonian by applying a magnetic field misaligned from the crystal axis of YFeO$_3$ and harnessing the resultant ultrastrong coupling between the quasi-FM and quasi-AFM modes of the Fe$^{3+}$ spins. As shown in Figure 49(a), an external magnetic field is applied at an intermediate angle $\theta$ between the $y$ axis and the $z$ axis, and a THz pulse whose magnetic field polarizes along the $x$ axis propagates through the crystal to probe the magnon frequencies and magnon–magnon coupling. The external field required to observe sizable magnon–magnon coupling should be as large as 30 T, which was supplied by a unique table-top pulsed magnet[124,357] coupled with a broadband optical cryostat.

We first explain qualitatively why magnon–magnon coupling is induced by a misaligned magnetic field. At zero field, or finite fields aligned along the $z$ axis ($\theta = 0$), the two-fold rotation around the $z$ axis (C$_{2z}$) is a symmetry operator of the crystal ($\Gamma_4$ phase). The operator $C_{2z}$ is therefore within the group of the Schrödinger equation, and any eigen-mode of the spin Hamiltonian must be characterized as either even or odd under the operation. The quasi-FM (quasi-AFM) mode is odd (even) under $C_{2z}$. Magnon–magnon coupling would not occur at $\theta = 0$ since they are both eigen-modes that diagonalize the Hamiltonian. When $\theta \neq 0$, the magnetic field causes a tilt in the spin structure that breaks $C_{2z}$. The quasi-FM and quasi-AFM modes, which according to definition should still retain their transformation properties under $C_{2z}$, are no longer well-defined eigen-modes, and thus would couple to one another. Makihara et al.[356] worked out the mathematics and showed that, when the equation of motion is written using the basis of the FM vector $\boldsymbol{F} = \boldsymbol{S}^A + \boldsymbol{S}^B$ and the AFM vector $\boldsymbol{G} = \boldsymbol{S}^A - \boldsymbol{S}^B$, where $\boldsymbol{S}^A$ and $\boldsymbol{S}^B$ are the two $Fe^{3+}$ sublattices [Equation (1) and (3)], the time dynamics of the fluctuations of the $\boldsymbol{F}$ and $\boldsymbol{G}$ vectors, $\delta \boldsymbol{F}$ and $\delta \boldsymbol{G}$, obey

$$\begin{bmatrix} \delta \dot{F}_x \\ \delta \dot{F}_y \\ \delta \dot{G}_x \\ \delta \dot{G}_y \end{bmatrix} = 2\gamma \sin \beta_z \begin{bmatrix} 0 & 2A_y & D_{yx} & 0 \\ -2A_x & 0 & 0 & -D_{xy} \\ D_{xy} & 0 & 0 & 2B_y \\ 0 & -D_{yx} & -2B_x & 0 \end{bmatrix} \begin{bmatrix} \delta F_x \\ \delta F_y \\ \delta G_x \\ \delta G_y \end{bmatrix}, \quad (52)$$

where $\gamma$ is the gyromagnetic ratio, $\beta_z$ is the static tilt angle of $\boldsymbol{S}^{A/B}$ from the a-b plane, and $A_x$, $A_y$, $B_x$, $B_y$, $D_{xy}$, and $D_{yx}$ are coefficients calculated from spin model parameters. The 4 by 4 matrix can be divided into four blocks of 2 by 2 matrices. Blocks involving $A_x$, $A_y$, $B_x$, and $B_y$ only couple components within $\delta \boldsymbol{F}$ or $\delta \boldsymbol{G}$, while off-diagonal blocks that contain $D_{xy}$ and $D_{yx}$ (which are functions of $\theta$) couple $\delta \boldsymbol{F}$ with $\delta \boldsymbol{G}$.

As shown in Figure 49(b), as $\theta$ progressively increases from 0 to $\pi/2$, the modes identified by experiment show stronger repulsions, that is, the coupled modes (red solid, labeled by UM and LM) show larger frequency separation from the uncoupled quasi-FM and quasi-AFM modes (black dashed). The uncoupled modes are calculated by setting $D_{xy} = D_{yx} = 0$ in Equation (52).

Counter-intuitive behavior emerges for $\theta = \pi/2$, where the upper mode (UM) becomes red-shifted, rather than blue-shifted from the coupled mode; see the red shaded region, which is not expected for the standard Hopfield Hamiltonian [Figure 44(c) and (d)]. Makihara et al.[356] pointed out that the reason is that the spin Hamiltonian, when mapped to the Hopfield Hamiltonian,

$$H = \hbar\omega_{0a}(\hat{a}^\dagger\hat{a} + \frac{1}{2}) + \hbar\omega_{0b}(\hat{b}^\dagger\hat{b} + \frac{1}{2}) + i\hbar g_1(\hat{a}\hat{b}^\dagger - \hat{a}^\dagger\hat{b}) + i\hbar g_2(\hat{a}^\dagger\hat{b}^\dagger - \hat{a}\hat{b}), \tag{53}$$

is anisotropic. Here, $\hat{a}$ ($\hat{a}^\dagger$) and $\hat{b}$ ($\hat{b}^\dagger$) are annihilation (creation) operators for the decoupled quasi-FM and quasi-AFM magnons, respectively, and $g_1$ and $g_2$ are coupling coefficients for the co-rotating and counter-rotating terms, respectively. The red-shifted UM originates from an exceptionally large $g_2$, for which the polariton frequencies are dominated by the vacuum Bloch–Siegert shift[325] that arises from the counter-rotating terms of the Hamiltonian. Analytical calculations of $g_1$ and $g_2$ versus field for various $\theta$ values are plotted in Figure 49(c). For all $\theta \neq 0$, $g_2 > g_1$, and $g_2 - g_1$ becomes more prominent when $\theta$ approaches $\pi/2$ and at stronger fields, which corroborates the observation of the anomalous frequency shifts in Figure 49(b). Figure 49(d) shows $g_1$ and $g_2$ versus $\theta$ at the zero-detuning magnetic fields, again exhibiting consistency with this understanding.

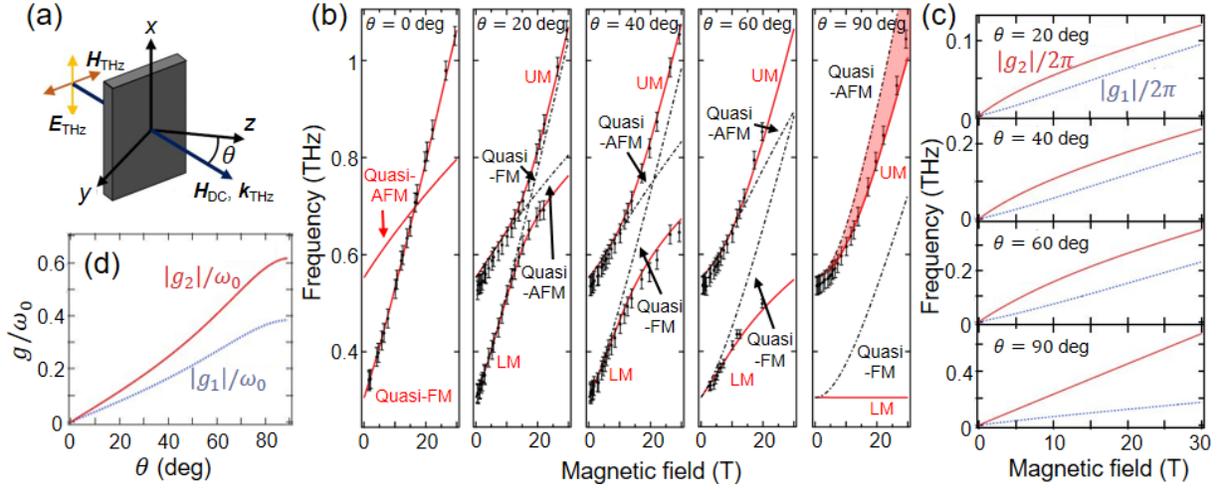

Figure 49. YFeO$_3$ as a quantum simulator of the anisotropic Hopfield Hamiltonian.[356] (a) Experimental geometry. (b) Experimental mode frequencies (with error bars) versus magnetic field for various tilt angles $\theta$. Theoretical mode calculations using the full Hamiltonian (red solid) and decoupled Hamiltonian (black dashed) are overlaid. (c) $g_1$ and $g_2$ versus field for various $\theta$. (d) $g_1$ and $g_2$ versus $\theta$ at the zero-detuning magnetic field. Reproduced with permission from [356].

Achieving the anisotropic Hopfield Hamiltonian [Equation (53)] possessing tunable $g_1$ and $g_2$ with $g_2 > g_1$ is a unique finding of this work, and is expected to open up opportunities to search for more exotic quantum phenomena. In particular, the exceptionally large counter-rotating terms will amplify the two-mode vacuum squeezing of the coupled ground state. Makihara et al.[356] defined a generalized magnon annihilation operator from $\hat{a}$ and $\hat{b}$

$$\hat{c}_{\phi,\psi} = \hat{a}\cos\phi + e^{i\psi}\hat{b}\sin\phi, \tag{54}$$

and a quadrature by this operator as

$$\hat{X}_{\phi,\psi,\varphi} = (\hat{c}_{\phi,\psi}e^{i\varphi} + \hat{c}^{\dagger}_{\phi,\psi}e^{-i\varphi})/2, \tag{55}$$

where $\phi$, $\psi$, and $\varphi$ are phase factors. With the help of the Hopfield-Bogoliubov transformation, the variance $\langle 0|(\hat{X}_{\phi,\psi,\varphi})^2|0\rangle$ with respect to the coupled ground state $|0\rangle$ was minimized to find the optimal $\phi$, $\psi$, and $\varphi$. They found that the fluctuation of the quadrature can be reduced by up to 5.9 dB at 30 T for $\theta = \pi/2$, which is a direct consequence of the large counter-rotating terms under these experimental conditions.

Building upon the same line of theory, Hayashida et al.[335] further discovered perfect intrinsic vacuum squeezing at a SRPT critical point. They started from a standard Dicke Hamiltonian, performed the Holstein-Primakov transformation, and defined the two-mode annihilation operator and the associated quadrature in a similar way to Equations (54) and (55). They tuned $\phi$, $\psi$, and $\varphi$ to minimize the quadrature for every light–matter coupling strength $\Omega_0$, and plotted the optimized quadrature versus $\Omega_0/\omega_{eg}$ as in Figure 50. At the SRPT critical point, the $\Omega_0/\omega_{eg} = 0.5$ point at which the lower-polariton soften to zero-frequency, the optimized quadrature vanishes, suggesting perfect squeezing. The single-mode quadrature associated with $\hat{X}_{0,0,\pi/2}$ also exhibits suppressed fluctuations, despite being incomplete. The product of the optimized quadrature with its orthogonal always gives ¼, satisfying the uncertainty principle.

The two-mode vacuum squeezing is useful for applications such as quantum metrology and decoherence-robust quantum information technology.[358] Its realization in Dicke-like models in thermal equilibrium (i.e., *instrinsic* squeezing) presents unique advantage compared to standard photon squeezing protocols, which usually rely on out-of-equilibrium approaches.[359,360] The squeezing should also be resilient against thermal fluctuations at finite temperatures. Future research awaits to explore these unprecedented functionalities in more depth.

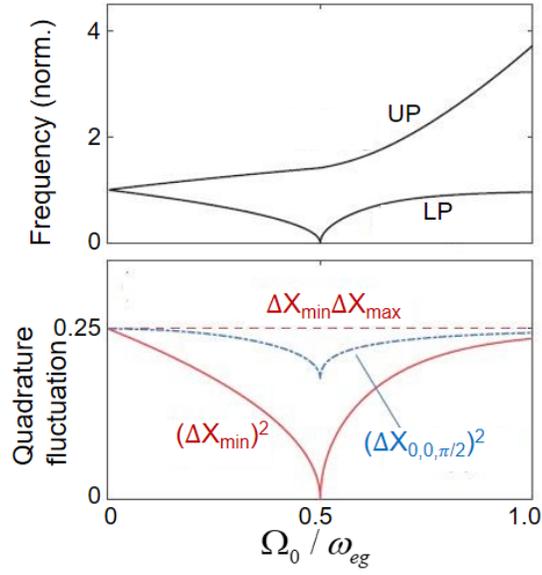

Figure 50. Perfect intrinsic squeezing at a SRPT critical point.[335] Minimized quadrature fluctuation (red) versus coupling strength shows perfect suppression at the SRPT transition point. Upper panel shows complementary polariton mode frequencies.

# 7 Summary and outlook

In this article, we reviewed THz-frequency phenomena associated with spin dynamics in the $RFeO_3$ class. We first discussed how one can use THz radiation to probe magnetic phase transitions in equilibrium. While the polarization selection rule of acoustic magnons is a useful indicator of spin reorientation transitions, electromagnons associated with dynamic magnetoelectric coupling appear as a unique class of excitation in noncentrosymmetric low-temperature magnetic phases. We then reviewed the scenario where laser radiation drives the system far from equilibrium, and a few emergent microscopic pathways for laser manipulation of magnetic order. Further, for antiferromagnetic spintronic applications, the ability to achieve coherent quantum control over magnon dynamics is important. Thus, we reviewed a variety of protocols to manipulate coherent magnons in time and space. Finally, new insights into the connection between dynamic magnetic coupling in condensed matter systems and ultrastrong light–matter coupling in the setting of cavity quantum electrodynamics were provided; this sets the stage for novel approaches of manipulating condensed matter phases through tools and concepts in quantum optics.

In the introduction section we mentioned that an advantage of presenting a comprehensive survey on various spintronic phenomena related to a single material class is that the connections between various physical phenomena can be identified more easily. Looking into the future, it is exactly this type of connection we seek that may enable the community to invent new protocols for active spin control and quantum phase engineering. For example, a direction emerging very recently explores the marriage of nonlinear phononics (Subsection 4.3) and nonlinear magnonics (Subsection 5.4), where multi-dimensional THz spectroscopy can be utilized to precisely characterize the strength of anharmonic phonon–phonon,[314] spin–spin,[319,361] or even spin–phonon interactions.[276] Secondly, a variety of analogies can be drawn from the idea of Floquet engineering of electronic structures (Subsection 4.5) and the notion of ultrastrong light–matter coupling in cavity quantum materials (Subsection 6.1),[72,292] since both investigate coherent light–matter interaction Hamiltonians; the former relies on external pump light while the latter on the cavity vacuum field. Being equipped with the knowledge of magnonic superradiant phase transitions, one might be interested in engineering unusual light-dressed magnetic states with quantum controllable counter-rotating interactions (Subsections 6.2 and 6.3). The third idea arises from the connection between electromagnons (Subsection 3.3) and the rare-earth pathway of laser manipulation of $Fe^{3+}$ order (Subsection 4.4). Electromagnons are only starting to be understood by using THz spectroscopic probes recently, but they have not been the subject of strong-field driving. However, since such an excitation unambiguously involves both the $Fe^{3+}$ and $R^{3+}$ subsystems (because symmetry can only be broken by $R^{3+}$ ordering within the matrix of $Fe^{3+}$ order), nonperturbative driving of this mode by intense THz radiation can potentially be even more efficient in controlling the $Fe^{3+}$ order than the method of anisotropy torque created by driving the crystal-field transitions of $R^{3+}$.

We believe that the above three examples are merely the tip of the iceberg of potential connections one can make within the existing demonstrations of THz-frequency spintronic phenomena in $RFeO_3$. We hope that the same ideology can apply to a wider material class, to trigger a broader range of interdisciplinary efforts for fostering the field of THz spintronics and ultrafast condensed matter physics.

**Disclosures**

None declared.

**Acknowledgements**

X.L. acknowledges support from the Caltech Postdoctoral Prize Fellowship and the IQIM.